\documentclass[final,authoryear,1p,times]{elsarticle}

\usepackage[utf8]{inputenc}
\usepackage{graphicx}    
\usepackage{amsmath,bm}  
\usepackage{amssymb}     
\usepackage{geometry}    
\usepackage{setspace}    
\usepackage{hyperref}    
\usepackage{float}
\usepackage{caption}
\usepackage{subcaption}
\usepackage[normalem]{ulem}
\usepackage[noabbrev]{cleveref}

\usepackage{xcolor}

\usepackage[justification=raggedright, textfont=it,font=normalsize]{caption}

\usepackage{tikz}
\usetikzlibrary{shapes.geometric, arrows.meta, positioning}
\usepackage{natbib}

\begin{document}

\begin{frontmatter}

\title{Averaging Molecular Dynamics simulations to study the slow-strain-rate behavior of metals}

\author[iitd]{Sarthok Kumar Baruah}
\ead{amy237541@am.iitd.ac.in}
\author[iitd]{Sabyasachi Chatterjee\corref{cor1}}
\ead{sabyasachi@am.iitd.ac.in}
\cortext[cor1]{Corresponding Author}
\author[cmu+na]{Amit Acharya}
\ead{acharyaamit@cmu.edu}
\author[cmu]{Gerald J. Wang}
\ead{geraldw@andrew.cmu.edu}

\address[iitd]{Department of Applied Mechanics, Indian Institute of Technology Delhi, Hauz Khas, New Delhi 110016, India}
\address[cmu]{Department of Civil and Environmental Engineering, Carnegie Mellon University, Pittsburgh, PA 15213}
\address[cmu+na]{Department of Civil and Environmental Engineering \& Center for Nonlinear Analysis, Carnegie Mellon University, Pittsburgh, PA 15213}
\address{}

\begin{abstract} 
The application of molecular dynamics (MD) simulations to quasi-static loading is severely limited by the large separation between atomic vibration timescales and experimentally relevant deformation rates. Even approaches aimed at accessing long timescales in MD, such as those involving Potential Energy Surface (PES) exploration and alteration, become prohibitively expensive with the increase in number of atoms, particularly for slow-rate applications. In this work, we employ the Practical Time Averaging (PTA) framework to overcome this limitation and enable atomistic simulations of crystalline solids under quasi-static loading conditions. PTA exploits the intrinsic separation of timescales by defining slow variables as time-averaged observables of the fast atomistic dynamics and their evolution in the slow loading timescale, thereby avoiding explicit integration of the fast dynamics. 

Using this approach, we simulate uniaxial deformation, in both tension and compression, of ($4$ to $20$) nanometer sized cubic specimens of a metallic nanocrystal (face-centered cubic Aluminum utilized here as a specific choice) at applied strain rates approaching quasi-static conditions ($10^{-4}s^{-1} - 10^{-3}s^{-1}$). We define slow variables as the averaged kinetic energy, potential energy and normal stress in the loading direction, and show their evolution in the slow time scale. The stress-strain curves show yield close to the theoretical yield stress for homogeneous nucleation, followed by successive load drops and rise, caused due to dislocation nucleation, motion and exit from free surfaces. The "smaller is harder" effect is evident from the stresss-strain response as well as from the variation of yield stress with the sample size. The serrations in the response are more pronounced for smaller samples. The effects of applied strain rate and initial temperature are studied, and the difference between results of lattice statics and PTA-MD quasi-static simulations is demonstrated. 
The PTA framework enables simulations at strain rates several orders of magnitude lower than those accessible to conventional MD, demonstrating significant speedup in computer time, while retaining full atomistic resolution. To the best of our knowledge, this is the first study in the literature involving a large scale MD system to study the mechanical response and underlying plastic deformation in metals under slow applied strain rate, without any alteration of the PES of the system. 
\end{abstract}



\end{frontmatter}


\section{Introduction}

Molecular dynamics (MD) simulations have become a powerful tool in understanding material behavior. However, a key difficulty in using MD for engineering applications is due to the extremely large separation between the timescales of atomic vibrations (on the order of femtoseconds) and the timescales of applied loading (on the order of seconds). A long-standing limitation in the use of molecular dynamics (MD) simulation is that it can only be applied directly to processes that take place on very short timescales. Many important processes in chemistry, physics and materials science take place on time scales that cannot be reached by molecular dynamics, which is limited to nanoseconds (or a few microseconds for very small systems). This restricts MD to extremely high strain rates of $10^8\,\text{s}^{-1}$ to $10^{10}\,\text{s}^{-1}$. 

Application of MD is therefore not feasible in many problems of practical interest such as materials under slow loading rates (such as quasi-static tension test), evolution of defects in materials such as voids and bubbles, twinning, phase-transformation, protein folding and many others. Radiation damage in structural materials used in nuclear reactors involves the evolution of an isolated collision cascade over picosecond (ps) time scales followed by long-term evolution of such defects which include both annihilation and aggregation. In order to effectively predict radiation-damage evolution, longer time-scale behavior of such defects must be simulated. Another potential application is investigating thin-film deposition and crystal growth where deposition events take place in the order of picoseconds but the time to next deposition is in the order of seconds. Most MD simulations employ deposition rates which are $10^8 - 10^{11}$ orders of magnitude higher than experimental values. However, thermally activated atomic processes with rates as low as one per second can have significant effects on thin-film microstructures. The high atom deposition rate required for conventional MD simulations cannot realistically model such processes and thus, alternative methods must be used. In the specific case of metal plasticity, dislocation motion is characterized by stick-and-slip motion where dislocations are stuck for long periods of time (with respect to MD time-scale) followed by sudden slip. 

Several MD studies have been made on the effect of size and strain rate on the mechanical behavior of materials like \cite{kabir2024size, CHANG2017348, yu2013study, wan2021size, komanduri2001molecular, vogl2021effect, uchic2004sample, van2003quantifying}.
In almost all the above studies, simulating material behavior at much slower strain rates still remains a significant challenge. Thus, traditional MD methods become computationally infeasible at such slow rates because of the unrealistically large number of time steps required to reach the characteristically large observed deformations.

Numerous techniques have been developed to address the challenge of accessing long timescales in MD simulations, which we briefly review here. A popular class of methods -- including hyperdynamics \cite{Voter1997}, metadynamics \cite{Laio2002}, Gaussian accelerated MD \cite{Miao2015}, the activation--relaxation technique (ART) \cite{Barkema1996}, and the dimer method \cite{Henkelman1999} -- broadly involves modifying the energy landscape to accelerate exploration of phase space. Of particular interest here is the Autonomous Basin Climbing (ABC) method, originally introduced by Yip and co-workers as a potential-energy-surface (PES)-exploration framework for atomistic processes occurring on timescales inaccessible to conventional molecular dynamics \cite{Kushima2009}. In this approach, the system is driven from one local minimum to a neighboring basin through the successive application of penalty functions; the associated barrier is then refined, typically using Nudged Elastic Band (NEB), the transition time is estimated through transition-state theory (typically introducing further assumptions, most notably harmonicity), and the configuration is advanced to the next loading state. For slow strain-rate problems, the strictly strain-rate-controlled formulation reviewed by  \cite{Yan2016Review} applies this procedure under a prescribed rate, such that deformation is represented as a sequence of activated barrier-crossing events while maintaining explicit control over the imposed strain rate. This framework is attractive in that it connects atomistic barrier crossing directly to the applied loading rate; however, it also becomes increasingly demanding in the slow-rate regime, since the PES is strain-dependent and must be re-explored after each increment, so that many successive searches may be required before appreciable deformation is accumulated \cite{Yan2016Review}. Of particular practical importance, the computational cost of ABC rapidly becomes prohibitive as the number of atoms (and thus the dimensionality of the PES) increases, due to the large number of penalty functions that must be added to the PES. In practice, as noted by the authors \cite{YanSharma2016}, this particular curse of dimensionality renders unrealistic the study of systems in excess of $\mathcal O(10^3)$ atoms, even with the use of modern computing resources. The review highlights applications of this strictly strain-controlled framework to dislocation--defect interactions in HCP Zr \cite{Fan2013PNAS} and to the strain controlled compression of a metallic Ni nanoslab/nanopillar \cite{YanSharma2016}; notably, the latter system was a two-dimensional nanoslab only a few nanometers in dimension and comprised around 400 atoms, studied down to strain rates of order $1\,\mathrm{s^{-1}}$. In contrast, in this work, we study a method that avoids this curse of dimensionality by focusing on a (substantially lower-dimensional) collection of macroscopic slow variables and only requires intermittent periods of traditional MD with an \emph{unmodified} PES.

Another broad class of methods -- including replica exchange MD \cite{Sugita1999} and temperature-accelerated dynamics \cite{Sorensen2000, Zamora2016} -- achieves this accelerated exploration by performing simulations at temperatures higher than the temperature of interest. A large number of approaches fall under the umbrella of pathway-focused methods. Such methods focus on sampling along specific reaction coordinates, with less or no effort invested in sampling in directions orthogonal to the chosen ones (see, e.g., reviews in \cite{Bolhuis2021, Mohr2024}). 

Coarse-graining spatial degrees of freedom is yet another common approach to access long timescales, covering an enormous number of methods (see, e.g., reviews in \cite{Shi2023, Noid2024}). The method discussed herein shares the most conceptual similarity with existing techniques that explicitly (or implicitly) leverage the separation of timescales between fast and slow dynamics, most notably Mori-Zwanzig-based methods (\cite{Izvekov2006, Hijn2010,mielke2025deriving}), methods based on adiabatic elimination of fast dynamics (\cite{Berezhkovskii2011}), or mathematical homogenization for Hamiltonian systems, see, e.g., (\cite{bornemann1997homogenization,bornemann1998homogenization,klar2021second}).

Each existing method presents certain drawbacks, which the method discussed herein avoids, either altogether or at least in ways that differ from the existing method. The energy-landscape-modification and elevated-temperature methods all fundamentally alter transition rates between states (and typically require some knowledge of and, even more typically, assumptions about the underlying kinetics in order to correct for these altered transition rates). Moreover, for mechanically driven problems, particularly under strain-controlled loading, repeated explicit exploration of the potential-energy surface is not always necessary and can become prohibitively expensive, since the underlying PES itself evolves with each load increment. Pathway-focused methods typically require prior knowledge of and/or assumptions about productive reaction coordinates along which to sample. Current methods driven by separation of timescales mentioned in the previous paragraph typically require a) expensive computation of and/or assumptions about a memory kernel that links the fast and slow dynamics or b) an assumption of the fast dynamics in question settling on an equilibrium point in phase space for fixed slow variables (adiabatic elimination or the Tikhonov scheme (see, e.g., \cite{artstein2002singularly, chatterjee2018computing}), which is not satisfied by MD (especially NVE) systems or c) the satisfaction of delicate resonance conditions (\cite[discussion of Takens chaos]{bornemann1998homogenization}, \cite{neishtadt2019mechanisms}).  Spatial coarse-graining has a long history of successes and challenges; since these methods act on spatial degrees of freedom, most of these methods could likely be integrated with the approach developed herein in a relatively straightforward manner.

The separation of timescales of atomic vibrations in MD and applied slow loading leads to singularly perturbed forms of the evolution equations. In this context, the coarse-graining scheme named Practical Time Averaging (PTA), originating in \cite{slemrod2012time, AS06} and given definitive form in \cite{chatterjee2018computing},  
 was developed to understand the behavior of nonlinear systems on a time scale much slower than that of the intrinsic dynamics. The technique deals with the averaging of \emph{singularly perturbed} differential equations, which involve a small parameter representing the ratio between the fast and slow fundamental time periods involved and the goal of PTA is to develop a tool to model the limiting behavior as the parameter tends to zero. For small values of the parameter, the direct simulation of the underlying nonlinear dynamics becomes infeasible due to the restriction on the time-step. Hence, instead of evolving the fast dynamics, \emph{slow} variables were introduced as averages of the state function of the fast dynamics. The scheme also provides the evolution of \emph{slow} variables, which gives a measurement of the underlying intrinsic fast dynamics. In \cite{chatterjee2018computing} a further improvement of the scheme is developed, describing a procedure to obtain the initial conditions of the fast trajectory of the state at a time in the future on the slow time scale, not accessible using a direct calculation with the fast dynamics. They also demonstrated the application of the scheme to problems that involve both conservative as well as dissipative microscopic dynamics such as slowly evolving fast oscillations, exponential decay, and even sharp jump (i.e., fast behavior) in the evolution of the \emph{slow} variable. 

The work of \cite{tan2014modeling} used long-(fast)time averages of state variables of the fine dynamics as slow variables at fixed slow loads, which can also be interpreted as the first moment of a fine state function w.r.t.~the invariant measure of the fine dynamics at the slow time that the load is fixed (see eqn.~(55) of that work). A primary result of the works \citep{slemrod2012time, chatterjee2018computing} is that such averages (see \eqref{eq:R_m_t} in this work)  are not guaranteed to execute slow response, even in slow time intervals where the Young measure of the fast dynamics evolves continuously. In this work, we improve on this important aspect, and work only with mathematically guaranteed slow variables - which require another layer of averaging (see \eqref{eq:v_t} in this work) - corresponding to fast MD, and systematically use their extrapolation rules (the coarse evolution equations) to evolve the slow state variables, following the theory and algorithmic procedures of \cite{chatterjee2018computing}. A two-dimensional lattice made of Nickel–Manganese undergoing detwinning and a three-dimensional atomic chain made of face-centered cubic (FCC) Nickel under uniaxial tension was studied in \cite{tan2014modeling}. Rough, in slow-time, macroscopic stress-strain curves with averages consistent with generic observed behavior for the systems involved were shown using a Lennard--Jones potential and relatively small system sizes of only about 1500 atoms. Significant time savings compared to conventional MD was observed. The reported tensile response of the atomic chain was correspondingly limited to necking and fracture, and the study was not aimed at more realistic three-dimensional metallic systems or at dislocation-mediated plastic deformation. In another work, \cite{chatterjee2020plasticity} used PTA to time-average fast Dislocation Dynamics (DD) and use the resulting slow-variables to replace constitutive phenomenological assumptions in Mesoscale Field Dislocation Mechanics (MFDM) continuum model of plasticity \cite{acharya_roy_2006_1,roy_acharya_2005,roy_acharya_2006_2,rarora_2020_jmps,rarora_2020_ijss}. The mechanical response of macroscopic samples at slow loading rates up to moderately large strains was computed with significant savings in computing time compared to conventional DD.

The objective of the present work is to demonstrate the application of PTA to more complex and realistic engineering problems. The system we consider is a molecular dynamical system of FCC Aluminum crystal undergoing uniaxial tension and compression. This is the first application of PTA (or, possibly, any MD based method) to three-dimensional FCC nanocuboids up to $20$ nanometers in side length and comprising up to millions of atoms, under \emph{quasi-static loading rates and up to appreciable values of strain}. We define \emph{slow} variables of interest and evolve them on the slow time-scale of applied loading. We also provide visualization of instantaneous snapshots of a measure of time-averaged atomic positions along the overall slow trajectories followed by our calculations. Of particular focus in this work is  understanding the effect of the underlying dislocation mediated plastic deformation at these length scales on the overall stress-strain response of the system under quasi-static loading. The results are found to be different from those obtained from lattice static calculations under the same loading protocol. 

\textit{To our knowledge, this is the first demonstration in the literature of the calculation of up to a half-million atom MD assembly under (physically-relevant) slow applied strain-rate loading protocols by a technique that, furthermore, does not alter the PES of the system.}


This paper is organized as follows. In Section 2, we discuss the PTA numerical scheme and algorithm. In Section 3, we discuss the problem setup. In Section 4, we show the results of our work followed by conclusion in Section 5. 

\section{Methodology}

In this section, we will discuss the PTA scheme and algorithm. A detailed discussion can be found in \cite{chatterjee2018computing}. Here, we provide a summary which includes the definition of the \emph{slow} variable and its evolution equation. This is followed by the algorithm of its implementation.  

\subsection{\emph{Singularly perturbed} differential equations}\label{sec:sing_pert}


A particular class of ODE which involves a split into fast and slow dynamics, coupled to each other, is of the form   
\begin{equation}\label{eq:sing_pert}
\begin{aligned}
    \frac{dx}{dt} &= \frac{1}{\epsilon} F(x,l) \\
    \frac{dl}{dt} &= L(x,l),
\end{aligned}
\end{equation}
with $x \in \mathbb{R}^n$ and $l \in \mathbb{R}^m$. Here, $x$ corresponds to the fast variable and its evolution is governed by the fast dynamics. Denoting $\sigma=\frac{t}{\epsilon}$ as time scale of the fast dynamics, the fast evolution equation becomes 
\begin{align}\label{eq:fast_dyn}
    \frac{dx}{d \sigma}= F(x,l). 
\end{align}

On the other hand, $l$ corresponds to the load, which evolves in the slow time scale and can be considered to be fixed in the fast dynamics in Eq. \eqref{eq:fast_dyn}. Such class of ODEs are called \emph{singularly perturbed} differential equations. In the context of MD, we can think of $x$ as the position and velocities of atoms with characteristic time period of $T_f$ which is in the order of the time period of atomic vibrations (femtoseconds). $l$ is the slow applied loading rate, with time period $T_f$ (typically a few to 1000 seconds). The small real parameter $\epsilon > 0$, represents the ratio between fast and slow time periods, i.e. $\epsilon = \frac{T_f}{T_s}$. In the case of MD under slow strain rates, it is of the order of $10^{-15}$ or even smaller, which shows the vast separation in time scales between fast and slow dynamics. This small parameter multiplies the highest order derivative in the governing equation, thus forming a \emph{singular perturbation} of the rest of the terms in the equation. Often, the limit behavior as $\epsilon \to 0$ is well-recovered simply by obtaining the solution to the problem by setting $\epsilon = 0$. However, in many cases of practical relevance, as in MD, this is no longer true and obtaining the slow limit behavior requires more delicate analysis and the use of such understanding in robust and successful computation of the slow behavior of such systems. As already mentioned, the primary challenge in the direct computation of a singularly perturbed evolution is that when $\epsilon$ becomes very small, the time-step needs to reduce significantly as well, making the computations impractical. In such cases, instead of obtaining a full solution, a practical strategy is to define \emph{slow} variables which give a measurement of the underlying dynamics. In the next section, we discuss the Practical Time Averaging (PTA) framework and algorithm, which precisely follows this idea. 

\subsection{Practical Time Averaging (PTA) - framework and algorithm}\label{sec:pta_algo}


In \cite{chatterjee2018computing}, a class of \emph{slow} variables called \emph{H-observables} (where $H$ stands for history) were defined as: 
\begin{equation}\label{eq:v_t}
\textnormal{v}_m(t) = \frac{1}{\Delta} \int_{t-\Delta}^{t} \int_{\mathbb{R}^n} m(x)\,\mu(s)(dx)\, ds,
\end{equation}
where $m(x)$ is a state function of the fast dynamics, $\mu(s)$ is the invariant measure (also called Young measure) which gives the probability density function of the fast trajectory and $\Delta$ is an interval in the slow time scale. 
Thus, \emph{H-observables} are averages over an interval in slow time of the moments of state functions with respect to the probability density function (Young measure) of the fast trajectory. Hence, such variables not only depend on the value of the measure at time $t$ but on the ``history'' of the measure in the interval $[t-\Delta,t]$. 

Differentiation of Eq. \eqref{eq:v_t} in time using Newton-Leibnitz rule gives the time-derivative of the slow variable in the form:
\begin{align}
\frac{d\textnormal{v}_m}{dt}(t) &= 
\frac{1}{\Delta} \left(
\int_{\mathbb{R}^n} m(x) \, \mu(t)(dx) 
- \int_{\mathbb{R}^n} m(x) \, \mu(t-\Delta)(dx)
\right) .
\label{eq:dvdt}
\end{align}


Given the initial conditions of the fast and slow variables i.e. $x(t_0)$ and $l(t_0)$, where $t_0=-\Delta$ is the initial time, we think of the calculations marching forward in slow time-scale in discrete steps (also called \emph{jumps}) of size $h$ in the slow time scale, with total time as $T_0 = nh$. Thus the variable $t$ below in description of our algorithm takes values of $0, h, 2h, ..., nh$. The interval $\Delta$ in the slow time-scale is a fraction of $h$ and its value is given in Table.~\ref{table_1}. The goal is to determine the successive values of the slow variable $\textnormal{v}(t)$ in the slow time, which gives a measure of the underlying fast dynamics. 

\begin{itemize}
\item \textbf{Step 1: Calculate the rate of change of slow variable}\\
We denote $\int_{\mathbb{R}^n} m(x) \, \mu(t)(dx)$ as $R^m_t$ and $\int_{\mathbb{R}^n} m(x) \, \mu(t-\Delta)(dx)$ as $R^m_{t-\Delta}$. In practice, we do not compute the Young Measure and then take the moment of the state function with respect to it, but instead obtain it as running time averages of the state function till it converges. However, the latter definition using the running time average is computationally efficient and is therefore utilized in our work. We denote it as  $R^m_t$, which is computed as
\begin{align}\label{eq:R_m_t}
    R^m_t = \frac{1}{N_t} \sum_{r=1}^{N_t} m(x(\sigma_r),l_t), 
\end{align}
where the successive values of $x(\sigma_r)$ are calculated by running the fast dynamics given by Eq. \eqref{eq:fast_dyn} while holding the load $l_t$ at slow time $t$ fixed. The initial conditions to run the fast dynamics is discussed in Step 4. Here, $N_t$ is the number of fine time steps required for $R^m_t$ to converge up to a specified tolerance. A discussion on the convergence criteria is provided in \ref{app:convergence}. In this work, we have three state functions (as defined later in Eq. \eqref{eq:state_variables}), and the tolerance for checking the convergence of the running time averages for each of them is provided in Table \ref{table_1}.

Substituting Eq. \eqref{eq:R_m_t} to the RHS of Eq. \eqref{eq:dvdt}, the time derivative of the slow variable is computed as: 
\begin{align}\label{eq:dvdt_algo}
    \frac{d\textnormal{v}_m}{dt}(t) =  \frac{1}{\Delta} \left( R^m_{t} - R^m_{t-\Delta} \right)
\end{align}

\item \textbf{Step 2: Find the value of slow variables} \\
The predicted value of slow variable using PTA at $(t+h)$ is given using the extrapolation rule: 
\begin{equation}
\textnormal{v}_m(t+h) = \textnormal{v}_m(t) + \frac{d\textnormal{v}_m(t)}{dt} h   
\label{eq:pta_pred}
\end{equation}
where $\frac{d\textnormal{v}_m(t)}{dt}$ is obtained from Eq. \eqref{eq:dvdt_algo}.

\item \textbf{Step 3: Accept the Measure}\\
The value of the slow variable at $(t+h)$ is defined as 
\begin{equation}
\label{eq:measure}
\textnormal{v}_d^m(t+h)
:= \frac{1}{N'} \sum_{r} m\!\left( x(\sigma_r) \right),
\quad 
\text{where} \quad
N' = \frac{\Delta}{\epsilon\, \Delta\sigma},
\end{equation}

and successive values $x(\sigma_r)$ are obtained by running the fast dynamics in Eq. \eqref{eq:fast_dyn}, using a suitable initial condition as described in Step 4. Note that the number of steps $N'$ used to calculate the slow variable $\textnormal{v}(t+h)$ above is different from the number of steps $N_t$ to determine the $R^m_t$ using Eq. \eqref{eq:R_m_t} (with $N' >> N_t$). This shows that both $\textnormal{v}(t+h)$ and $R^m_t$ are running time averages but are computed over different number of fast time steps. Moreover, $N'$ is fixed but $N_t$ depends on how fast the running time average converges at slow time $t$. 
In practice, it is not feasible to run the fast dynamics over the period $\Delta$ when $\epsilon$ becomes very small, since $N'$ becomes exceedingly large. Hence, we approximate it using Simpson's rule as
\begin{equation}
\textnormal{v}_d^m(t+h) \approx \frac{1}{6} \left( R^m_{t+h-\Delta} + 4R^m_{t+h-\frac{\Delta}{2}} + R^m_{t+h} \right)
\label{eq:simp}
\end{equation}
where $R^m_{t+h-\Delta}$, $R^m_{t+h-\frac{\Delta}{2}}$, and $R^m_{t+h}$ are the converged values of the running time averages (as defined in Eq. \eqref{eq:R_m_t}) at the slow times given by their respective subscripts. The initial conditions to run the fast dynamics in order to obtain these averages are provided in Eq. \eqref{eq:x_guess_step3} in Step 4. 

The relative error  between the values of the slow variable $\textnormal{v}(t+h)$ computed using Eq. \eqref{eq:pta_pred} and Eq. \eqref{eq:simp} is denoted as $e^{\textnormal{v}}_{t+h}$ and given by
\begin{equation}
\label{eq:rel_err}
e^{\textnormal{v}}_{t+h} = \left| \frac{\textnormal{v}_m(t+h) - \textnormal{v}_d^m(t+h)}{\textnormal{v}_d^m(t+h)} \right|. 
\end{equation} 
 Since the fast dynamics at play is not known to satisfy the ergodicity assumption for each fixed value of applied load, the invariant measures of the limit (as $\epsilon \to 0$) dynamics at any slow time depends on the initial condition used to generate the fast trajectory - this is also borne out by our practical experience. The relative error $e^{\textnormal{v}}_{t+h}$ does not address that component of the error. Addressing and quantifying that component of the error is challenging because it requires knowing the solution of the fast dynamics over impossibly large (fast) times. 
 
We say there is a match in the value of the slow variable at $t+h$ if $e^{\textnormal{v}}_{t+h} < tol_{\textnormal{v}}$, where $tol_{\textnormal{v}}$ is a specified tolerance for slow variable $v$. Since we have 3 different slow variables in our work (as defined in the discussion following Eq. \eqref{eq:state_variables}), we have three tolerances corresponding to each of them, which are listed in Table \eqref{table_1}.
 In that case, we accept the value of the slow variable predicted by PTA at $t+h$ as defined in Eq. \eqref{eq:pta_pred} and move on to the next slow time.  

If not, we check whether there is a jump in the measure at $t + h$. Since we do not compute the measure but instead obtain the moments of state functions with respect to it, a jump in such moments identifies a jump in the measure indirectly. Hence, we check if the following inequality holds: 
\begin{equation}\label{eq:jump_err}
\left| \frac{e_1}{e_2} - 1 \right| \geq tol_j \, , \quad \textnormal{where} \qquad e_1 =\left|\frac{R_{t+h}^{m} - R_{t}^{m}}{R_{t}^{m}}\right|, \quad e_2 = \left|\frac{R_{t}^{m} - R_{t-h}^{m}}{R_{t-h}^{m}}\right|. 
\end{equation} 
The above inequality is satisfied when $R^m_{t+h}$ is significantly different from $R^m_t$ and $R^m_{t-h}$, which occurs when there is a jump in the measure at time $t+h$. 

In Eq. \eqref{eq:jump_err}, we use the assumption that the measure does not undergo a jump at time $t$. However, if a jump in the measure is already detected at time $t$, we detect if there is a jump in the measure at time $t+h$ by checking if $e_1 \geq tol_j$, where $e_1$ is given in Eq. \eqref{eq:jump_err}. 

If a jump in the measure is detected, $\textnormal{v}_d^m(t+h)$ defined in Eq. \eqref{eq:measure} is accepted as the value of the slow variable at $t+h$ and the prediction of the slow variable given by the extrapolation rule in Eq. \eqref{eq:pta_pred} is discarded. 

\item \textbf{Step 4: Obtain the fine initial conditions} \\
Step 1 and Step 3 requires obtaining the values of $R^m$ at different slow times. To obtain $R^m_t$, we need to run MD with an initial guess of the fast variable, $x^{guess}_t$, which is
 chosen as a fine state at a fast time corresponding to $t_{prev}$ for which $R^m_{t_{prev}}$ was considered to have converged. We denote such a state as $x^{conv}_{t_{prev}}$. As mentioned in Eq. \eqref{eq:R_m_t}, it takes $N_{t_{prev}}$ fine time steps for $R^m_{t_{prev}}$ to converge. Hence, $x^{conv}_{t_{prev}}$ is equivalent to $x(\sigma_{N_{t_{prev}}})$. 

Based on the above argument, the initial guess at slow times $t-\Delta$ and $t$, used to obtained $R^m_{t-\Delta}$ and $R^m_t$ respectively in Eq. \eqref{eq:dvdt_algo} in Step 1 are obtained as: 
\begin{equation}
\label{eq:x_guess_step1}
\begin{aligned}
x^{guess}_{t-\Delta} &= x^{conv}_{t-h} \\
x^{guess}_t &= x^{conv}_{t-\frac{\Delta}{2}} 
\end{aligned}
\end{equation}

Similarly, the initial guess used to obtained $R^m_{t+h-\Delta}$, $R^m_{t+h-\frac{\Delta}{2}}$ and $R^m_{t+h}$ in Eq. \eqref{eq:simp} in Step 3, are obtained as: 
\begin{equation}
\label{eq:x_guess_step3}
\begin{aligned}
x^{guess}_{t+h-\Delta} &= x^{conv}_{t} \\
x^{guess}_{t+h-\frac{\Delta}{2}} &= x^{conv}_{t+h-\Delta} \\
x^{guess}_{t+h} &= x^{conv}_{t+h-\frac{\Delta}{2}} 
\end{aligned}
\end{equation}


If the running time average obtained by running the fast dynamics with initial condition $x^{guess}_t$ does not converge even after running a reasonably large number of fast time steps, then we cannot obtain $R^m_t$ using that $x^{guess}_t$.
In such cases, we have to try different values of $x^{guess}_t$ and repeat Steps 3 and 4. 

While the protocol defined above appears to produce plausible results for the applications we are interested in, as shown later in Section \ref{sec:results}, defining these guesses for generating the approximate measures at well-separated slow times is a fundamental problem. The challenge here is that the initial condition for generating the measure at any slow time has to to be within the basin of attraction of the invariant measure(s) corresponding to the limit flow at that slow time. Estimating such a location in phase space with some plausibility from well-separated, in slow-time, spurts of fast computation, as employed by our scheme requires some special consideration. One line of reasoning was provided by the closest-point-projection approach as described in  \cite[Step 3 and Step 5 of Section 9]{chatterjee2018computing}.  There, the initial condition was obtained by taking the closest point projection of a point in the support of the invariant measure at a slow time with respect to the measure at a previous slow time. Then linearly extrapolating these points gave the initial condition for the fast trajectory at the next slow time (under the condition that the measure evolves slowly and does not undergo a jump in this  interval of slow time). Applying that protocol here did not result in success, most likely due to the lack of a adapted metric w.r.t which the required projection can be executed in phase space. Based on the naive choices made, the guesses obtained for initial conditions seemed to have excessive kinetic energy, resulting in non-convergence of the $R^m$ values over the typical time intervals that were employed to perform the averages.
 
At a fundamental level, the plausibility of our obtained results (up to validation) does not rule out the fact that the measure that we work with, say at slow time $t$, is not an approximation of any of the invariant measures of the limit dynamics we are trying to probe. A full answer to this (very) difficult question of predicting the slow dynamics of a fast evolution \emph{with guarantees} and furthermore, with computational efficiency, awaits further study. A natural idea in this regard would be to evolve the slow variables $(\mbox{v}_{\bm{r}^\alpha}, \mbox{v}_{\bm{v}^\alpha})$ defined in Sec.~\ref{sec:MD} below (i.e., the slow variables corresponding to the atomic positions and velocities of an MD asembly)  and consider these at a discrete slow time, $t$, as the initial condition for the fine dynamics at time $t$ to generate the various $R^m_t$.

In the meanwhile, we proceed with our strategy laid out above which produces plausible outcomes.

\end{itemize}


\subsection{Molecular Dynamics}\label{sec:MD}
In this work, the fast dynamics is the evolution of a molecular dynamical system consisting of a block of FCC Aluminum atoms, as discussed later in Section \ref{sec:problem_setup}. We have performed MD simulations with the software Large-scale Atomic/Molecular Massively Parallel Simulator (LAMMPS) to study the tensile and compressive behaviors of metals under slow strain rates. The simulations were carried out on one Skylake GPU node with 8 MPI processes and one NVIDIA Tesla V100 GPU. We used embedded-atom method (EAM) potential for this work. Specifically, we used the \cite{PhysRevB.59.3393} aluminum potential. It is converted in LAMMPS format by C. A. Becker (2008). The total potential energy of the system is represented as
\begin{equation}
U = \frac{1}{2} \sum_{\alpha} \sum_{\substack{\beta \\ \beta \neq \alpha}} V(r^{\alpha\beta}) + \sum_{\alpha} B(\rho^{\alpha})
\label{eq:pe}
\end{equation}
Here, $V$ is the pair potential which is a function of the distance $r^{\alpha\beta}$ between atoms $\alpha$ and $\beta$ and $B$ is the embedding energy which is a function of the atomic electron density $\rho$. The latter is given by 
\[
{\rho}^\alpha = \sum_{\substack{\beta \\ \beta \neq \alpha}} \rho(r^{\alpha\beta})
\]
In MD the atoms are treated as classical Newtonian particles. Thus, the total force on particle $\alpha$ is expressed as
\begin{equation}
\label{eq:f=ma}
\bm{F}^{\alpha} = m_a \, \bm{a}^\alpha  = \bm{F}^{\text{ext},\alpha} + \bm{F}^{\text{int},\alpha}   
\end{equation}
Here, $m_a$ is the mass of each atom (note the difference between this and the state function defined previously as $m(x)$ in Eq. \eqref{eq:v_t}) and $\bm{a}^\alpha$ is the acceleration of atom $\alpha$.  
$\bm{F}^{\text{ext},\alpha}$ represents the external force acting on the atom due to external fields and atoms outside the system, while $\bm{F}^{\text{int},\alpha}$ represents the internal force arising from interactions with other atoms within the system. The internal force on atom $\alpha$ can be expressed as
\begin{equation}
\label{eq:f_int}
\bm{F}^{\text{int},\alpha} = \sum_{\substack{\beta \\ \beta \neq \alpha}} \bm{F}^{\alpha\beta} = - \frac{\partial}{\partial\bm{r}^\alpha} U
\end{equation}
In the above the expression of potential energy $(U)$ is substituted from Eq. \eqref{eq:pe}. To evolve the fast dynamics of the sample LAMMPS uses the Velocity-Verlet (VV) algorithm \cite{lammps_nve,lammps_verlet}. Given, at fast time $\sigma$ the positions and velocities of the particles are $\bm{r}^\alpha (\sigma)$ and $\bm{v}^\alpha (\sigma)$, respectively. As discussed in Section \ref{sec:pta_md_app}, they form the state of the system. The steps of the algorithm are as follows:
\begin{itemize}
\item \textbf{Step 1: Calculate the velocity at fast time $\sigma + \frac{\Delta \sigma}{2}$} \\
We calculate a half-step update to the velocities as

\begin{equation}
\label{eq:vv_algo_pos_pred_half}
    \bm{v}^\alpha \bigg(\sigma + \frac{\Delta \sigma}{2}\bigg) = \bm{v}^\alpha (\sigma)+ \frac{{\bm F}^\alpha({\bm r}^\alpha(\sigma))}{m}\frac{\Delta \sigma}{2}
\end{equation}

\item \textbf{Step 2: Calculate the position at fast time $\sigma + \Delta \sigma$} \\
We calculate a full-step update to the positions using the half-step update to the velocities as
\begin{equation}
\label{eq:vv_algo_pos_pred}    
\bm{r}^\alpha (\sigma + \Delta \sigma)
= \bm{r}^\alpha (\sigma)
+ \bm{v}^\alpha \bigg(\sigma + \frac{\Delta\sigma}{2}\bigg)\Delta \sigma
\end{equation}

\item \textbf{Step 3: Calculate the velocity at fast time $\sigma + \Delta \sigma$} \\
Finally, we calculate one additional half-step update to the velocities as
\begin{equation}   
\label{eq:vv_algo_vel_pred}
\bm{v}^\alpha (\sigma + \Delta \sigma) = \bm{v}^\alpha \bigg(\sigma + \frac{\Delta \sigma}{2}\bigg) + \frac{{\bm F}^\alpha({\bm r}^\alpha(\sigma + \Delta \sigma))}{m}\frac{\Delta \sigma}{2}
\end{equation}
\end{itemize}



\subsection{Application of PTA to Molecular Dynamics}\label{sec:pta_md_app}

In this section, we discuss the application of PTA to a specific problem in which the fast dynamics is given by Molecular Dynamics, as discussed the previous section. This requires the identification of the load and fast variable, the state functions and the slow variable, which we discuss next. 

\subsubsection{Problem Setup}\label{sec:problem_setup}
\begin{figure}[h!] 
\centering \includegraphics[width=\textwidth]{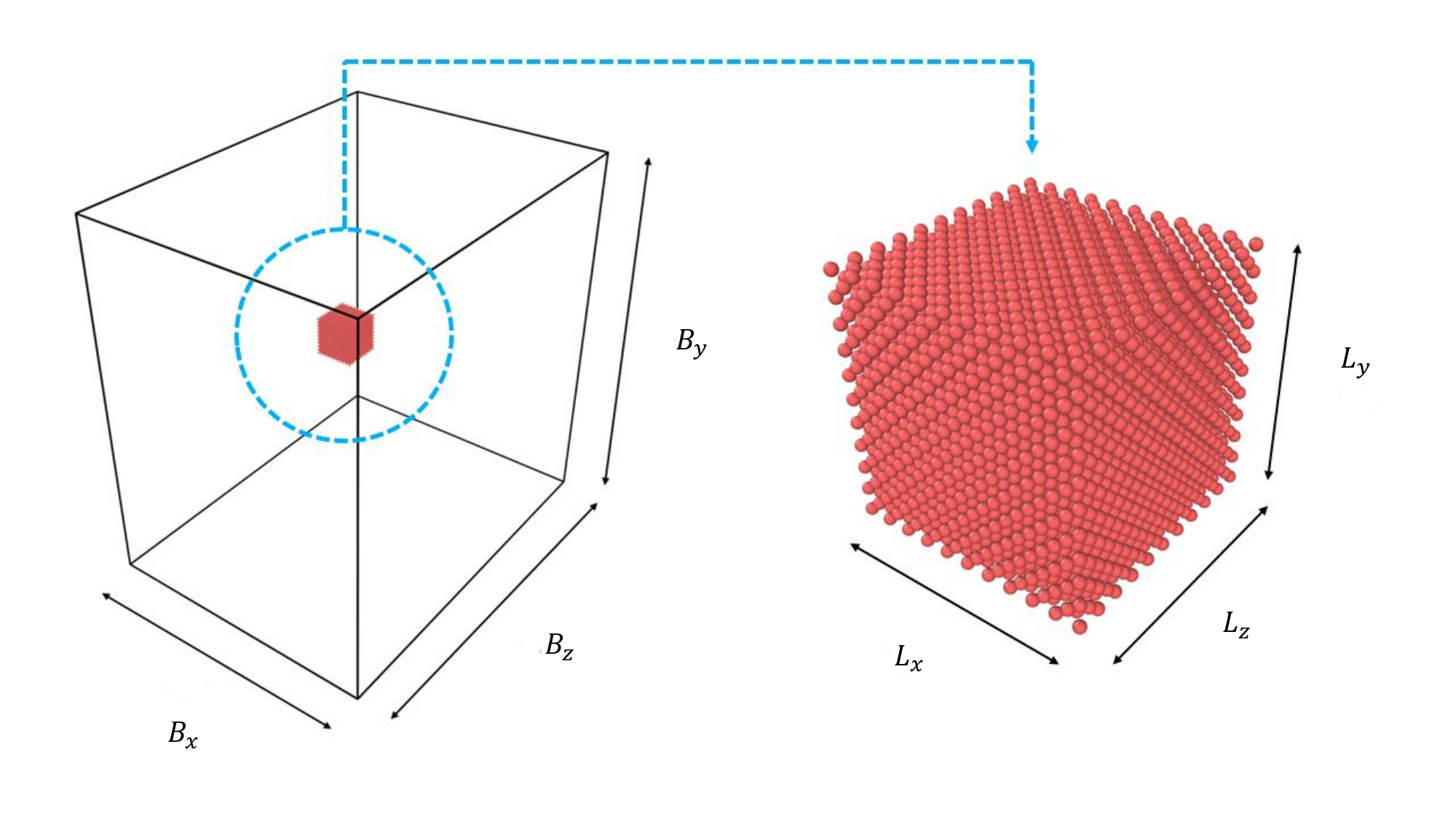}
\caption{MD Box and Sample.} 
\label{fig:box_sample}
\end{figure}

\begin{figure}[h!]
    \hspace{0.8 cm} 
    \includegraphics[width=1.0\textwidth]{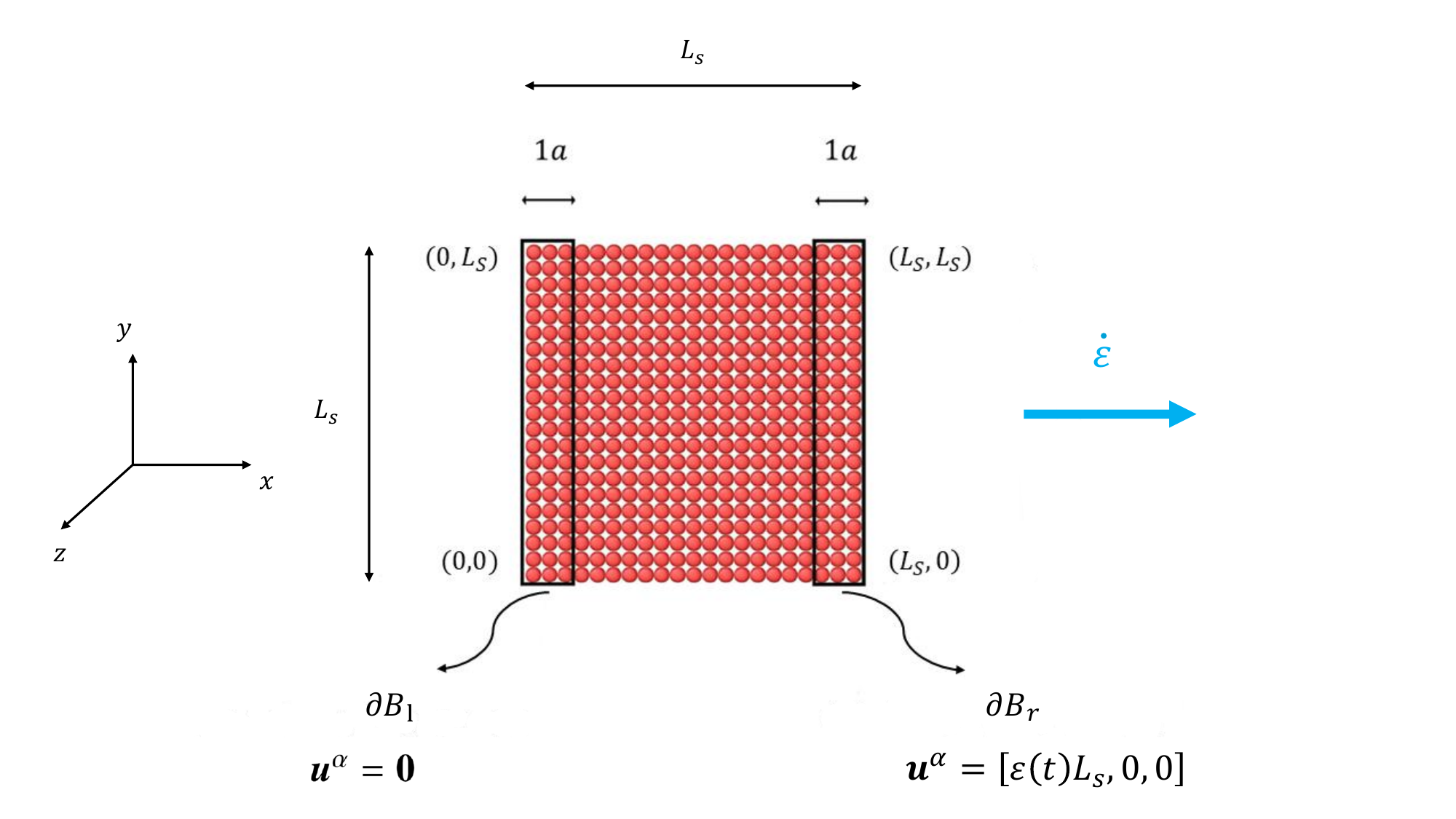}
    \caption{Boundary conditions (BCs) applied on the sample corresponding to uniaxial tension/compression in the $x$ direction. To obtain the converged running time average $R^m_t$ at slow time $t$, MD is run with the left boundary atoms $\partial B_l$ fixed at zero displacement while the right boundary atoms $\partial B_r$ are fixed at $u_x(t)=\varepsilon(t) \, L_s=\dot{\varepsilon} \, t \, L_s$ while $u_y=u_z=0$, where $\dot{\varepsilon}$ is the applied strain rate and $t$ is the slow time.}
    \label{fig:bc_sample}
\end{figure}

A simulation box of constant volume is defined within which a sample of Aluminum with FCC lattice structure, with lattice parameter $a$ is created. The dimensions of the box are $B_x$, $B_y$ and $B_z$ (whose values are specified in Table \ref{table_1}) while that of the sample are $L_x$, $L_y$ and $L_z$, respectively, as shown in Fig. \ref{fig:box_sample}. In this work, we have used cubic samples, hence $L_x=L_y=L_z=L_s$, where $L_s$ is the size of the cubic sample, which ranges from $4 \, nm$ to $30 \, nm$, as specified in Table \ref{table_1}. Since the dimensions of the simulation box are much larger than those of the sample, they need not be changed during the deformation of the sample. Hence, the system is evolved using NVE ensemble (where $\textnormal{V}$ is the simulation box volume) with EAM potential as defined in Eq. \eqref{eq:pe}. The global coordinate axes $x$, $y$, and $z$ are aligned along the lattice directions [100], [010], and [001] respectively of the sample.  The MD box is subjected to fixed boundary condition in all directions. As shown in Fig. \ref{fig:bc_sample}, we define the left boundary atoms of the sample as the set of atoms which belong to the left boundary region $\partial B_l = \{ (x,y,z) \in \mathbb{R}^3: 0 \leq x \leq a, 0 \leq y \leq L_y, 0 \leq z \leq L_z \}$. Similarly, the right boundary atoms of the sample are defined as the set of atoms which belong to the right boundary region $\partial B_r = \{ (x,y,z) \in \mathbb{R}^3: (L_x - a) \leq x \leq L_x, 0 \leq y \leq L_y, 0 \leq z \leq L_z \}$. 

The state of the system comprises the set of atomic positions and velocities and is denoted as 
\begin{equation}\label{eq:state}
\begin{aligned}
\bm s =\{ \bm r^\alpha, \bm v^\alpha \}  &= \left( \bm r^1, \bm r^2, ..., \bm r^n, \bm v^1, \bm v^2, ... \bm v^n \right) \\ 
      & =\left(x^1, y^1, z^1, ..., x^n, y^n, z^n, v^1_x, v^1_y, v^1_z, ... v^n_x, v^n_y, v^n_z \right), 
\end{aligned}
\end{equation}
where the system consists of $n$ atoms and $\bm r^\alpha$ and $\bm v^\alpha$ are the position and velocity of atom $\alpha$. Thus, the state $\bm s$ is a vector of size $6n$, where $n$ is the number of atoms in the system. The state $\bm s$ evolves on the fast time scale $\sigma$, hence it is denoted as $\bm s(\sigma)$. This serves as the fast variable in our work, similar to $x(\sigma)$ in Section \ref{sec:sing_pert} and \ref{sec:pta_algo}. The load $l(t)$ corresponds to the applied displacement on the system under quasi-static loading rate. A detailed discussion on the boundary conditions is provided next. 

\subsubsection{Boundary conditions}
In this section, we discuss the boundary conditions applied to the sample at different stages of the simulation. Fig. \ref{fig:bc_sample} shows the boundary conditions applied to the sample. The left and right boundary atoms, $\partial B_l$ and $\partial B_r$, and the corresponding constraints on the atomic positions, velocities, and forces are indicated. First, we generate the initial state $\bm s(-\Delta)$. The initial positions of the atoms are the lattice positions; velocities are initialized from a Maxwell-Boltzmann distribution corresponding to the temperatures $T_0$. During this step, the left boundary atoms are fixed at their reference lattice positions by setting $\bm v^\alpha(\sigma)=\bm 0, \, \bm F^\alpha(\sigma)=\bm 0$. Note that this is the standard procedure to fix atoms in LAMMPS, unlike in mechanics, where we cannot simultaneously specify both velocity and force. However, setting $\bm v^\alpha(\sigma)=\bm 0, \, \bm F^\alpha(\sigma)=\bm 0$ in Eq. \eqref{eq:vv_algo_pos_pred_half}-\eqref{eq:vv_algo_vel_pred} ensures that the position $\bm r^\alpha(\sigma)$ for atom $\alpha$ which belongs to the left boundary $\partial B_l$ remains unchanged during the Velocity-Verlet update in MD. The right boundary atoms are allowed to move freely in the $x$ direction until the resultant reaction force on the right boundary atoms in the $x$ direction, $R_x(\sigma)$,  vanishes. The sample thus undergoes thermal relaxation to achieve an initial stress-free state. 


After thermal relaxation, the simulation proceeds by a combination of two steps. Mechanical loading is applied to the sample corresponding to the applied strain rate and the slow time. After the load is applied, the right boundary atoms are fixed at that displacement, and the running time average of the state functions is calculated. The steps we followed are summarized below:

\begin{enumerate}
    \item We first displace the right boundary atoms to $u_x(t-\Delta) = \dot{\epsilon}(t-\Delta)L_s$ and then fix them at that position (by setting $\bm v^\alpha(\sigma)=\bm 0, \, \bm F^\alpha(\sigma)=\bm 0$, similar to the procedure used to fix the left boundary atoms as discussed above). At time $t-\Delta$, we run MD simulations until $R^m_{t-\Delta}$ converges, following Eq.~\eqref{eq:R_m_t}. 

    \item We then displace the right boundary atoms by $u_x(t-\frac{\Delta}{2}) - u_x(t-\Delta) = \dot{\epsilon}\frac{\Delta}{2}L_s$ and fix them at that position. At time $t-\frac{\Delta}{2}$, we run MD simulations until $R^m_{t-\frac{\Delta}{2}}$ converges, following Eq.~\eqref{eq:R_m_t}. 

    \item We then displace the right boundary atoms by $u_x(t) - u_x(t-\tfrac{\Delta}{2}) = \dot{\varepsilon}\tfrac{\Delta}{2}L_s$ and fix them at that position. At time $t$, we run MD simulations until $R^m_t$ converges, following Eq.~\eqref{eq:R_m_t}. 
    
    \item We calculate the rate of change of the slow variable using Eq.~\eqref{eq:dvdt_algo}, and obtain the extrapolated value $\textnormal{v}^m(t+h)$ using the extrapolation rule in Eq.~\eqref{eq:pta_pred}. Finally, we displace the right boundary atoms by $u_x(t+h-\Delta) - u_x(t) = \dot{\varepsilon}(h-\Delta)L_s$.

    \item We similarly fix the right boundary atoms at that position corresponding to $t+h-\Delta$, $t+h-\frac{\Delta}{2}$, and $t+h$. Using the initial guess for the state as defined in Eq. \eqref{eq:x_guess_step3}, we run MD until convergence to calculate $R^m_{t+h-\Delta}$, $R^m_{t+h-\frac{\Delta}{2}}$, and $R^m_{t+h}$, respectively, following Eq.~\eqref{eq:R_m_t}. 
    
    \item We then compare the value of the slow variable $\textnormal{v}^m_d(t+h)$ calculated using Eq.~\eqref{eq:simp} with the extrapolated value of the slow variable at slow time $t+h$. We accept the measure if the two values are close to each other, up to a specified tolerance, using  Eq.~\eqref{eq:jump_err}. 

    \item If the values of the slow variables are different, we check if there is a jump in the measure using Eq. \ref{eq:jump_err}. If there is a jump in the measure, we discard $\textnormal{v}$ obtained using the extrapolation rule in Eq. \eqref{eq:pta_pred} and set the value of the slow variable as the value $\textnormal{v}^m_d$ obtained using Eq.~\eqref{eq:simp}.  

    \item We then proceed to the next slow time $t+h$, where $h$ is the jump-size in slow time, and repeat the above steps. 
\end{enumerate}

\subsubsection{State functions and \emph{slow} variables}
The state functions (denoted as $m$ in the discussion following Eq. \eqref{eq:v_t}) that we choose in our work are the instantaneous kinetic energy $K(\sigma)$, potential energy $U(\sigma)$ and normal stress (obtained as resultant reaction force divided by the cross-sectional area) in the $x$ direction, $T_x(\sigma)$, which are defined as 
\begin{equation}
\label{eq:state_variables}
\begin{aligned}
    K(\sigma) &=  \frac{1}{2} \sum_{\alpha=1}^n m_a \, \bm{v}^{\alpha} (\sigma)\cdot \bm{v}^{\alpha} (\sigma) \\
    U(\sigma) &= \frac{1}{2} \sum_{\alpha=1}^n \sum_{\substack{\beta \\ \beta \neq \alpha}} V(r^{\alpha\beta}(\sigma)) + \sum_{\alpha} B(\rho^{\alpha}(\sigma)) \\
    T_x(\sigma) &= \frac{\bm R(\sigma) \cdot \hat{\bm e}_x}{A_0} = \frac{\left(-\sum_{\alpha \in \partial B_r}  \bm{F}^{int,\alpha} (\sigma) \right) \cdot \hat{\bm{e}}_x}{L_s^2}. 
\end{aligned}
\end{equation}
The forms of the potential energy of the system $U(\sigma)$ and the internal force $\bm F^{int,\alpha}(\sigma)$ on atom $\alpha$ have been discussed previously in Eq.~\eqref{eq:pe} and Eq.~\eqref{eq:f_int} respectively. The right boundary region $\partial B_r$ is defined in Section~\ref{sec:problem_setup}.
Note that the state functions depend on the state $\bm s(\sigma)$ as defined above in Eq. \eqref{eq:state} through the set of atomic positions $\bm r^\alpha$ and atomic velocities $\bm v^\alpha$. 

\emph{The slow variables (denoted as $\textnormal{v}_m$ in Eq. \eqref{eq:v_t}) corresponding to the state functions defined in Eq. \eqref{eq:state_variables} in this work are $\textnormal{v}_K(t)$, $\textnormal{v}_U(t)$ and $\textnormal{v}_{T_x}(t)$. For simplicity, hereafter we denote them with overhead bars as $\overline{K}$, $\overline{U}$ and $\overline{T}_x$ and refer to them as averaged kinetic energy, averaged potential energy and averaged normal stress respectively.} 


In addition to the slow variables defined above, we also track the average atomic positions of each atom in the system to visualize the evolution of the microstructure at different slow times $t$. The averaged position of atom $\alpha$ is calculated as 
\begin{equation}
\tilde{\bm{r}}^\alpha (t)= \frac{1}{N_t} \sum_{i} \bm{r}^\alpha (\sigma_i),
\label{eq:mean_pos}
\end{equation}
where $N_t$ is defined in the discussion following Eq. \eqref{eq:R_m_t}. Note that $\tilde{\bm{r}^\alpha} (t)$ is not evolved as a slow variable (and neither is it the same as $\mbox{v}_{\bm{r}^\alpha}$) but is only meant for the purpose of visualization of a measure of a type of averaged atomic positions and dislocation microstructure in slow time. 


\section{Results}\label{sec:results}

\begin{table}[H]
\centering
\caption{Simulation parameters.} 

\begin{tabular}{|c|l|c|}
\hline
\textbf{Symbol} & \textbf{Description} & \textbf{Value} \\
\hline
$a$             & Lattice parameter of aluminum & $4.05$ \AA \\
\hline
$\Delta\sigma$  & MD timestep 
& $1$ femtosecond \\
\hline
$B_x \times B_y \times B_z$ & Dimension of box     & $202.5\,nm \times 90\,nm \times 90\,nm$ \\
\hline
$L_s$ & Size of cubic sample     & $4\,\textnormal{to}\,30\,nm$ \\
\hline
$\Delta$        & Time interval in Eq.~\eqref{eq:v_t} & $h/5$ \\
\hline
$N_{\max}$      & Maximum MD runs to check for convergence& $50 \times 10^3$ \\
\hline
$tol_{m}$       & Tolerance for convergence check in Eq.~\eqref{eq:R_m_conv} & $5\% (\overline{U}, \overline{K}, \overline{R_x})$ \\
\hline
$tol_{j}$     & Tolerance for jump check in Eq.~\eqref{eq:jump_err} & $1\%(\overline{U}), 8\%(\overline{K}), 5\%(\overline{R_x})$ \\
\hline
$tol_{\textnormal{v}}$     & Tolerance for acceptance of measure in Eq.~\eqref{eq:rel_err} & $5\%(\overline{U}), 1\%(\overline{K}), 1\%(\overline{R_x})$ \\
\hline
\end{tabular}
\label{table_1}
\end{table}

\subsection{Uniaxial tension test} 
\label{sec:tension_test}
In this section, we discuss the results for uniaxial tension simulations performed on samples of sizes $4$, $8$, $20$, $25$ ~\text{and}~ $30 \, nm$. 
The boundary conditions are described in Figure~\ref{fig:bc_sample} and we have used the applied strain rate of $10^{-3}~s^{-1}$. 

\begin{itemize}
\item \textbf{Evolution of averaged kinetic energy, averaged potential energy, and averaged normal stress:} 
The evolution of the slow variables - averaged potential energy $\overline{U}$, averaged kinetic energy $\overline{K}$ and averaged normal stress $\overline{T}_{x}$ along with the relative errors between these quantities and the corresponding values obtained from the measure, $\overline{U}_d$, $\overline{K}_d$, and $\overline{T}_{d_{x}}$, for the $8 ~\mathrm{nm}$ sample are shown in Fig.~\ref{fig:pe_tension}, \ref{fig:ke_tension}, and \ref{fig:stress_tension}, respectively. We observed that the relative error for all the slow variables are within 5$\%$ at all strains.

The increase in potential and kinetic energy due to application of mechanical load is visible. This is expected as energy is supplied to the system which leads to bond stretching (increase of potential energy) and increased lattice vibrations (increase of kinetic energy). The stress strain curve has an elastic part up to around 5$\%$ strain at which point dislocation nucleation and motion occurs. The resulting lattice waves lead to a sudden increase in kinetic energy. It also causes slip and plastic deformation and hence the stress-strain curve drops suddenly. Thus, dislocation nucleation events have a distinct signature in our simulations on slow time scales when the invariant measures jump, which is further reflected as jumps in the slow variables. Once dislocations exit the free-surface, the stress rises again due to dislocation starvation, followed by subsequent drop. These serrations are a characteristic feature of uniaxial tension and compression of samples of sizes ranging from a few nanometers to tens of nanometers (the sample sizes in this work range from $4~nm$ to $30~nm$).

The yield stress of around 5 GPa is reasonable since we are starting from a defect-free sample and homogeneous nucleation of dislocations is expected. The theoretical shear stress for the same is $G/10=2.5 \, GPa$ (where $G=25 \, GPa$ is the shear modulus of Aluminum) 
\cite{frenkel1926theorie}. 
The shear stress multiplied with the Schmid Factor $\sqrt{6}$ (for $\{111\}<110>$ slip systems in FCC crystal under uniaxial tension) gives around $6 \, GPa$, which is close to our model prediction.  

However, the primary contribution and key novelty of our approach is that the stress-strain curve shown till a strain of 20 $\%$ is obtained with applied strain rate of $10^{-3}/s$, which is almost 10 orders of magnitude smaller than the conventional rates applied in MD simulations. This is important for studying the mechanical response of samples at quasi-static loading rates using MD. To the best of our knowledge, MD simulations using such small rates and up to high strains have not been conducted yet. 

Another significant improvement is that the size effect that we observe does not involve calibration of any fitting parameter and instead is an emergent behaviour that is caused by the underlying microstructure evolution on the scale of applied loading at quasistatic loaidng rates.

\begin{figure}[tbh!]
    \centering
    \begin{subfigure}{0.5\linewidth}
        \centering
        {\includegraphics[width=\linewidth]{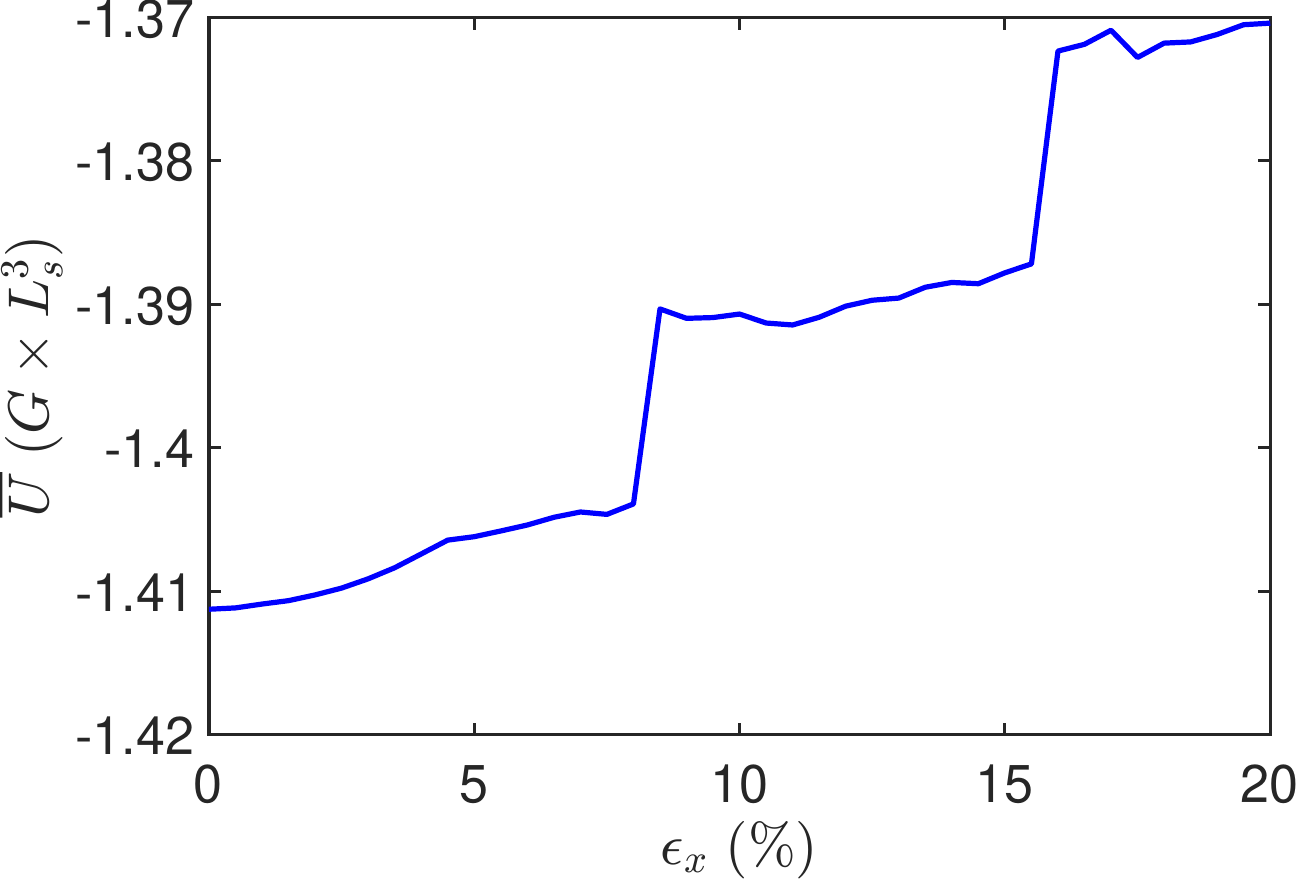}}
        \caption{}
    \end{subfigure}
    \hfill
    \begin{subfigure}{0.49\linewidth}
        \centering
        {\includegraphics[width=\linewidth]{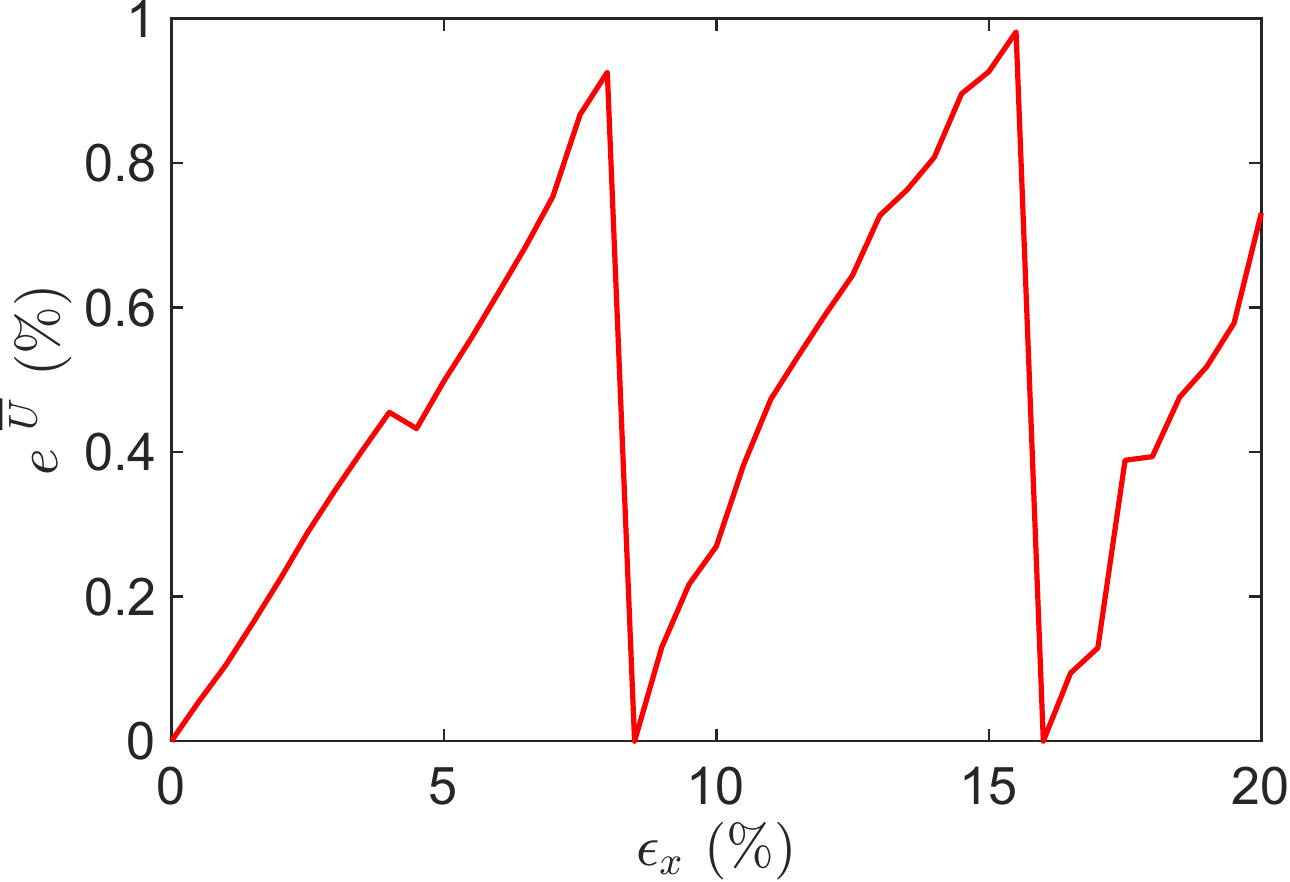}}
        \caption{}
    \end{subfigure}
    \caption{Evolution of averaged potential energy $(\overline{U})$ and its relative error $(e^{\overline{U}})$ with strain $(\epsilon_{x})$ for $8 \, nm$ sample in uniaxial tension. $L_s$ in the y-axis denotes the size of the sample.}
    \label{fig:pe_tension}
\end{figure}

\begin{figure}[tbh!]
    \centering
    \begin{subfigure}{0.5\linewidth}
        \centering
        {\includegraphics[width=\linewidth]{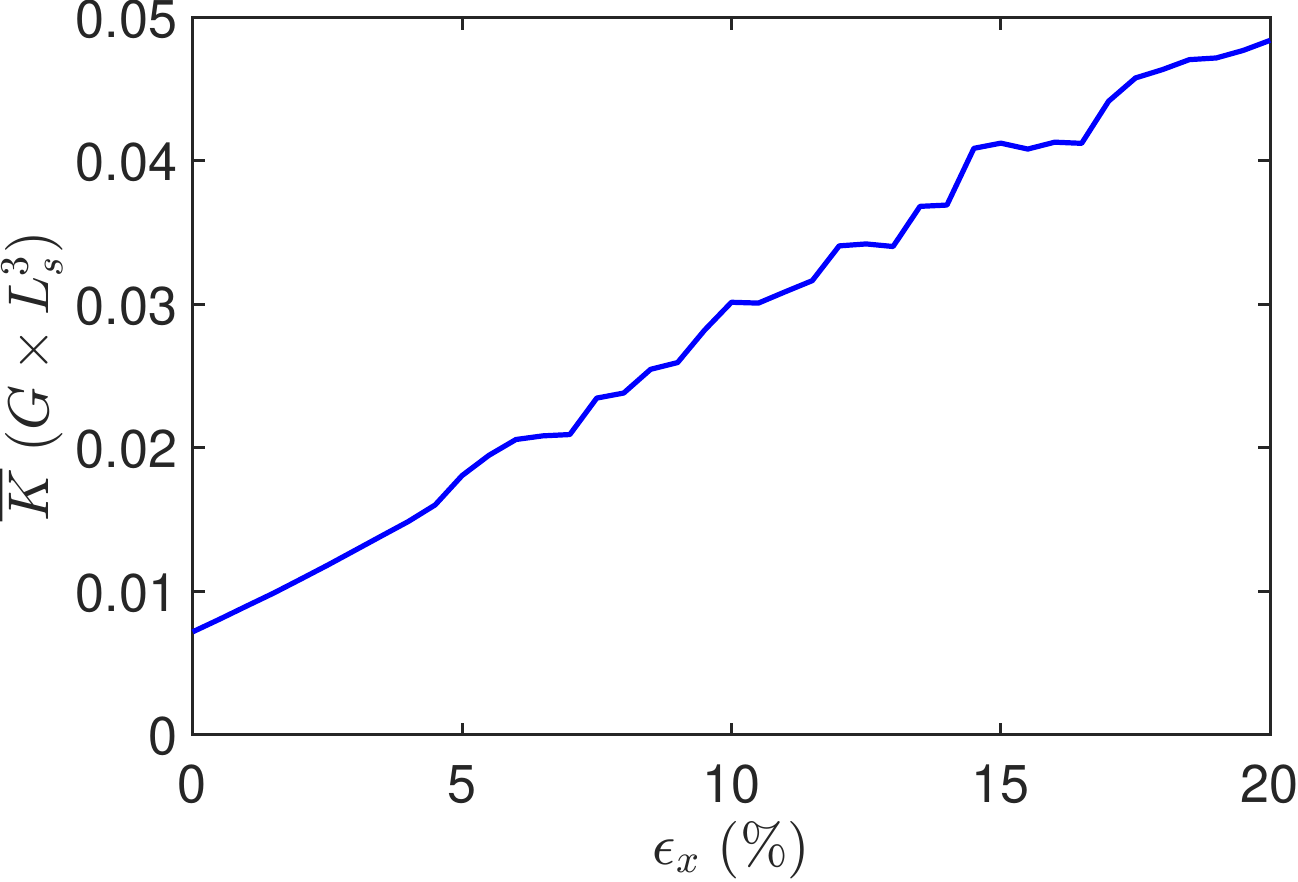}}
        \caption{}
    \end{subfigure}
    \hfill
    \begin{subfigure}{0.48\linewidth}
        \centering
        {\includegraphics[width=\linewidth]{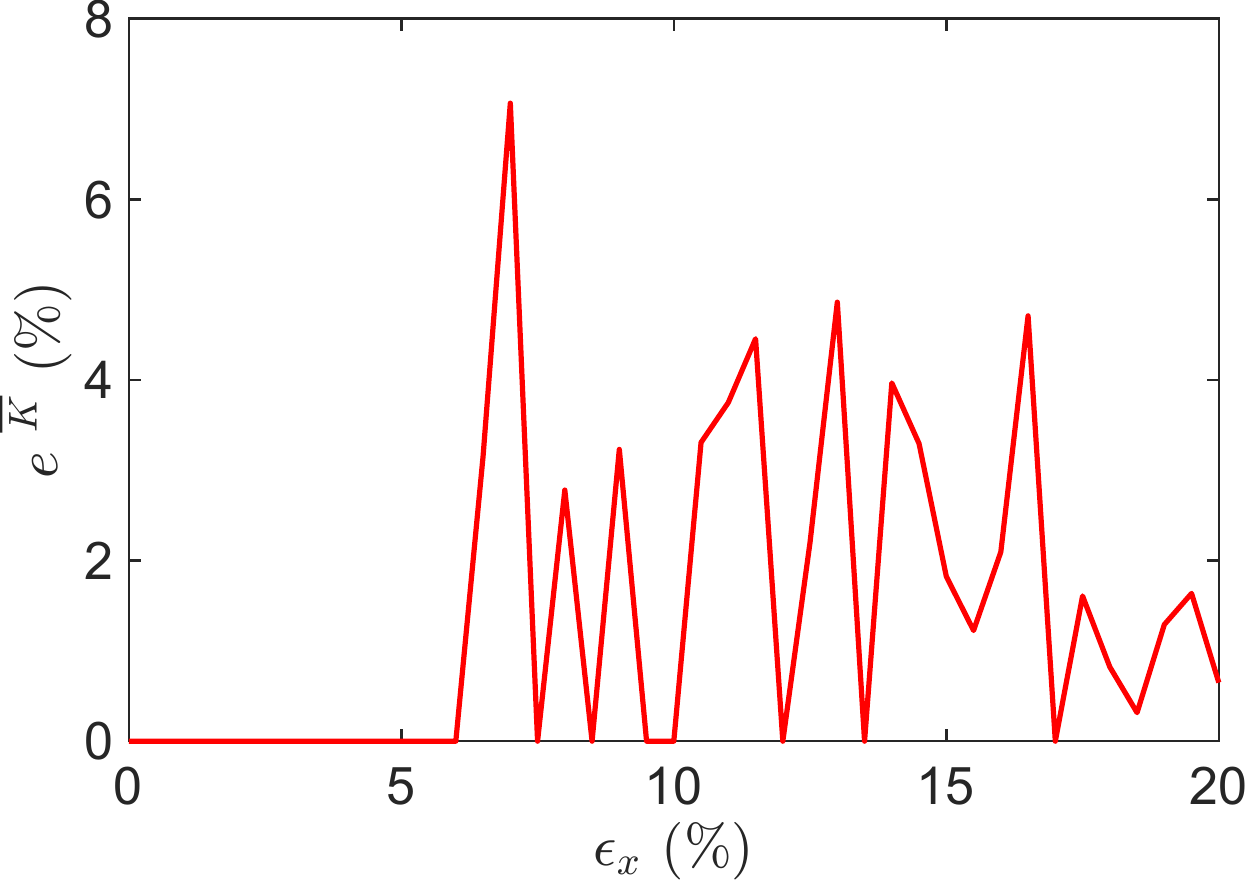}}
        \caption{}
    \end{subfigure}
    \caption{Evolution of averaged kinetic energy $(\overline{K})$ and its relative error $(e^{\overline{K}})$ with strain $(\epsilon_{x})$ for $8 \, nm$ sample in uniaxial tension.}
    \label{fig:ke_tension}
\end{figure}

\begin{figure}[tbh!]
    \centering
    \begin{subfigure}{0.5\linewidth}
        \centering
        {\includegraphics[width=\linewidth]{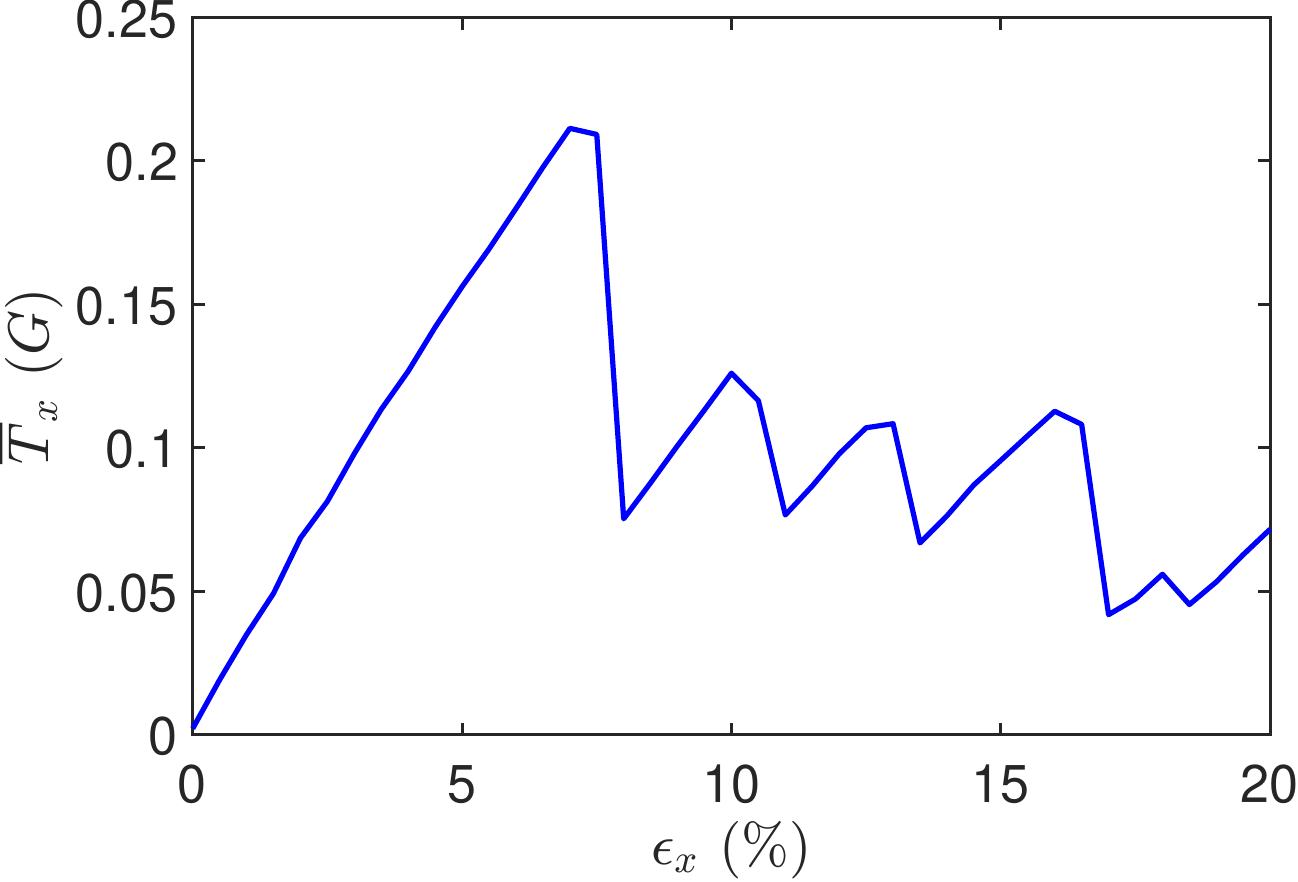}}
        \caption{}
    \end{subfigure}
    \hfill
    \begin{subfigure}{0.49\linewidth}
        \centering
        {\includegraphics[width=\linewidth]{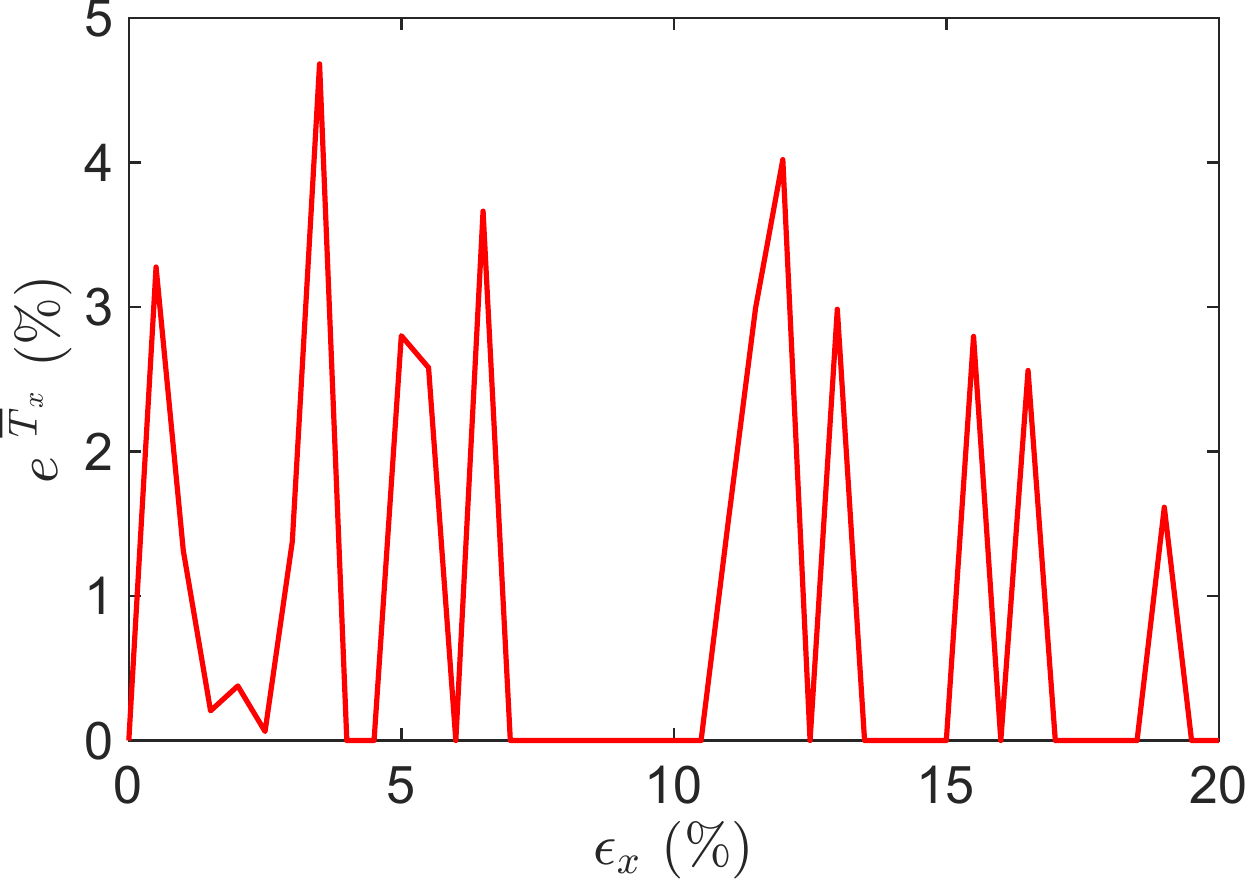}}
        \caption{}
    \end{subfigure}
    \caption{Evolution of averaged normal stress $(\overline{T}_{x})$ and its relative error $(e^{\overline{T}_{x}})$ with strain $(\epsilon_{x})$ for $8 \, nm$ sample in uniaxial tension.}
    \label{fig:stress_tension}
\end{figure}

\item \textbf{Effect of sample size on the stress–strain curve:} 
To ensure statistical reliability, five simulations results with different initial atomic velocity distributions are performed for both $8 \, nm$ and $20 \,nm$ blocks. For each block size, the mean of these simulation results is used to analyze the size effect. The corresponding stress–strain curves are shown in Figure~\ref{fig:tension_size_effect}. Under the same loading conditions, the $8 \, nm$ block exhibits a higher stress response than the $20 \, nm$ block, indicating a pronounced size effect.
Size effect in micron to submicron sized samples has been widely reported in micropillar compression tests and torsion tests of nanowires \cite{uchic2004sample}  
It is caused by the limited availability of dislocation sources in smaller samples, which also require higher stress to nucleate. Secondly, the strain gradients in smaller samples are higher, which leads to the generation of Geometrically Necessary Dislocations (GNDs) to maintain lattice compatibility. GNDs are a source of hardening in metals. The size effect observed in our simulations agrees qualitatively with the experimental results of \cite{uchic2004sample}, discussed later in Section~\ref{sec:exp_uchic}. 
\begin{figure}[tbh!]
    \centering
    \begin{subfigure}{0.48\linewidth}
        \centering
        \includegraphics[width=\linewidth]{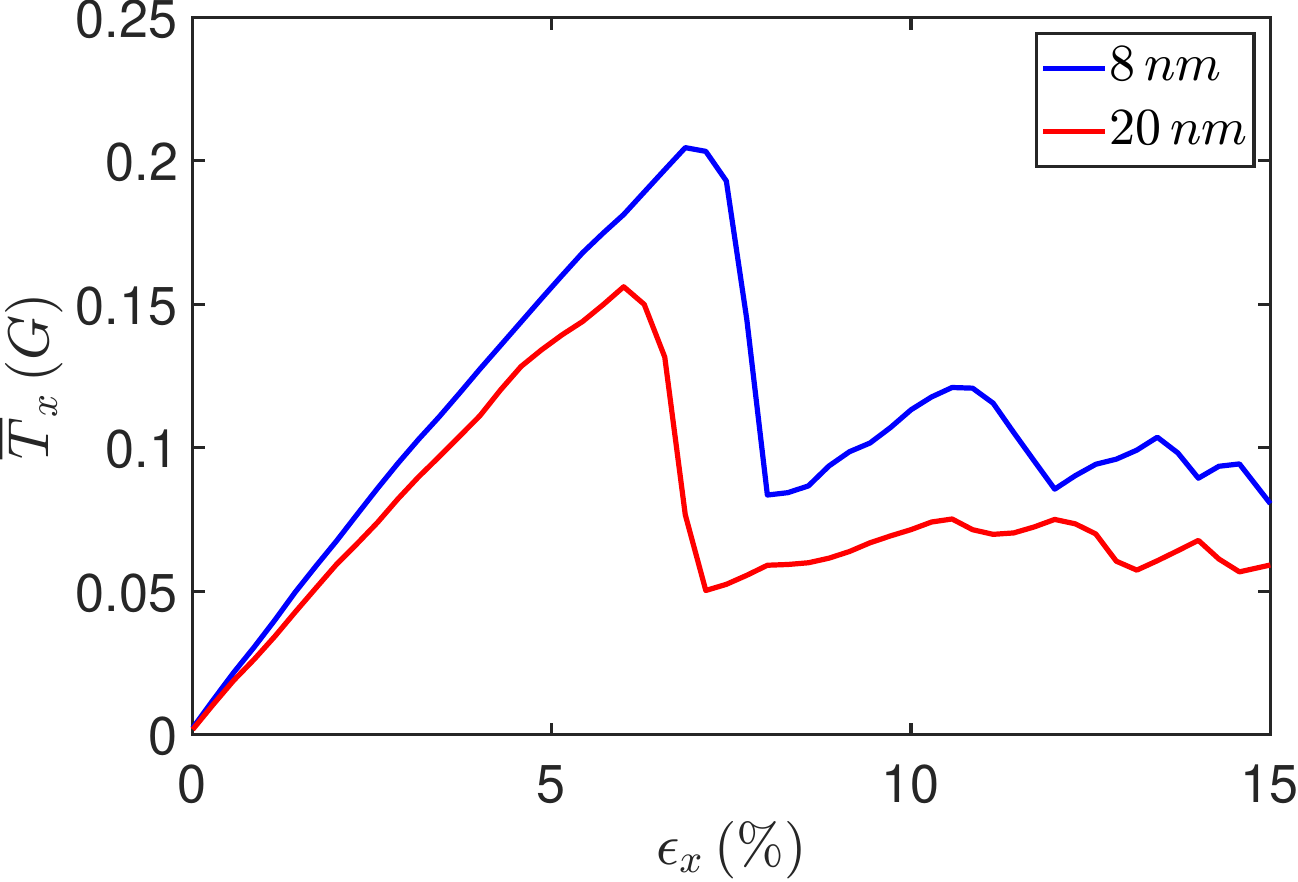}
        \caption{}
    \end{subfigure}
    \hfill
    \begin{subfigure}{0.48\linewidth}
        \centering
        \includegraphics[width=\linewidth]{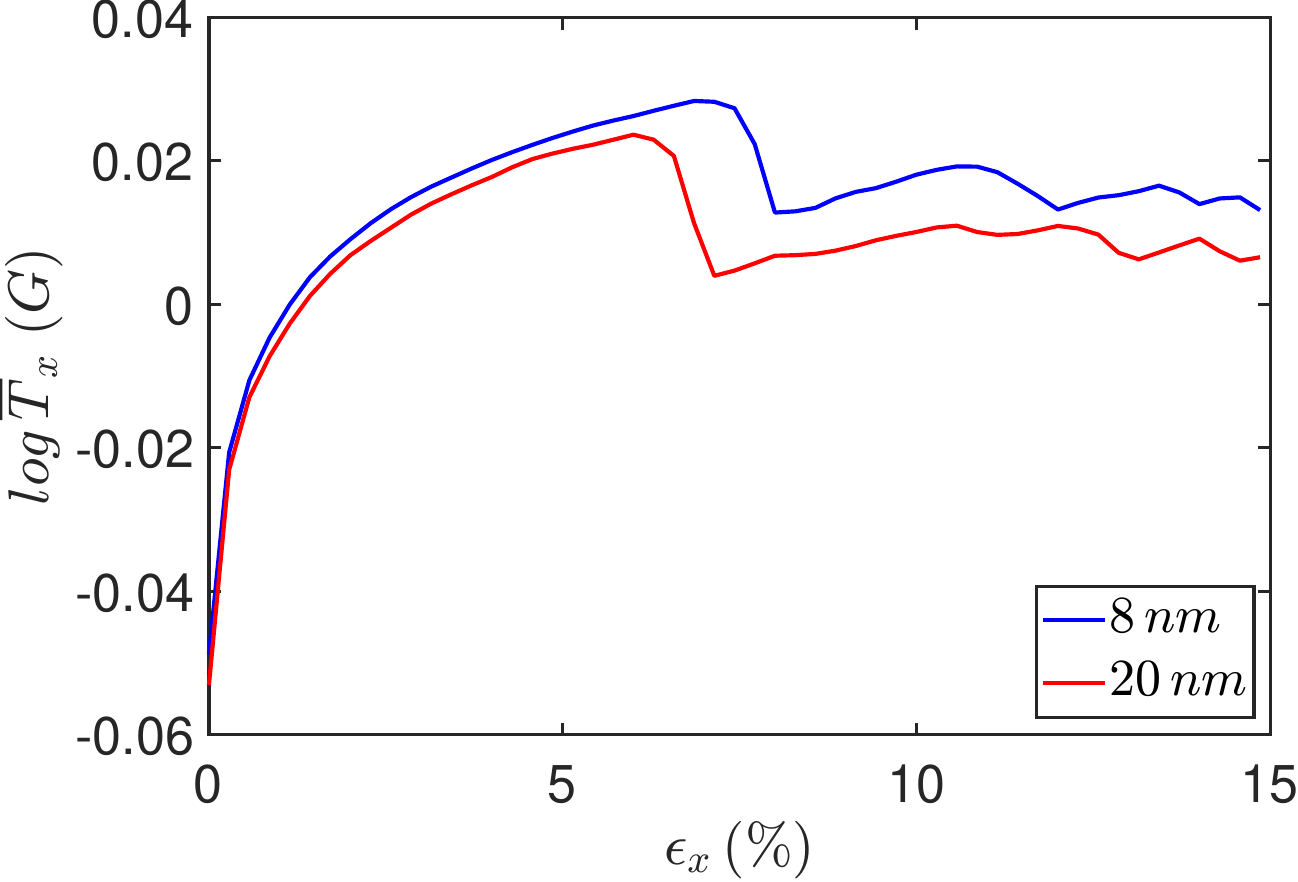}
        \caption{}
    \end{subfigure}
    \caption{Evolution of averaged normal stress $(\overline{T}_{x})$ with strain $(\epsilon_{x})$ for different sample sizes under uniaxial tension.}
    \label{fig:tension_size_effect}
\end{figure}

\item \textbf{Effect of size on standard deviation of mean stress:} 
We carry out five simulations with different initial atomic velocity distributions for both $8 \, nm$ and $20 \, nm$ blocks. At a certain strain value, the mean and standard deviation of stress across those simulations are then calculated for both sizes. In Figure~\ref{fig:tension_envelope}, we show the variation of the standard deviation of stress for different sizes. It can be observed that the smaller $8 \, nm$ block shows a higher standard deviation compared to the larger $20 \, nm$ block. To quantify this, we calculate the mean of the standard deviation. The values are $0.3549 \, GPa$ and $0.2188 \, GPa$ for the $8 \, nm$ and $20 \, nm$ sizes, respectively. After yielding, the shaded regions corresponding to the $8 \, nm$ and $20 \, nm$ sizes appear to overlap. The higher standard deviation for smaller sample is reasonable. In smaller samples, fewer dislocations are nucleated and once nucleated, they exit the surface quickly, followed by a period of dislocation starvation, leading to pronounced serrations. However, in larger samples, higher number of dislocations are present at any given time (post-yielding), hence dislocation nucleation and exit events do not have such pronounced effect on the stress-strain curve.

\begin{figure}[tbh!]
    \centering
    \includegraphics[width=0.8\textwidth]{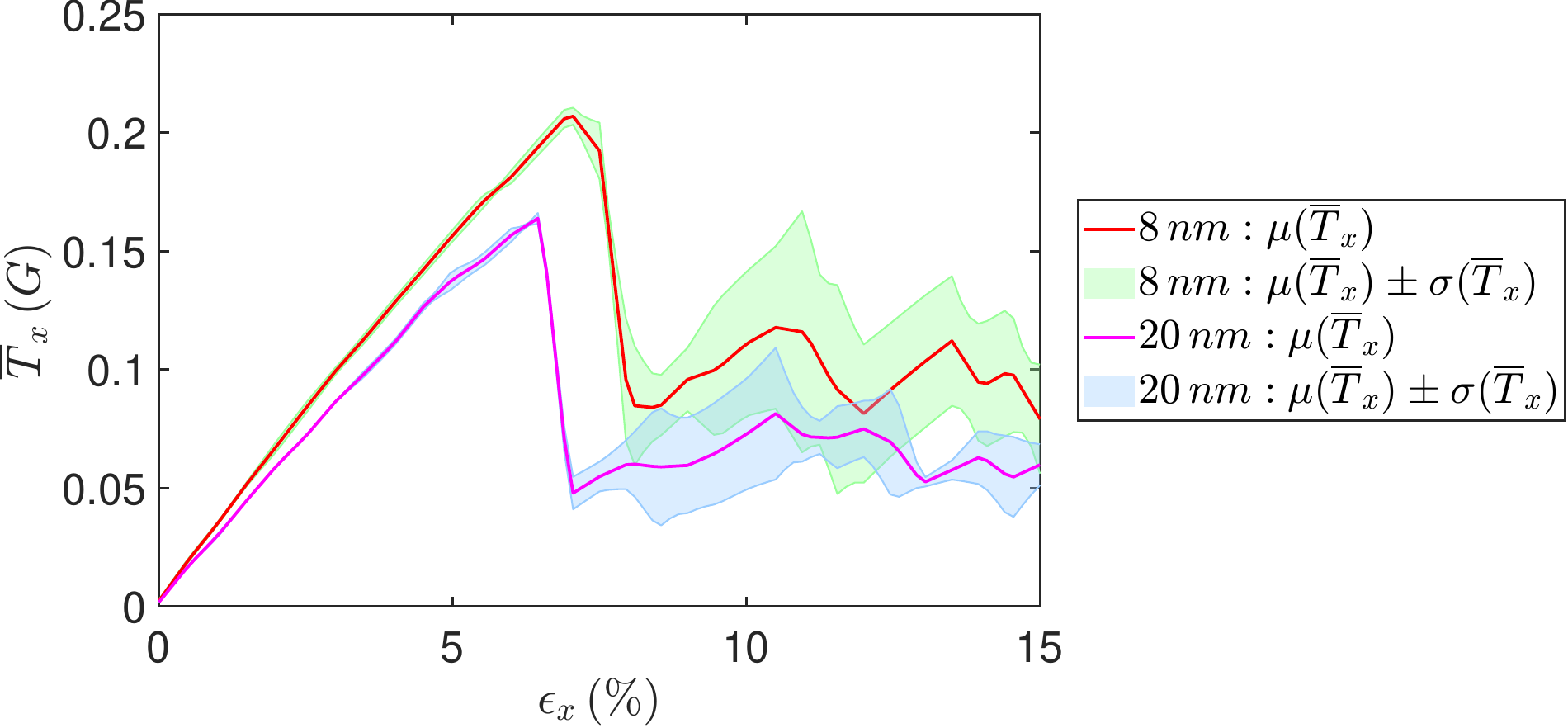}
    \caption{Evolution of mean and standard deviation of $\overline{T}_{x}$ across different runs corresponding to different initial velocity distributions and for different sample sizes with strain $(\epsilon_{x})$. The mean averaged stress $\mu(\overline{T}_{x})$ is marked using the solid curve while the shaded envelope around it shows the range of $\mu(\overline{T}_{x}) \pm \sigma (\overline{T}_{x})$, where $\sigma (\overline{T}_{x})$ is the standard deviation of the averaged stress.} 
    \label{fig:tension_envelope}
\end{figure}

\item \textbf{Effect of size on the tensile properties of the sample:} 
We performed uniaxial tension tests on samples of various sizes $L_s = \{4, 8, 20, 25, 30\} \, nm$ to understand the effect of sample size on the mechanical properties of the material. We then fit a linear curve to the stress–strain data in the $0~  \text{to}~ 1 \, \%$ strain range and the slope of this fitted line is taken as the Young's modulus ($E$). The yield strength ($\sigma_{ys}$) is taken as the value of the averaged normal stress ($\overline{T}_{x}$) at which the first dislocation nucleation in slow time is observed. This is done by tracking the mean atomic positions of every atom calculated using Eq.~\eqref{eq:mean_pos}. Figure~\ref{fig:tension_elastic_mod} shows the elastic modulus plotted as a function of size. It is observed that the Young's modulus decreases significantly as the size increases. Similarly, Figure~\ref{fig:tension_yield_strength} shows the variation of yield strength with size, indicating a decreasing trend with increasing sample size. 

The increase in Young's modulus with decreasing sample size is because of the increase in the surface to volume ratio. Surface atoms have higher stiffness compared to bulk atoms, which increases the Young's modulus. The decrease in yield strength with increasing sample size is because larger samples have more surface area, and hence higher number of potential sites for dislocation nucleation (which are usually parts of the sample with higher local stress concentration such as edges and corners), compared to smaller samples. Although done for very different sample sizes, the trends observed in our simulation results align with the experimental observations of \cite{uchic2004sample}, discussed later in Section~\ref{sec:exp_uchic}. 
              
\begin{figure}[tbh!]
    \centering
    \begin{subfigure}{0.5\linewidth}
        \centering
        \includegraphics[width=\linewidth]{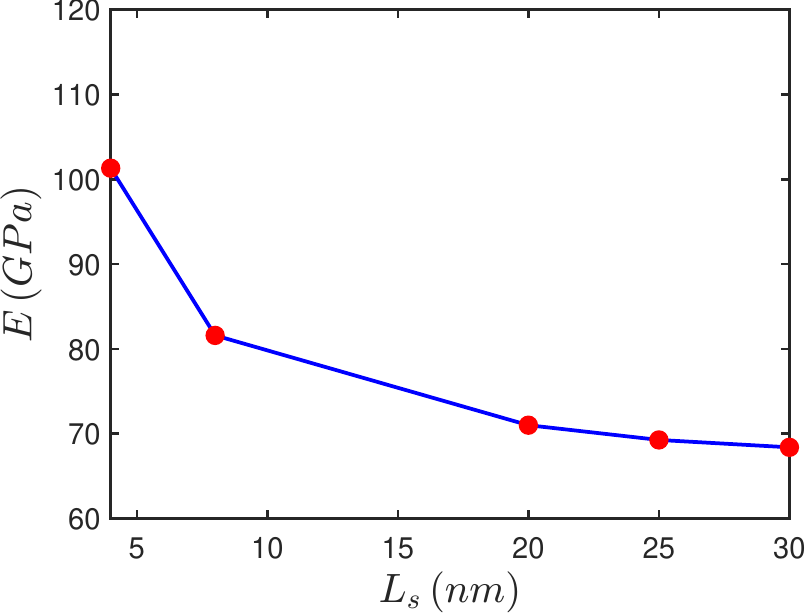}
        \caption{Variation of Young's modulus ($E$) with sample size $(L_s)$.}
        \label{fig:tension_elastic_mod}
    \end{subfigure}
    \hfill
    \begin{subfigure}{0.48\linewidth}
        \centering
        \includegraphics[width=\linewidth]{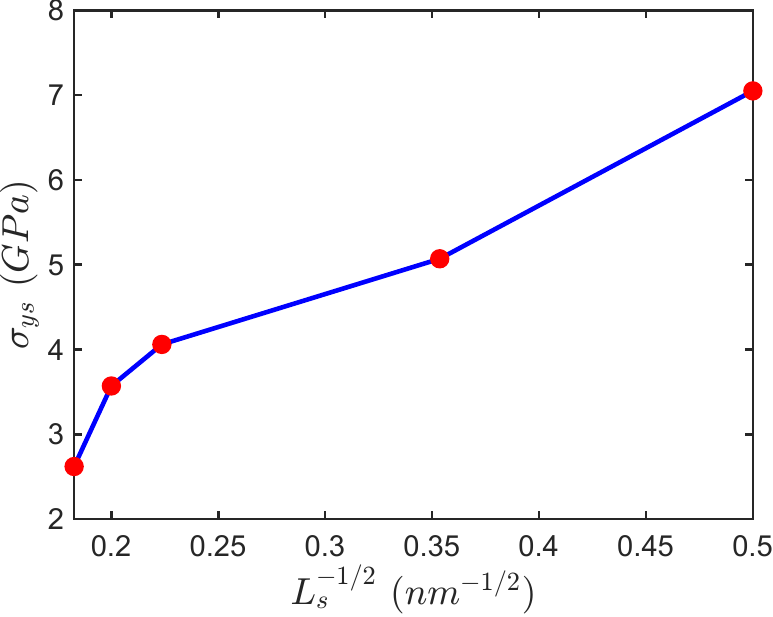}
        \caption{Variation of Yield strength ($\sigma_{ys}$) with sample size $(L_s^{-1/2})$.}
        \label{fig:tension_yield_strength}
    \end{subfigure}
    \caption{Dependence of mechanical properties on sample size.}
    \label{fig:tension_tensile_property}
\end{figure}

\item \textbf{Effect of initial state of system on the stress-strain curve:}   
We carry out five simulations with different initial atomic velocity distributions at the same initial temperature of $T_0=300 \, K$ for the $8 \, nm$ and $20 \, nm$ blocks, each. In Fig.~\ref{fig:tension_seed_effect}, we have shown the variation of stress response for different initial states of the system. It is observed that although the initial temperature of the system remains the same, different initial velocity distributions of atoms lead to variation in stress response. This is the primary reason for the stochasticity of the response observed in our simulations. As observed and discussed previously, a smaller sample size leads to harder responses and higher serrations in the stress-strain curve. 
  
\begin{figure}[tbh!]
    \centering
    \includegraphics[width=0.7\textwidth]{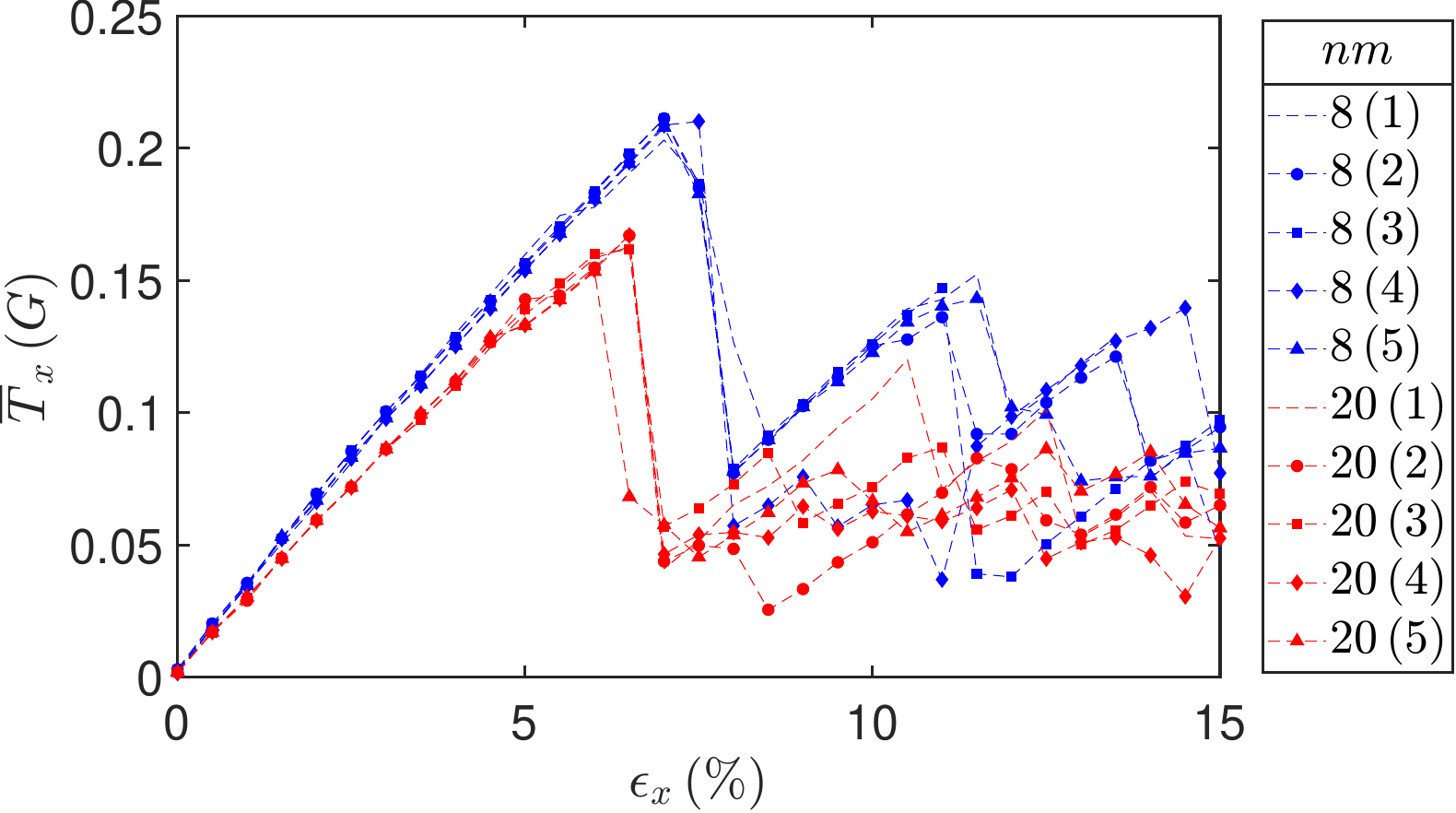}
    \caption{Averaged normal stress $(\overline{T}_{x})$ with strain $(\epsilon_{x})$ with different initial atomic velocity and same temperature for different sample sizes under uniaxial tension. Curves labeled 1–5 in brackets represent independent simulations for the same sample size but with different initial atomic velocity distributions. Blue curves correspond to the $8~nm$ sample, while red curves correspond to the 20 nm sample.}
    \label{fig:tension_seed_effect}
\end{figure}

\item \textbf{Effect of applied strain rate on stress-strain curve:}
We perform uniaxial tension simulations on the 20 nm block under applied strain rates of $10^{-4} \, s^{-1}$ and $10^{-3} \, s^{-1}$. Figure~\ref{fig:tension_rate_effect} shows the variation of stress with the applied strain rate. It is observed that although the elastic slopes are nearly identical, the plastic deformation behavior, governed by dislocation nucleation and evolution, depends on the applied strain rate. In particular, the stress levels after yielding remain consistently above those at a lower strain rate.

At higher strain rates, the material has less time for dislocations to nucleate and move, so higher stresses are required to activate plastic deformation mechanisms. Conversely, at lower strain rates, dislocations have more time to multiply and glide, leading to lower flow stresses. Plastic deformation involves overcoming energy barriers for dislocation nucleation and motion. Faster deformation (higher strain rate) means less time for thermal fluctuations to assist these processes, requiring larger applied stresses.

\begin{figure}[h!]
    \centering
    \includegraphics[width=0.55\textwidth]{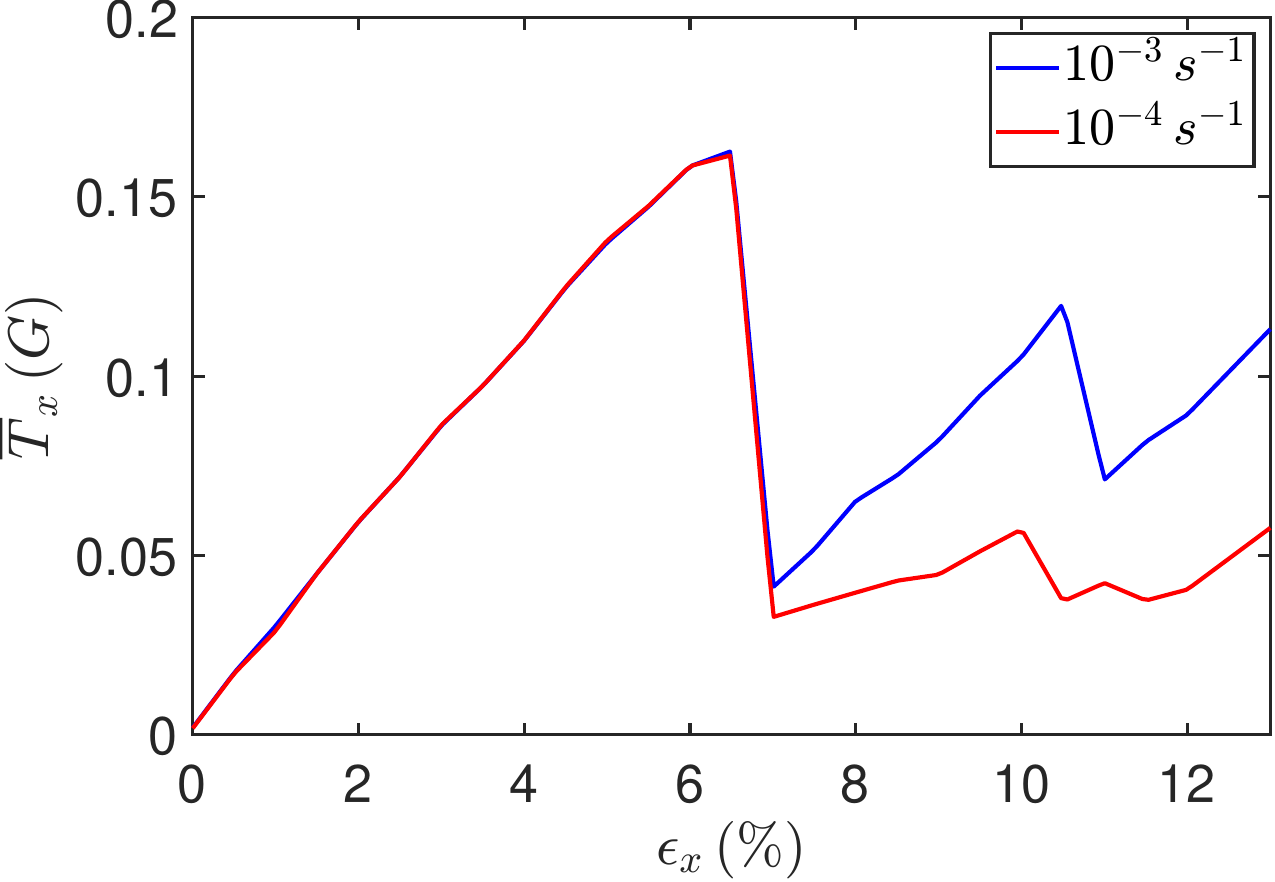}
    \caption{Evolution of averaged normal stress $(\overline{T}_{x})$ with strain $(\epsilon_{x})$ for $20 \, nm$ sample under uniaxial tension with different applied strain rates of $10^{-4}~s^{-1}$ and $10^{-3}~s^{-1}$.}
    \label{fig:tension_rate_effect}
\end{figure}   

\item \textbf{Effect of initial sample temperature on stress-strain curves:}

We initialize the atomic velocities corresponding to different initial temperatures. For each initial temperature, we run 3 simulations from different initial velocity distribution corresponding to the same initial temperature. The evolution of the mean kinetic energy and mean potential energy from the 3 simulations for each initial temperature are shown in Fig. \ref{fig:ke_strain_temp} and Fig. \ref{fig:pe_strain_temp} respectively. It is evident that there is a rise in both kinetic and potential energy with increasing initial temperature, at almost all strain values. The rise in kinetic energy is due to the increased initial kinetic energy supplied to the system at higher temperature. The rise in potential energy is because atoms vibrate at larger distance from their mean position due to increased kinetic energy.  Fig. \ref{fig:stress_strain_temp} shows the evolution of the stress (averaged over 3 simulations for each initial temperature) with strain. The initial yielding occurs at lower stress with increasing temperature as expected. Till about $7.5\%$ strain, the stress-strain curves show a drop with increasing temperature. However, at higher strains, the serrations due to stochasticity of dislocation motion and exit, particularly for smaller samples, cause stress at the same strain to lower at some strains with lowering of the initial temperature. Hence the serrations dominate over the effect of increased atomic vibrations at higher strains. 

For comparison, the results of the corresponding lattice statics calculations are provided (denoted as $0 \,K$), showing significant differences between them and the slow-strain rate MD results at finite temperatures, both considered as models for real quasi-static mechanical response. The lattice statics calculations were performed in the usual manner - load increments were provided and the nearest local equilibrium of the potential energy of the system, comprising the interatomic interactions and the potential energy of the loads, was sought by (local) energy minimization.

\begin{figure}[h!]
    \centering
    \begin{subfigure}{0.48\textwidth}
        \centering
        \includegraphics[width=\linewidth]{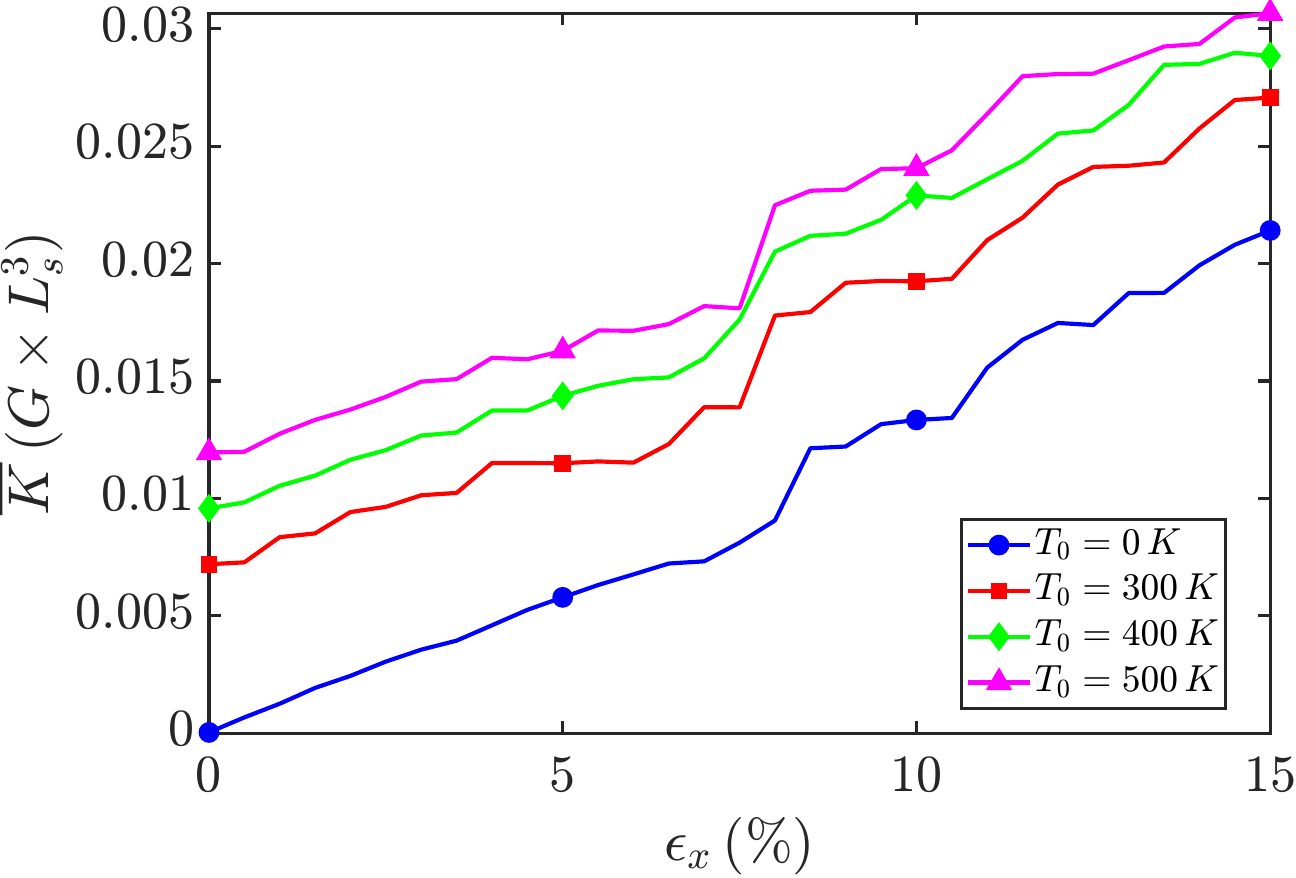}
        \caption{Averaged kinetic energy $(\overline{K})$}
        \label{fig:ke_strain_temp}
    \end{subfigure}
    \hfill
    \begin{subfigure}{0.48\textwidth}
        \centering
        \includegraphics[width=\linewidth]{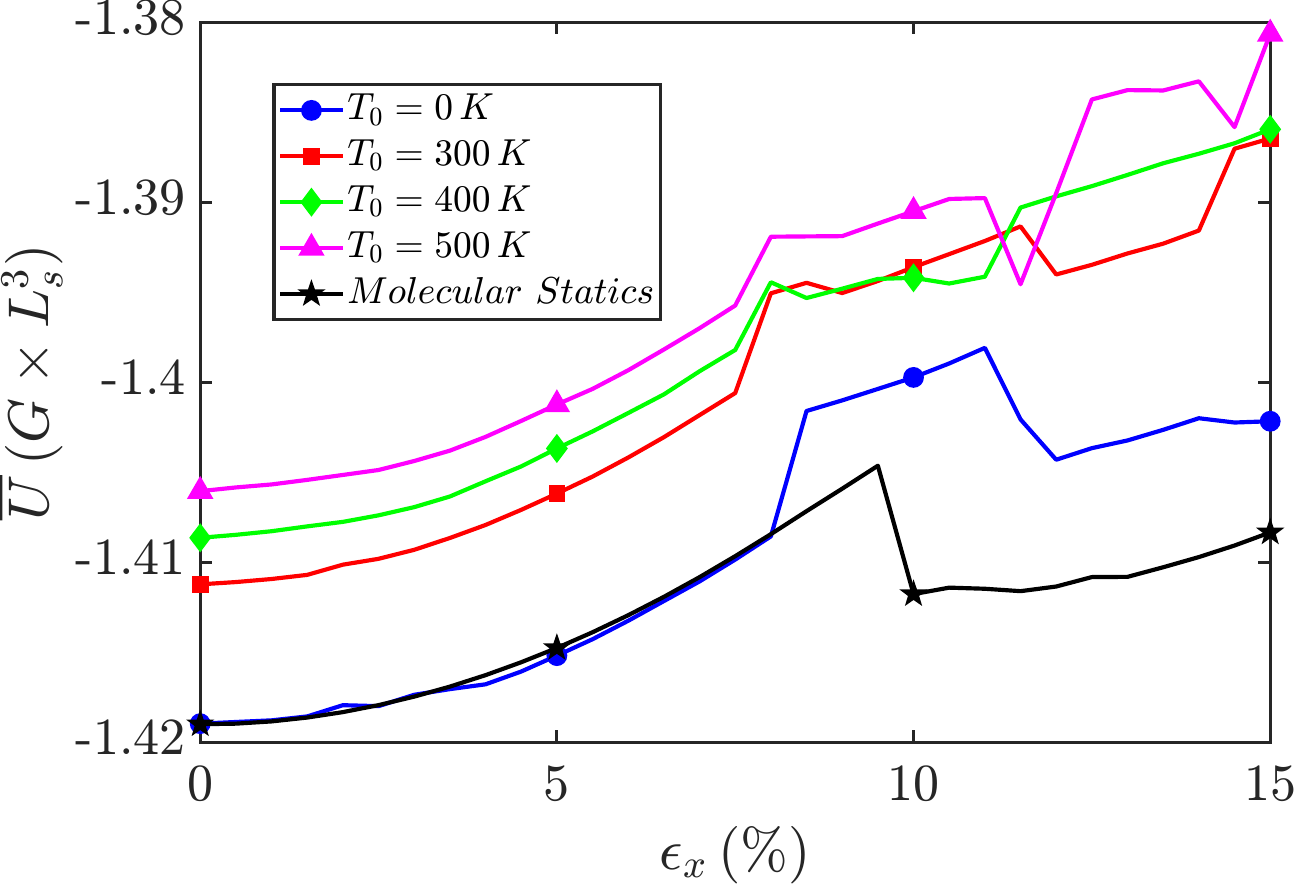}
        \caption{Averaged potential energy $(\overline{U})$}
        \label{fig:pe_strain_temp}
    \end{subfigure}
    
    \caption{Evolution of averaged kinetic $(\overline{K})$ and potential $(\overline{U})$ energy (each curve represents the mean over 3 different simulations for the same initial temperature)  with strain for different initial temperatures for $8 \, nm$ sample in uniaxial tension.}
    \label{fig:energy_strain_temp}
\end{figure}

\begin{figure}[h!]
    \centering
    \includegraphics[width=0.6\textwidth]{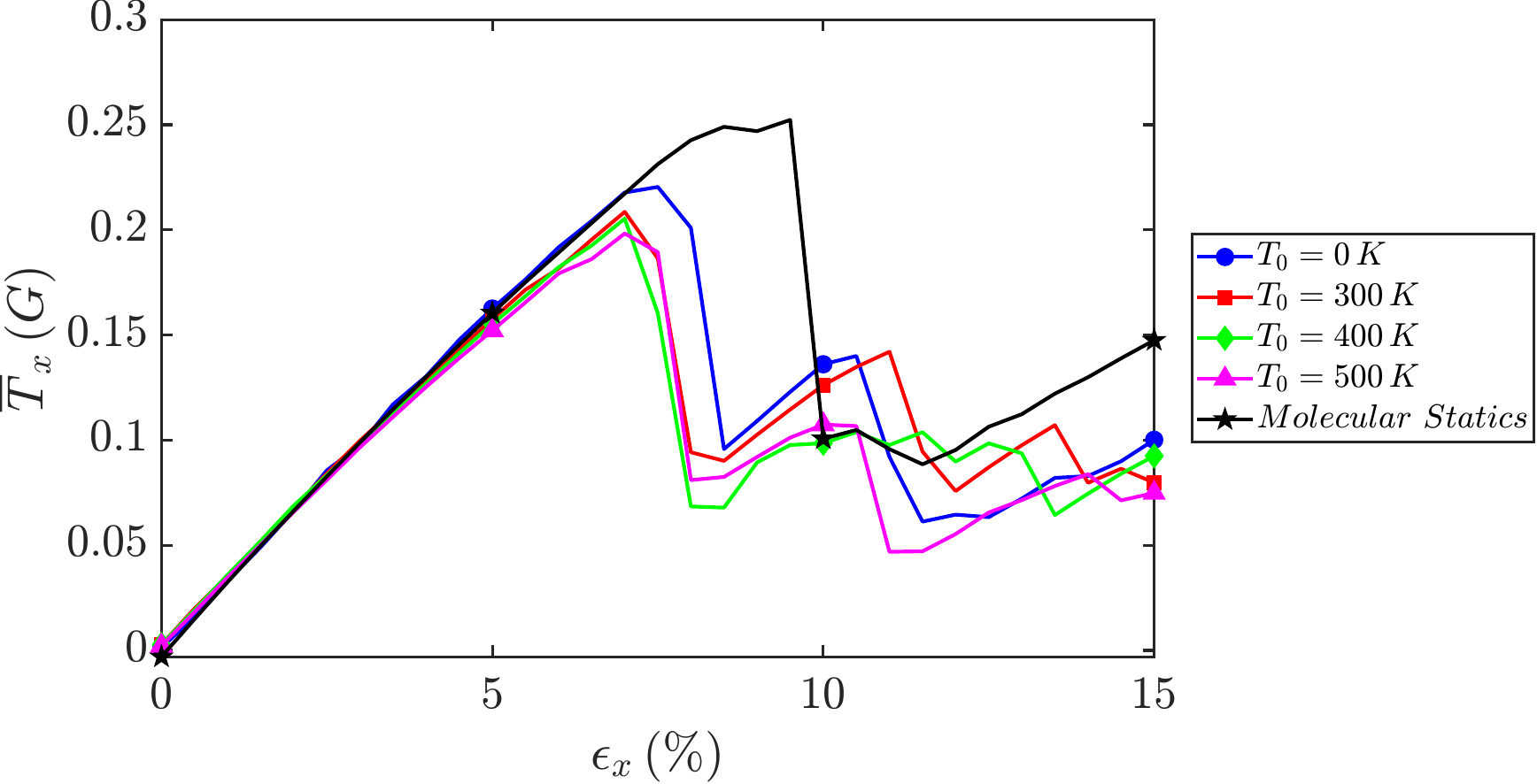}
    \caption{Evolution of averaged normal stress $(\overline{T}_x)$ (each curve represents the mean over 3 different simulations for the same initial temperature) with strain for different initial temperatures for $8 \, nm$ sample in uniaxial tension.}
    \label{fig:stress_strain_temp}
\end{figure}

\item \textbf{Evolution of the microstructure:} 
Tracking the evolution of averaged atomic positions at different slow times allows us to visualize the evolution of the dislocation microstructure in slow time. The dislocation microstructure at different strains for a $20 \, nm$ sample are generated using Dislocation Analysis DXA in OVITO software \cite{ovito}, as shown in Fig. \ref{fig:dxa_snap_tension}. We observe that dislocations nucleate for the first time at $6.5 \, \%$ strain and the dislocation network evolves with applied strain. The networks consists of predominantly Shockley partials and Stair-rod dislocations. Dislocations exit the free surface and this leads to creation of slip steps, which can be seen particularly in Fig. \ref{subfig:dxa_snap_tension_c} and Fig. \ref{subfig:dxa_snap_tension_d}.  

Fig. \ref{fig:mean_atom_tension} show the mean atomic positions at $7 \, \%$ strain. A defective zone of the crystal is zoomed in to show plastic deformation. Fig. \ref{fig:atom_tension_colored} shows the atoms in colors which are categorized into FCC, HCP and other crystal structures (using DXA analysis in OVITO). Only FCC atoms are in their perfect lattice positions while the HCP and other atoms are out of their regular lattice positions, caused by plastic deformation due to dislocation motion.  

\begin{figure}[ht!]
    \centering
    \begin{subfigure}{0.4\linewidth}
        \centering
        {\includegraphics[width=\linewidth]{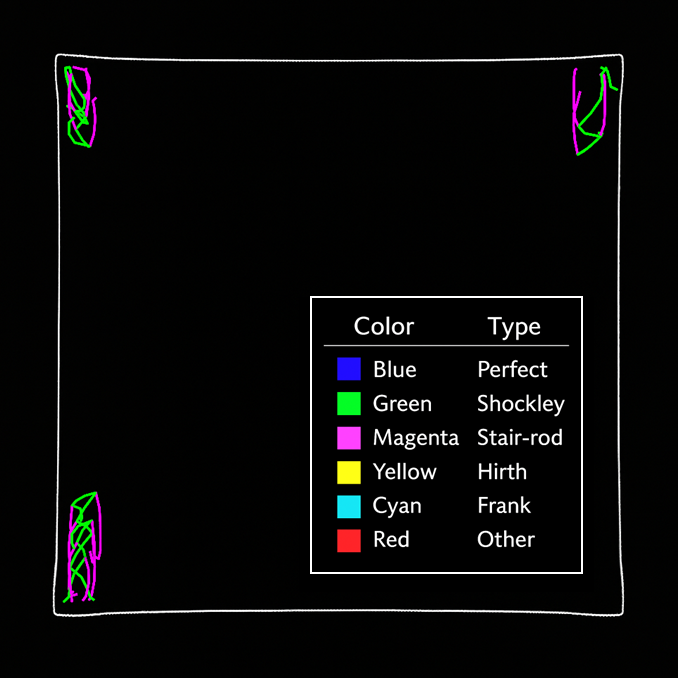}}
        \caption{$6.5 \, \%$ strain.}
    \end{subfigure}
    \hspace{1cm}
    \begin{subfigure}{0.4\linewidth}
        \centering
        {\includegraphics[width=\linewidth]{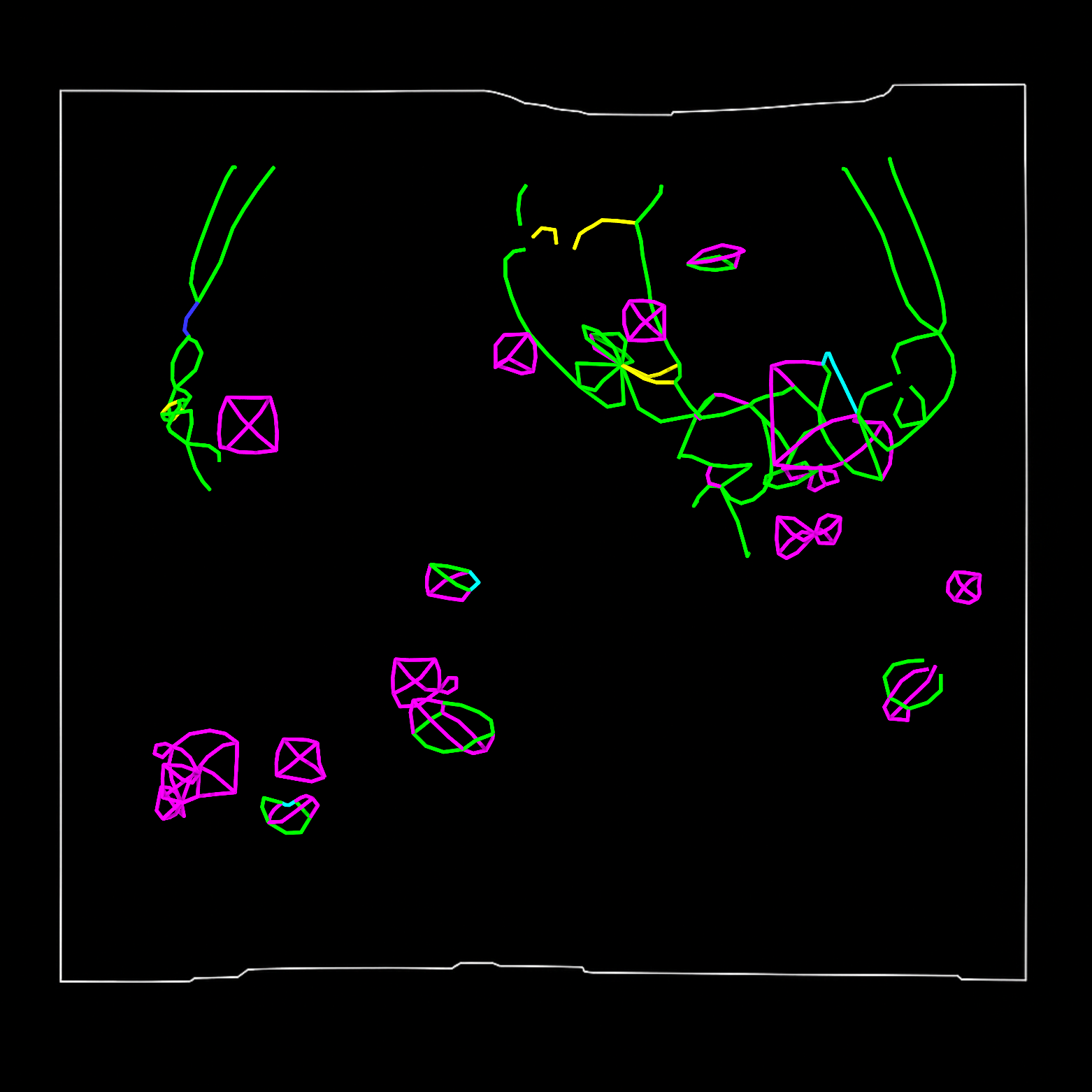}}
        \caption{$13 \, \%$ strain.}
    \end{subfigure}
    \hspace{1cm}
    \begin{subfigure}{0.4\linewidth}
        \centering
        {\includegraphics[width=\linewidth]{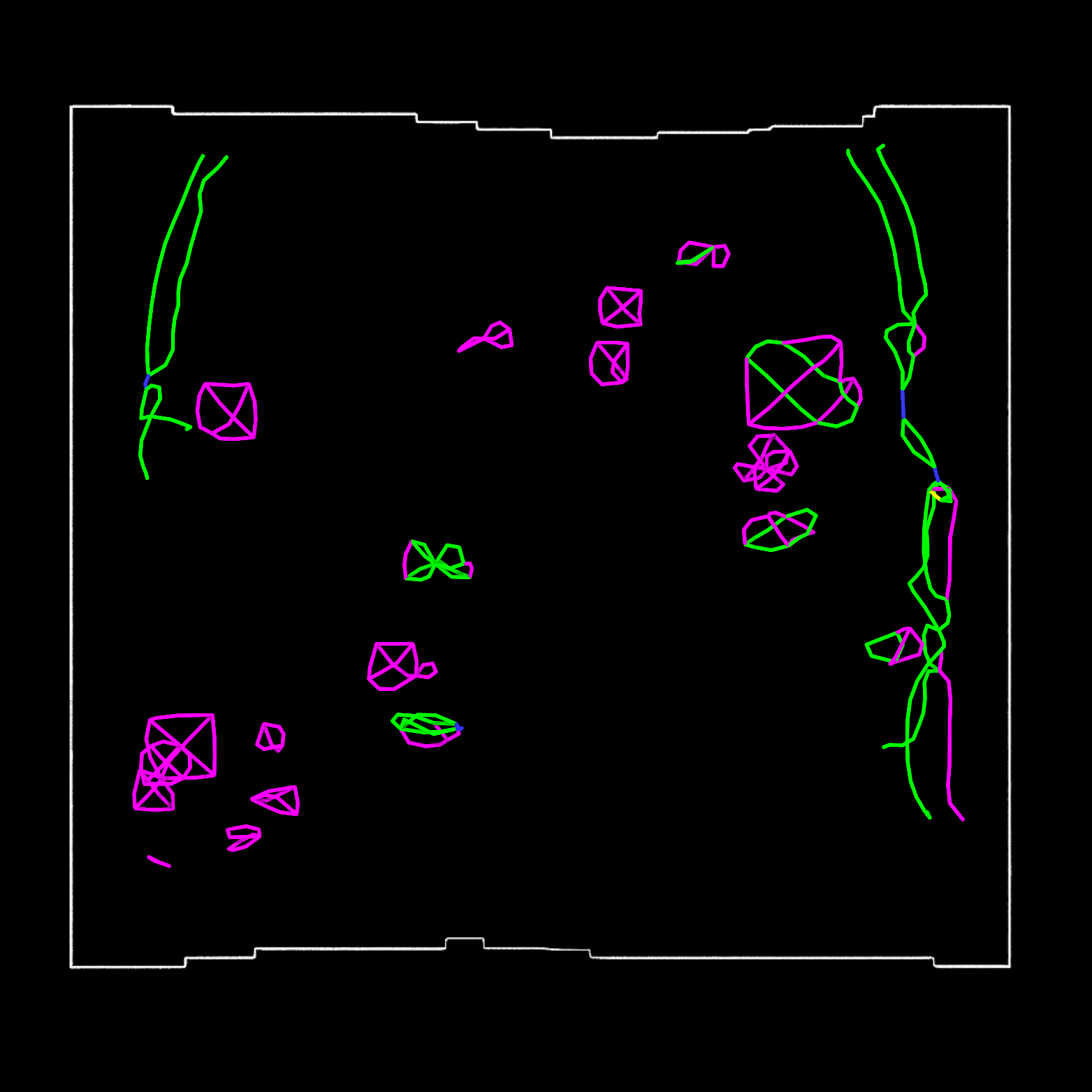}}
        \caption{$13.5 \, \%$ strain.}
        \label{subfig:dxa_snap_tension_c}
    \end{subfigure}
    \hspace{1cm}
    \begin{subfigure}{0.4\linewidth}
        \centering
        {\includegraphics[width=\linewidth]{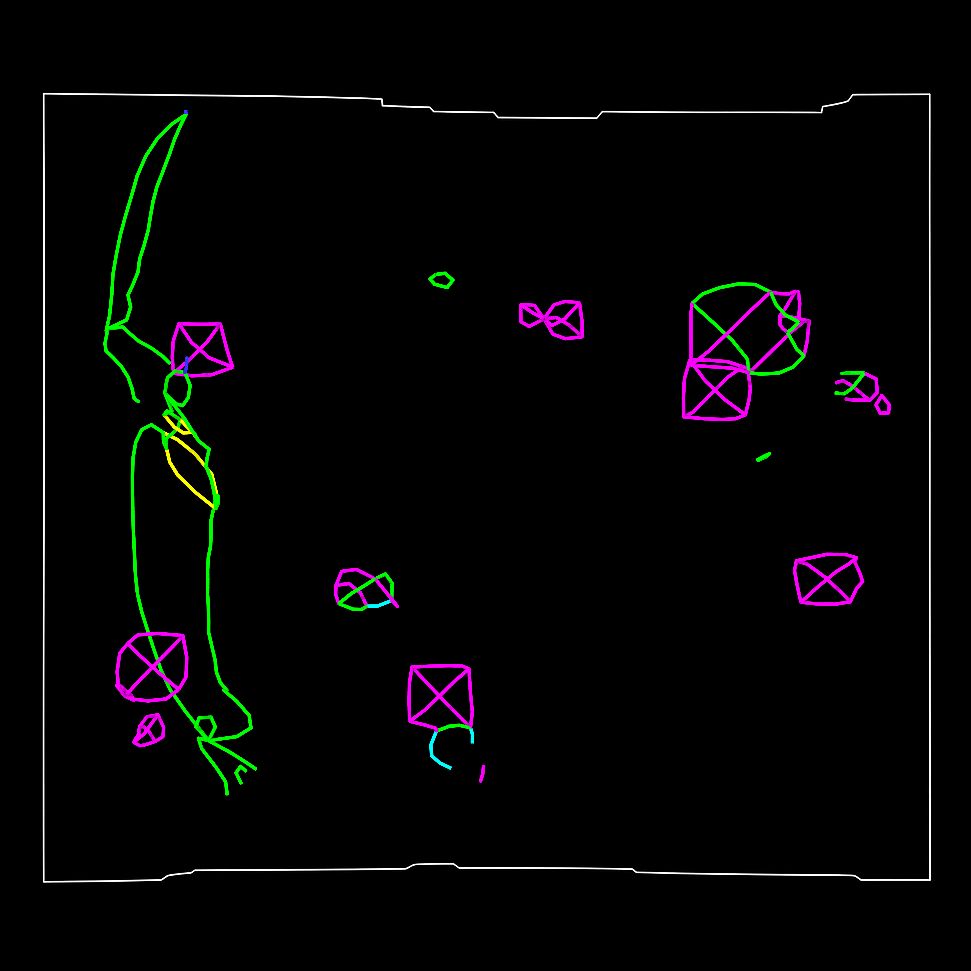}}
        \caption{$14.5 \, \%$ strain.}
        \label{subfig:dxa_snap_tension_d}
    \end{subfigure}
    \caption{Evolution of dislocation microstructure in slow time, under uniaxial tension at different strains for $20 \, nm$ sample. Different types of dislocations are indicated using the colors shown in the legend. The slip steps created by exiting dislocations can be observed clearly at higher strains.}
\label{fig:dxa_snap_tension}
\end{figure}

\begin{figure}[ht!]
    \centering
    \includegraphics[width=0.8\linewidth]{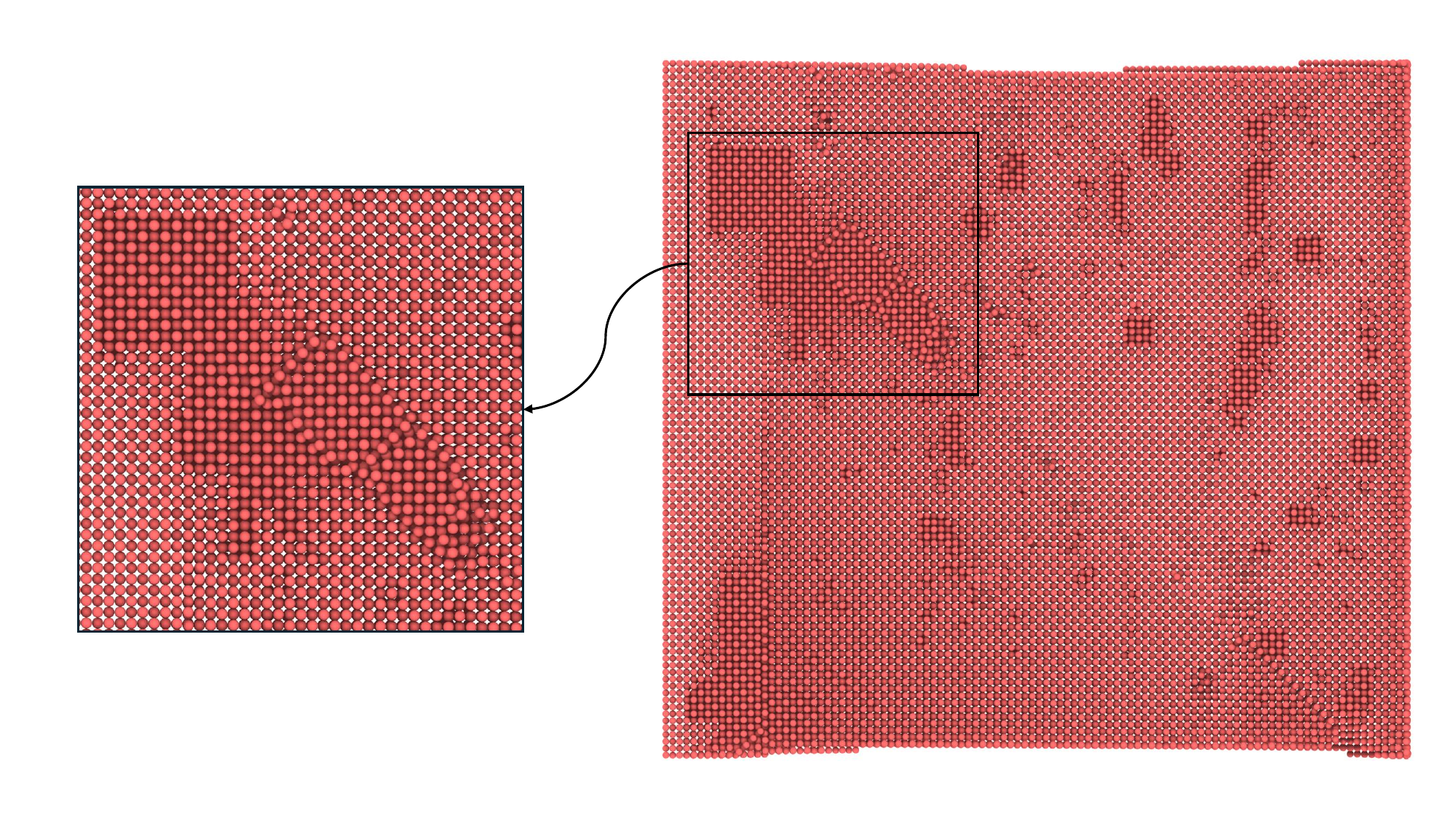}
    \caption{Mean atomic positions at $7 \, \%$ strain for $20 \, nm$ sample in uniaxial tension. A defective zone of the sample is zoomed in to show slip and plastic deformation.}
    \label{fig:mean_atom_tension}
\end{figure}
\begin{figure}[ht!]
    \centering
    \includegraphics[width=0.8\linewidth]{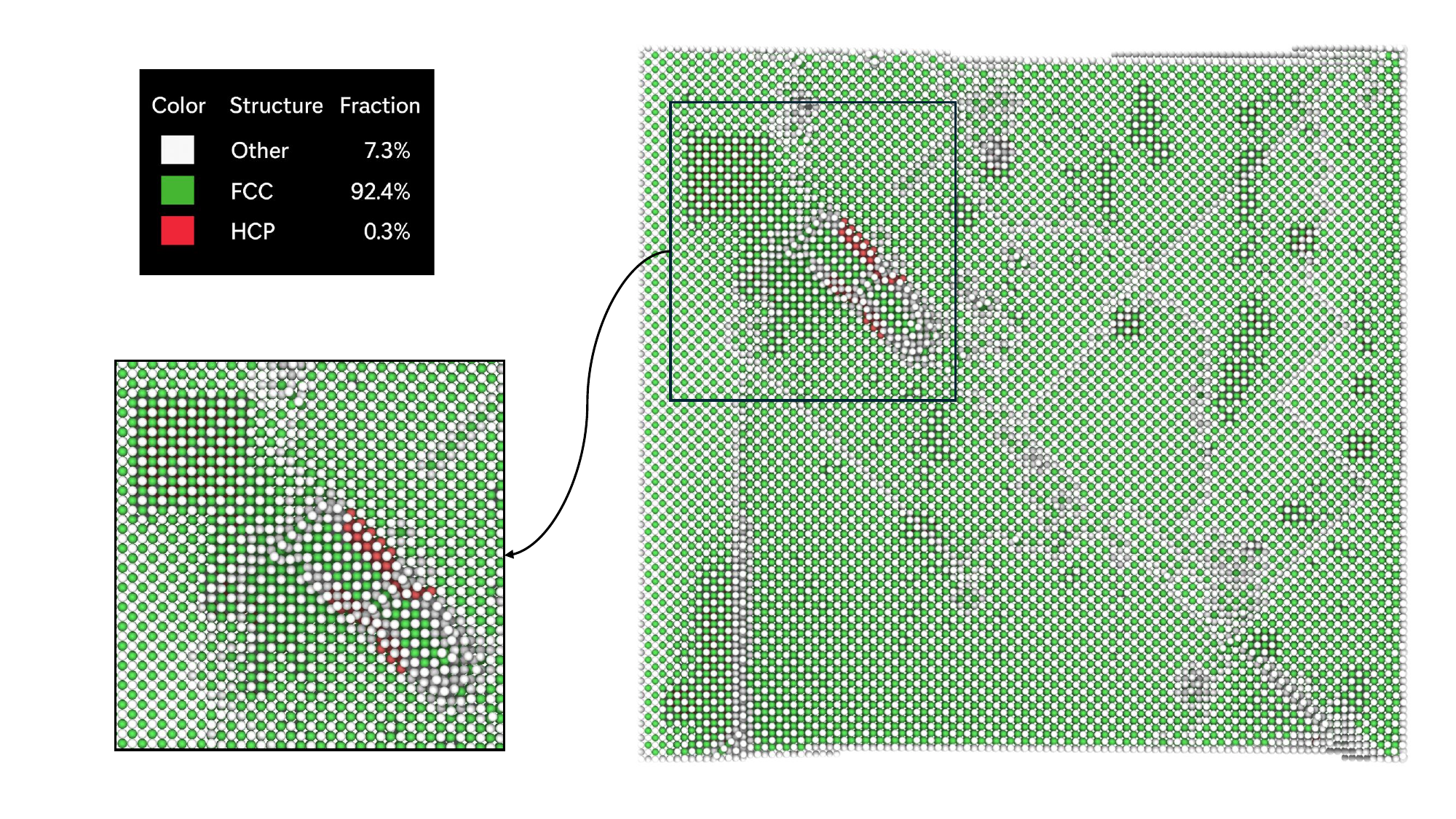}
    \caption{Mean atomic position at $7 \, \%$ strain for $20 \, nm$ sample in uniaxial tension. Atoms are colored as FCC, HCP or others. Only FCC atoms in green are at their regular lattice positions. Atoms which are not in their regular lattice positions such as the HCP atoms in red and other atoms in white result from slip caused by dislocation motion. A defective zone of the crystal is zoomed in to show the atoms which are out of their lattice positions.}
    \label{fig:atom_tension_colored}
\end{figure}

\end{itemize}  

\subsection{Uniaxial compression test}
\label{sec:comp_test}
Since compression tests are more common for micron sized samples, we have performed uniaxial compression tests on samples of sizes of 8 and 20 nm blocks to demonstrate the applicability of our method. We have used the applied strain rate of $(-10^{-3})~s^{-1}$. 

\begin{itemize} 

\item \textbf{Evolution of averaged kinetic energy, averaged potential energy and averaged normal stress:}
The evolution of the slow variables - averaged potential energy $\overline{U}$, the averaged kinetic energy $\overline{K}$ and the averaged normal stress $\overline{T}_{xx}$ and the relative errors between them and the corresponding values of the measure $\overline{U}_d$, $\overline{K}_d$ and $\overline{T}_{d_{xx}}$ for $8 \, nm$ sample are shown in Fig. ~\ref{fig:pe_comp},~\ref{fig:ke_comp} and~\ref{fig:stress_comp} respectively. We observed that the relative error for the slow variables, averaged potential energy, averaged kinetic energy and averaged normal stress are within 1$\%$, 8$\%$ and 5$\%$, respectively at all strains. A discussion on the physical reasoning behind these observations is provided in Section~\ref{sec:tension_test}.

\begin{figure}[h!]
    \centering
    \begin{subfigure}{0.49\linewidth}
        \centering
        {\includegraphics[width=\linewidth]{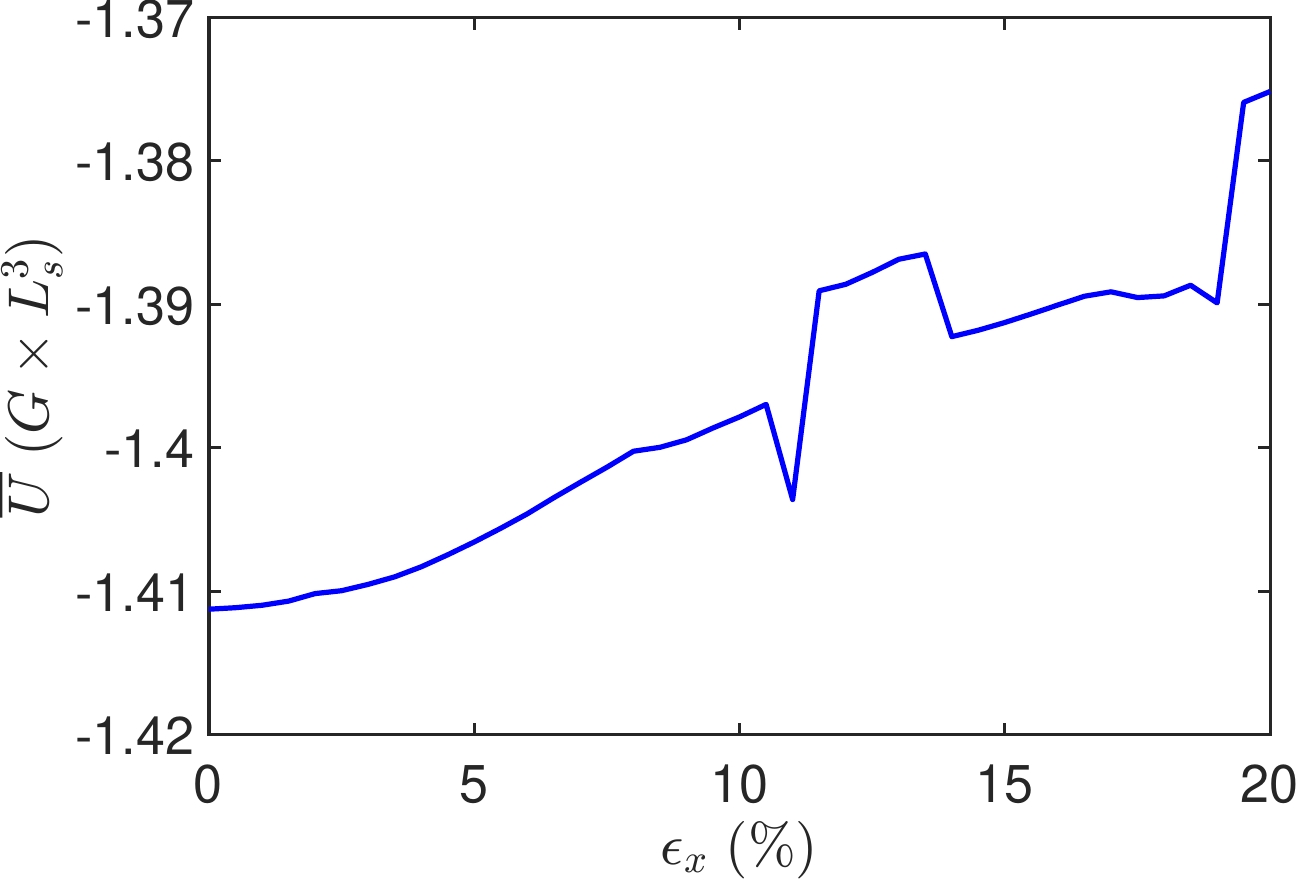}}
        \caption{}
    \end{subfigure}
    \hfill
    \begin{subfigure}{0.48\linewidth}
        \centering
        {\includegraphics[width=\linewidth]{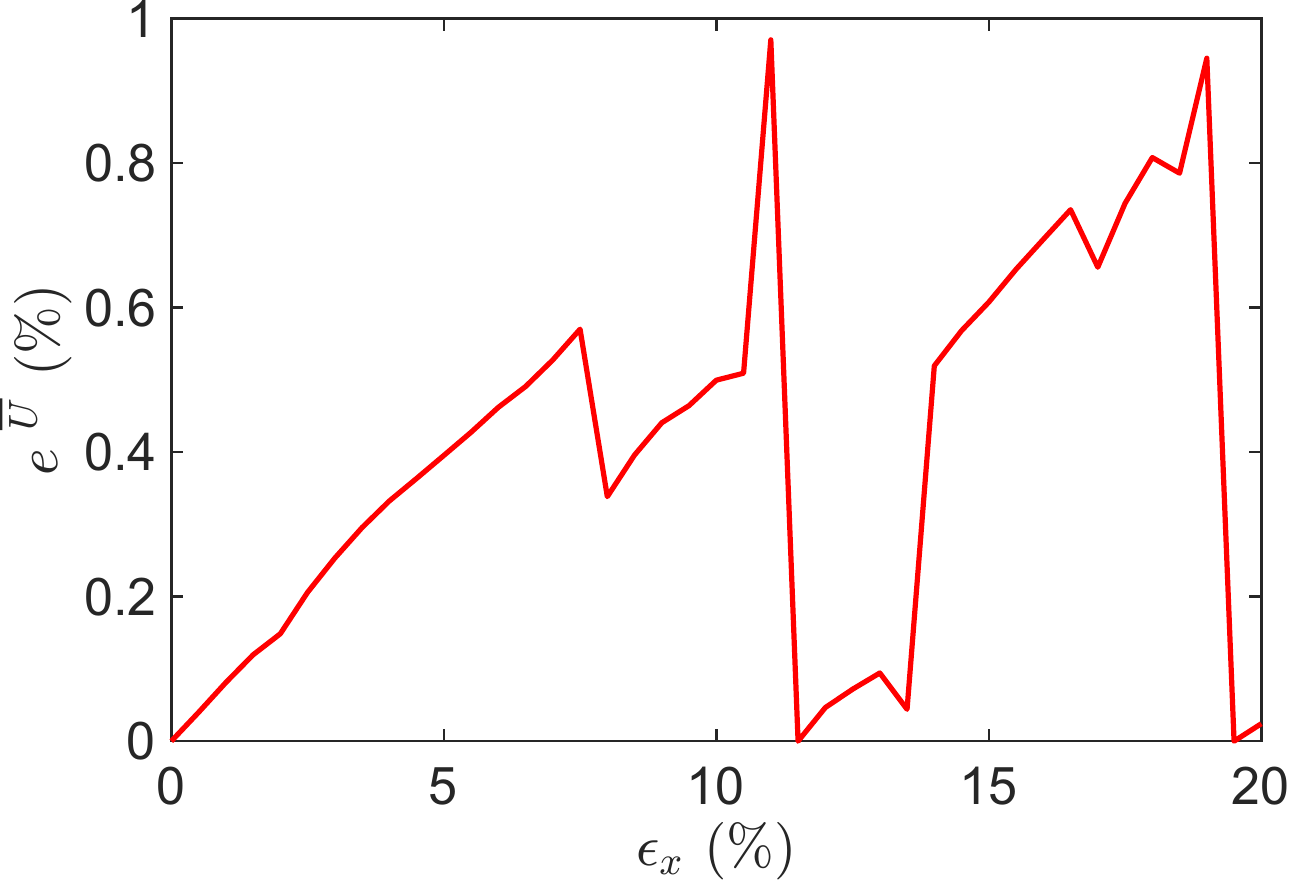}}
        \caption{}
    \end{subfigure}
    \caption{Evolution of averaged potential energy $(\overline{U})$ and relative error $(\epsilon^{\overline{U}})$ with strain $(\epsilon_{x})$ for $8 \, nm$ sample under uniaxial compression. $L_s$ in the y-axis denotes the size of the sample.}
    \label{fig:pe_comp}
\end{figure}
\begin{figure}[h!]
    \centering
    \begin{subfigure}{0.5\linewidth}
        \centering
        {\includegraphics[width=\linewidth]{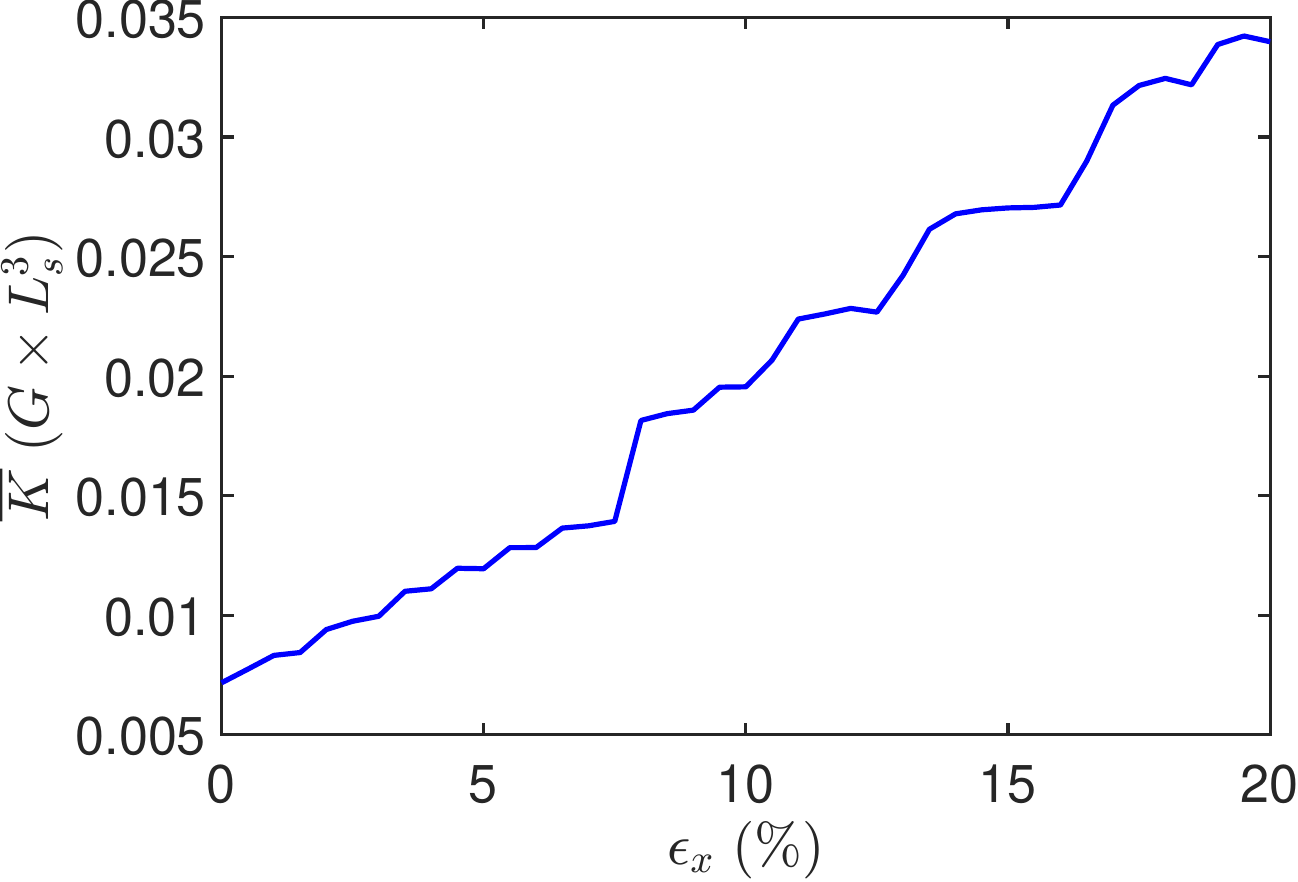}}
        \caption{}
    \end{subfigure}
    \hfill
    \begin{subfigure}{0.48\linewidth}
        \centering
        {\includegraphics[width=\linewidth]{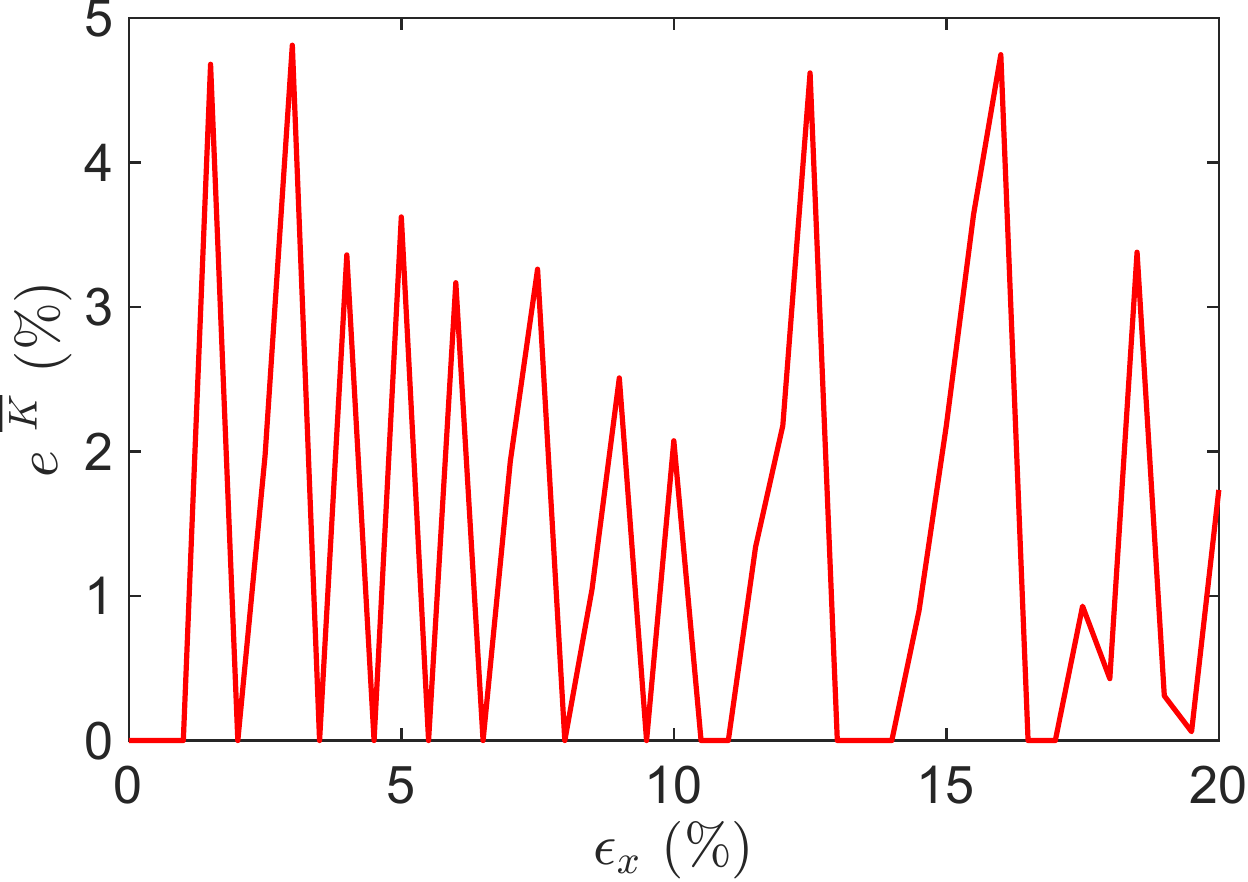}}
        \caption{}
    \end{subfigure}
    \caption{Evolution of averaged kinetic energy $(\overline{K})$ and relative error $(\epsilon^{\overline{K}})$ with strain $(\epsilon_{x})$ for $8 \, nm$ sample under uniaxial compression.}
    \label{fig:ke_comp}
\end{figure}
\begin{figure}[h!]
    \centering
    \begin{subfigure}{0.48\linewidth}
        \centering
        {\includegraphics[width=\linewidth]{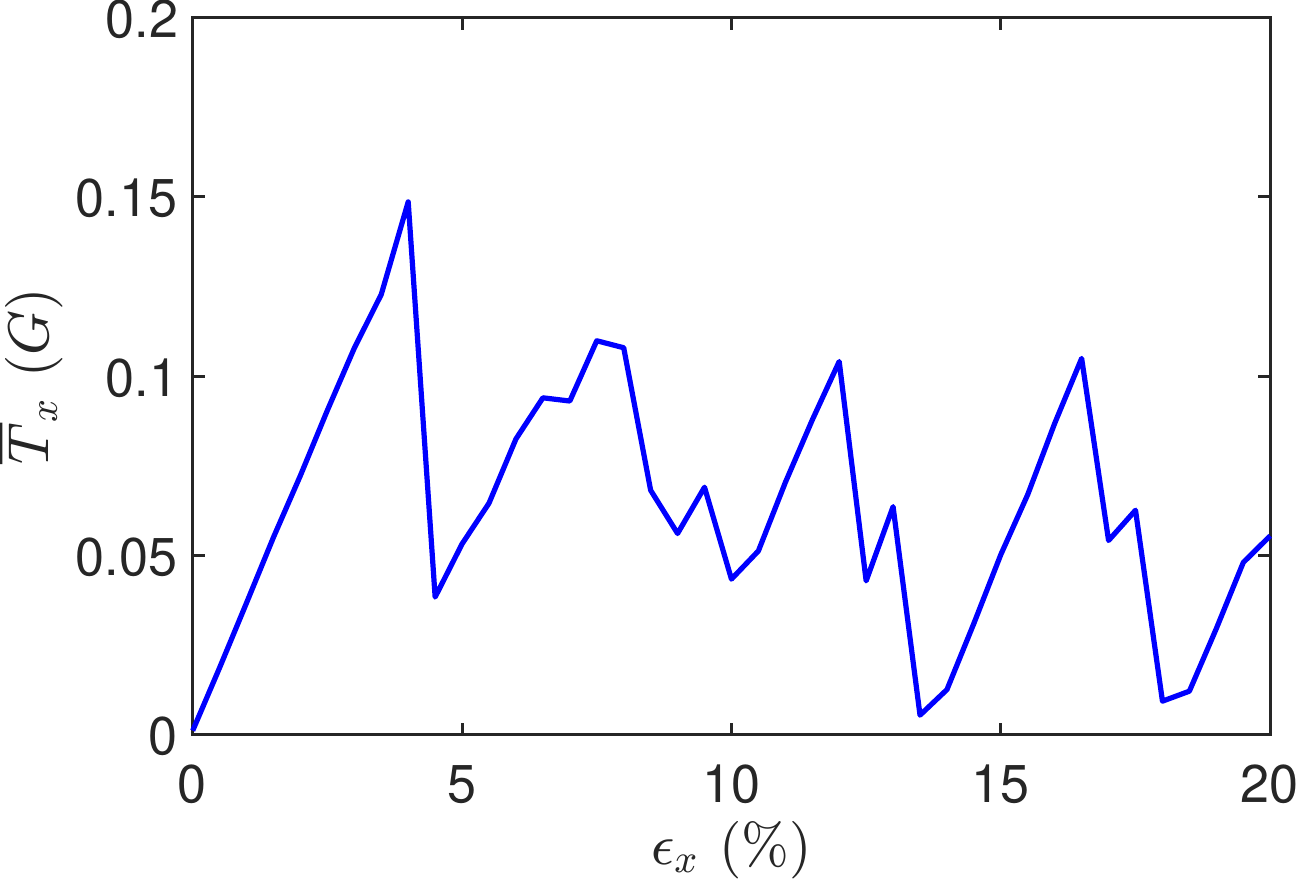}}
        \caption{}
    \end{subfigure}
    \hfill
    \begin{subfigure}{0.48\linewidth}
        \centering
        {\includegraphics[width=\linewidth]{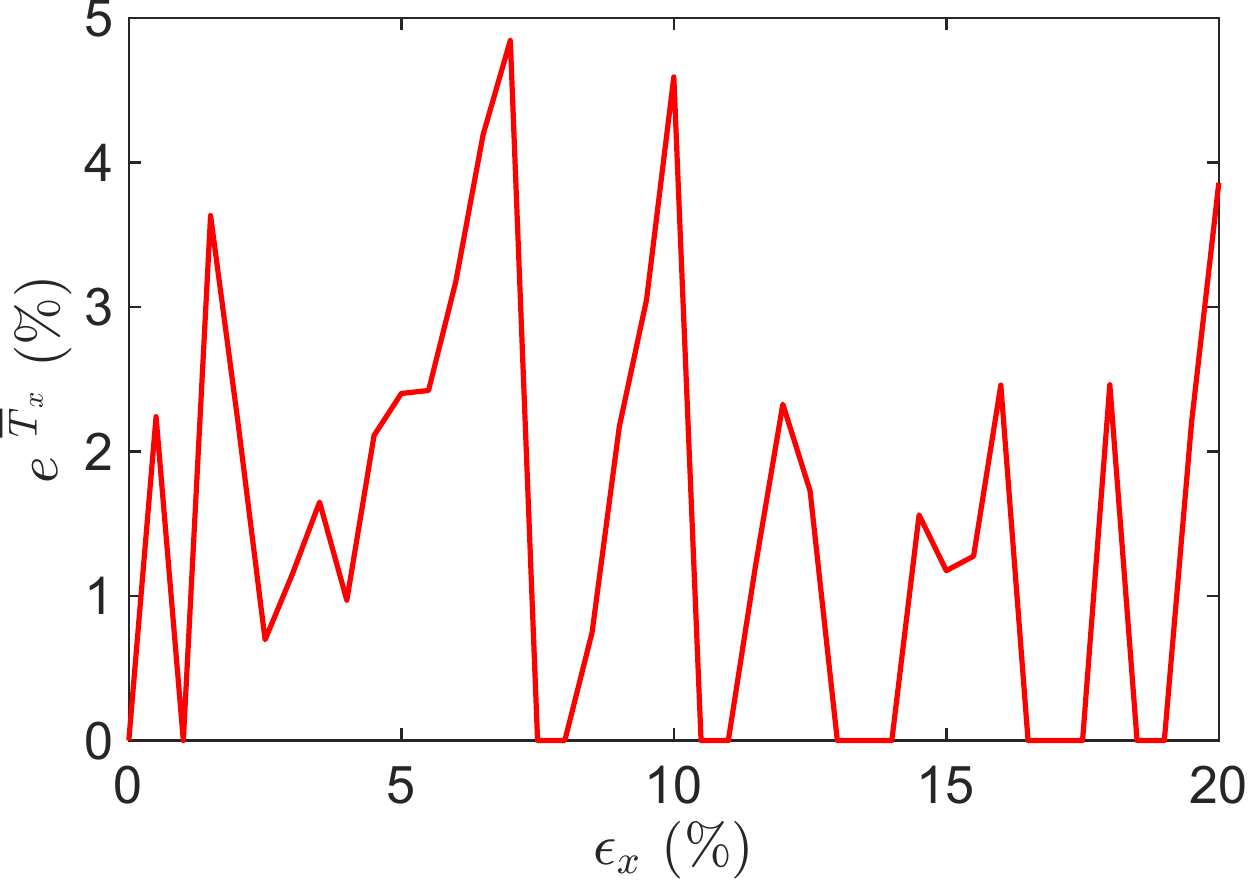}}
        \caption{}
    \end{subfigure}
    \caption{Evolution of averaged normal stress $(\overline{T}_{x})$ and relative error $(\epsilon^{\overline{T}_{x}})$ with strain $(\epsilon_{x})$ for $8 \, nm$ sample under uniaxial compression.}
    \label{fig:stress_comp}
\end{figure}

\item \textbf{Effect of sample size on the stress-strain curve:}  
To ensure statistical reliability, five and four simulations with different initial atomic velocity distributions are performed for the $8 \, nm$ and $20 \, nm$ blocks, respectively. For each sample size, the mean of the stress-strain curves are shown in Figure~\ref{fig:comp_size_effect}. Under the same applied loading rate, the $8 \, nm$ sample exhibits a higher stress response than the $20 \, nm$ sample, indicating a pronounced size effect. The discussion on the cause of size effect is provided in Section~\ref{sec:tension_test}. However, we observe that the elastic slope in compression is not dependent on the sample size, unlike in the tension case presented earlier. The yield stress of around $4 \, GPa$ is also lower than than $5 \, GPa$ which is observed in uniaxial tension. The stress-strain response is also softer in compression compared to tension as upon comparison of Fig. \ref{fig:tension_size_effect} and Fig. \ref{fig:comp_size_effect}. In general, a specimen is more stable in tension than compression because buckling (in slender specimens) and barrelling instabilities are avoided in the latter case. Although the sample sizes differ, the trends observed in our simulations are in qualitative agreement with the experimental observations of \cite{uchic2004sample}, as discussed later in Section~\ref{sec:exp_uchic}.
                       
\begin{figure}[h!]
    \centering
    \begin{subfigure}{0.48\linewidth}
        \centering
        \includegraphics[width=\linewidth]{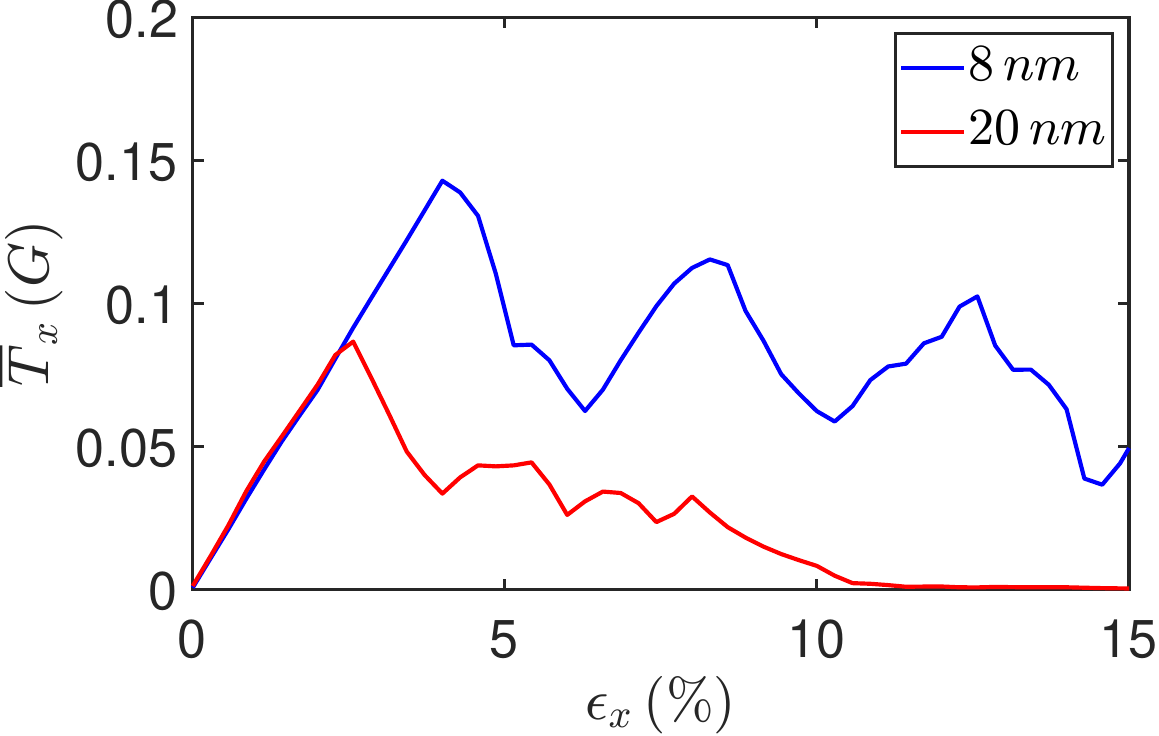}
        \caption{}
    \end{subfigure}
    \hfill
    \begin{subfigure}{0.48\linewidth}
        \centering
        \includegraphics[width=\linewidth]{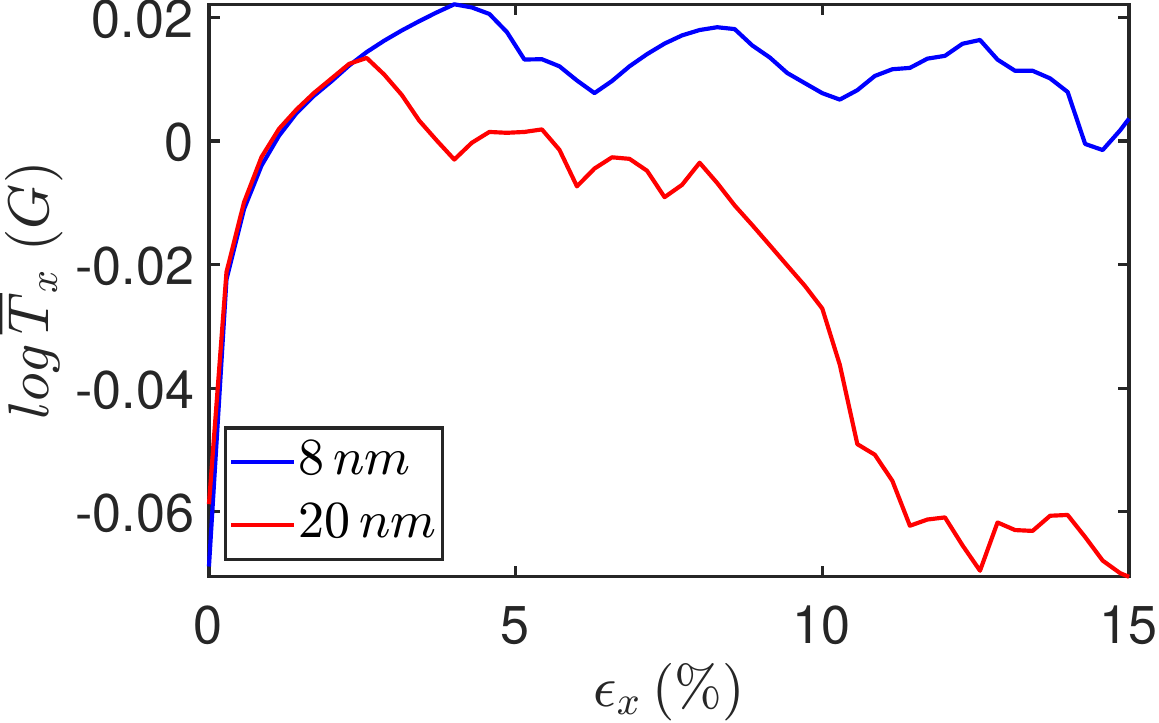}
        \caption{}
    \end{subfigure}
    \caption{Evolution of averaged normal stress $(\overline{T}_{x})$ with strain $(\epsilon_{x})$ for different sample sizes under uniaxial compression.}
        \label{fig:comp_size_effect}
\end{figure}

\item \textbf{Effect of size on standard deviation of mean stress:}
We carry out four and three simulations with different initial atomic velocity distributions for the $8 \, nm ~\text{and}~ 20 \, nm$ blocks, respectively. At a certain strain value, the mean and standard deviation of stress across those simulations are then calculated for both sizes. In Figure~\ref{fig:comp_envelope}, we show the variation of the standard deviation of stress for different sizes. It can be observed that the smaller $8 \, nm$ sample shows a higher standard deviation compared to the larger $20 \, nm$ block. To quantify this, we calculate the mean of the standard deviation. The values are $0.4285 \, GPa ~\text{and}~ 0.1087 \, GPa$ for the $8 \, nm ~\text{and}~ 20 \, nm$ sizes, respectively. A justification for this size effect is provided in Section~\ref{sec:tension_test}.
              
\begin{figure}[h!]
    \centering
    \includegraphics[width=0.8\textwidth]{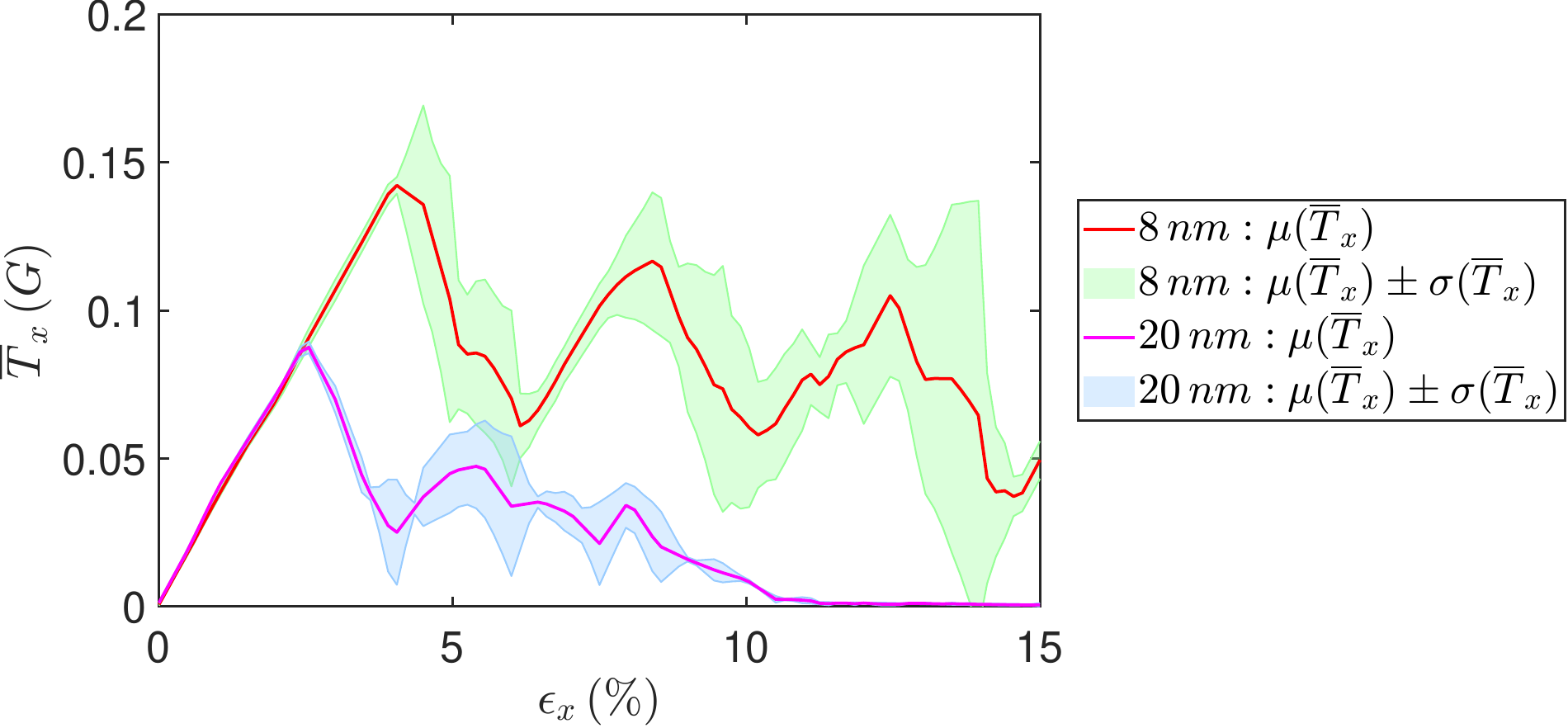}
    \caption{Evolution of mean and standard deviation of $\overline{T}_{x}$ across different runs corresponding to different initial velocity distributions and for different sample sizes with strain $(\epsilon_{x})$. The mean averaged stress $\mu(\overline{T}_{x})$ is marked using the solid curve while the shaded envelope around it shows the range of $\mu(\overline{T}_{x}) \pm \sigma (\overline{T}_{x})$, where $\sigma (\overline{T}_{x})$ is the standard deviation of the averaged stress.} 
    \label{fig:comp_envelope}
\end{figure}  

\item \textbf{Effect of the initial state of the system on the stress-strain curve:}  
We carry out four simulations with different initial atomic velocity distributions at same initial temperature of $T_0=300 \, K$ for the $8 \, nm$ and $20 \, nm$ blocks, each. In Fig.~\ref{fig:comp_size_effect} we have shown the variation of stress response for different initial state of the system. It is observed that although the initial temperature of the system remains the same, different initial velocity distributions of atoms lead to variation in stress response, similar to our observation in uniaxial tension case. 
              
\begin{figure}[h!]
    \centering
    \includegraphics[width=0.7\textwidth]{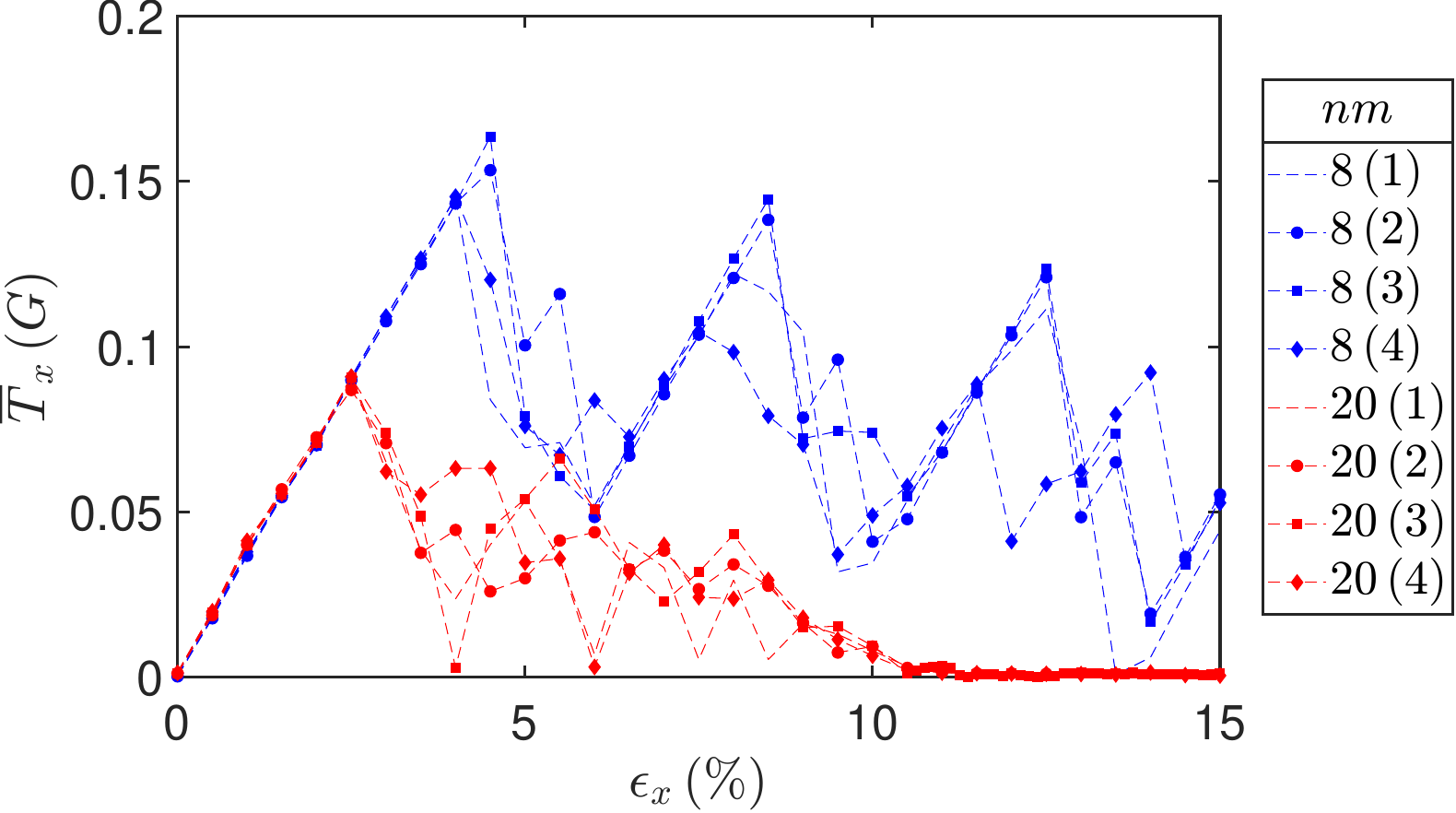}
    \caption{Averaged normal stress $(\overline{T}_{x})$ with strain $(\epsilon_{x})$ for different sample sizes with different initial atomic velocity and same temperature. Curves labeled 1–4 in brackets represent independent simulations for the same sample size but with different initial atomic velocity distributions. Blue curves correspond to the $8~nm$ sample, while red curves correspond to the 20 nm sample.}
    \label{fig:comp_seed_effect}
\end{figure}

\item \textbf{Evolution of the microstructure:} 
The dislocation microstructure at different strains, generated using Ovito as mentioned earlier, are shown in Fig. \ref{fig:snap_dxa_compression}. We observe that dislocations nucleate for the first time at 3\% strain and the dislocation network evolves with applied strain. The networks consists of predominantly Shockley partials along with some  Hirth, Stair-rod and full (Perfect) dislocations.

Fig. \ref{fig:mean_atoms_comp} show the mean atomic positions at 5\% strain. A defective zone of the crystal is zoomed in to show plastic deformation. Fig. \ref{fig:atoms_comp} shows the mean positions of atoms for a $20~nm$ sample under uniaxial compression at different values of strain. We observe that at sufficiently large strains ($\approx 11\%$ in this case), the sample changes its state and undergoes liquefaction, which is not observed in the case of uniaxial tension. Fig. \ref{fig:atoms_comp_colored} shows the atoms in colors which are categorized into FCC, HCP and other crystal structures (using DXA analysis in OVITO). Only FCC atoms are in their perfect lattice positions while the HCP and other atoms are out of their regular lattice positions, similar to our observation for the tension case presented earlier.   
\end{itemize}

\begin{figure}[h!]
    \centering
    \begin{subfigure}{0.35\linewidth}
        \centering
        {\includegraphics[width=\linewidth]{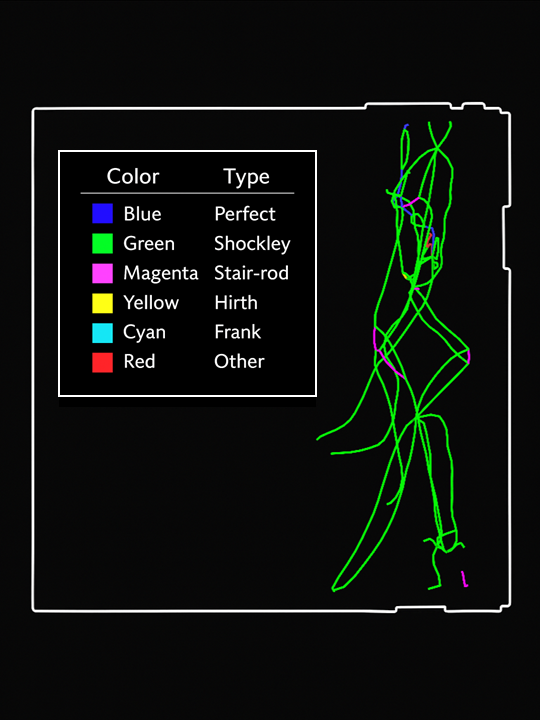}}
        \caption{$3 \, \%$ strain.}
    \end{subfigure}
    \hspace{1cm}
    \begin{subfigure}{0.35\linewidth}
        {\includegraphics[width=\linewidth]{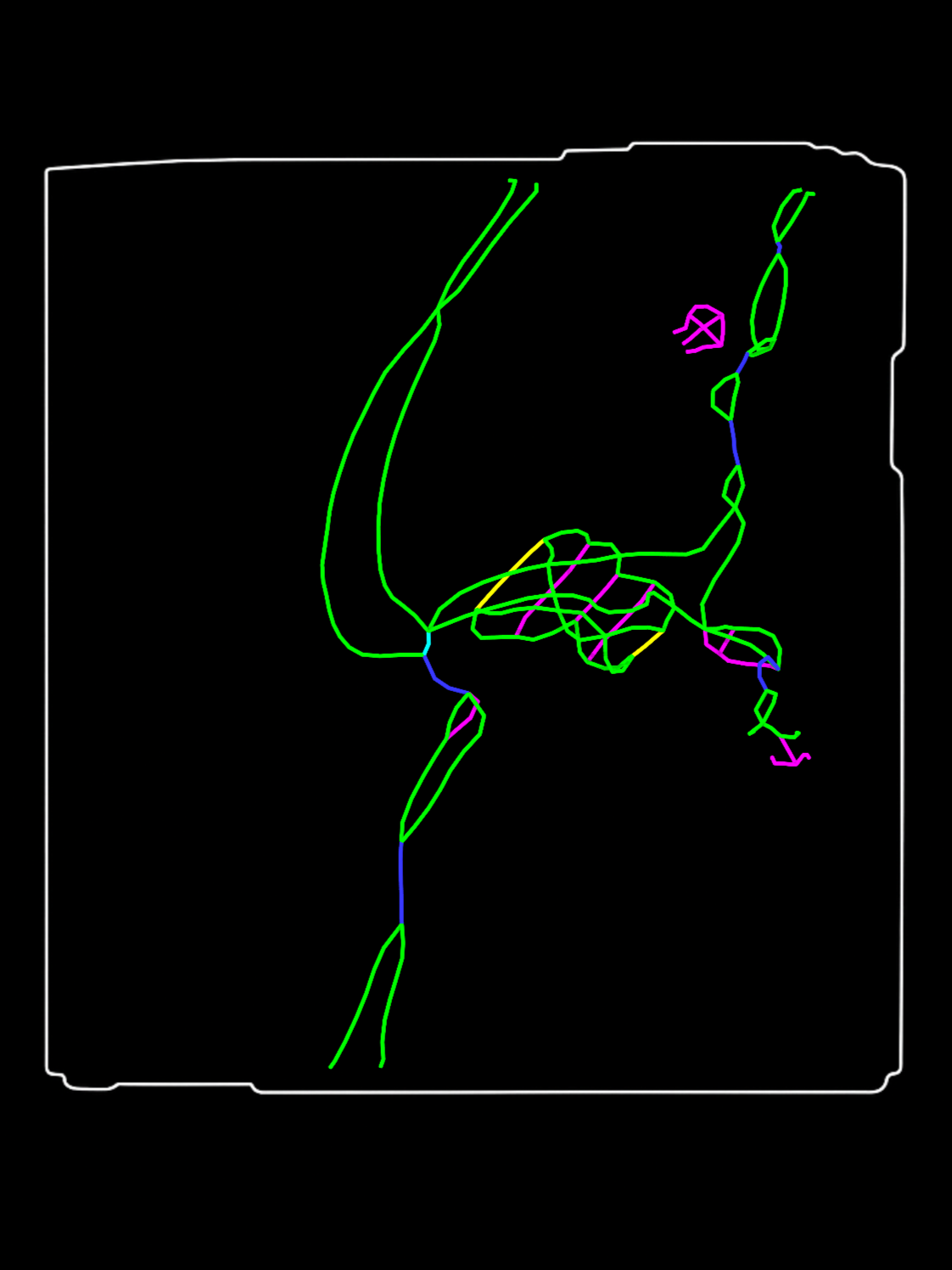}}
        \caption{$4 \, \%$ strain.}
    \end{subfigure}
    \begin{subfigure}{0.35\linewidth}
        {\includegraphics[width=\linewidth]{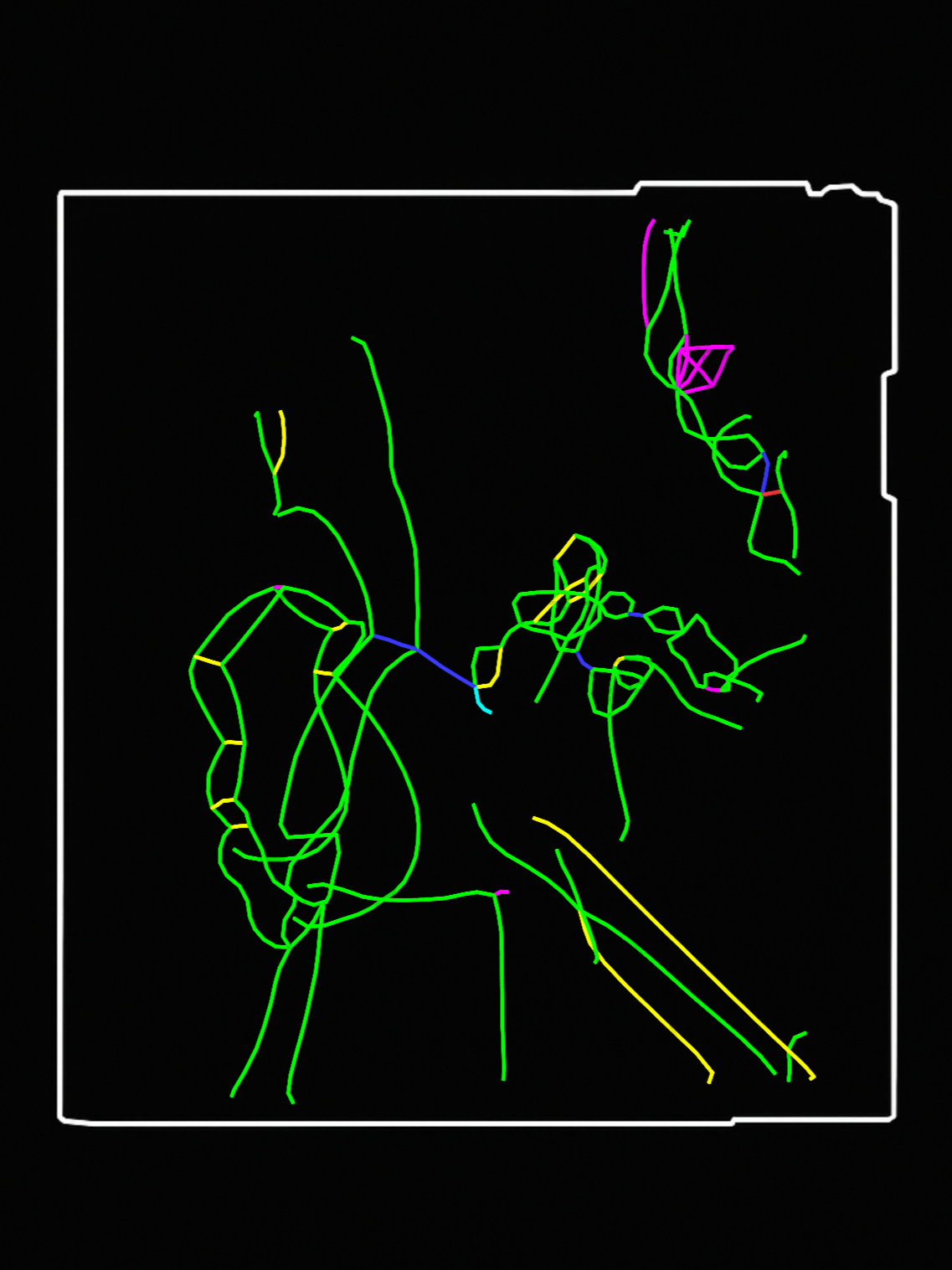}}
        \caption{$5.5 \, \%$ strain.}
    \end{subfigure}
    \hspace{1cm}
    \begin{subfigure}{0.35\linewidth}
        {\includegraphics[width=\linewidth]{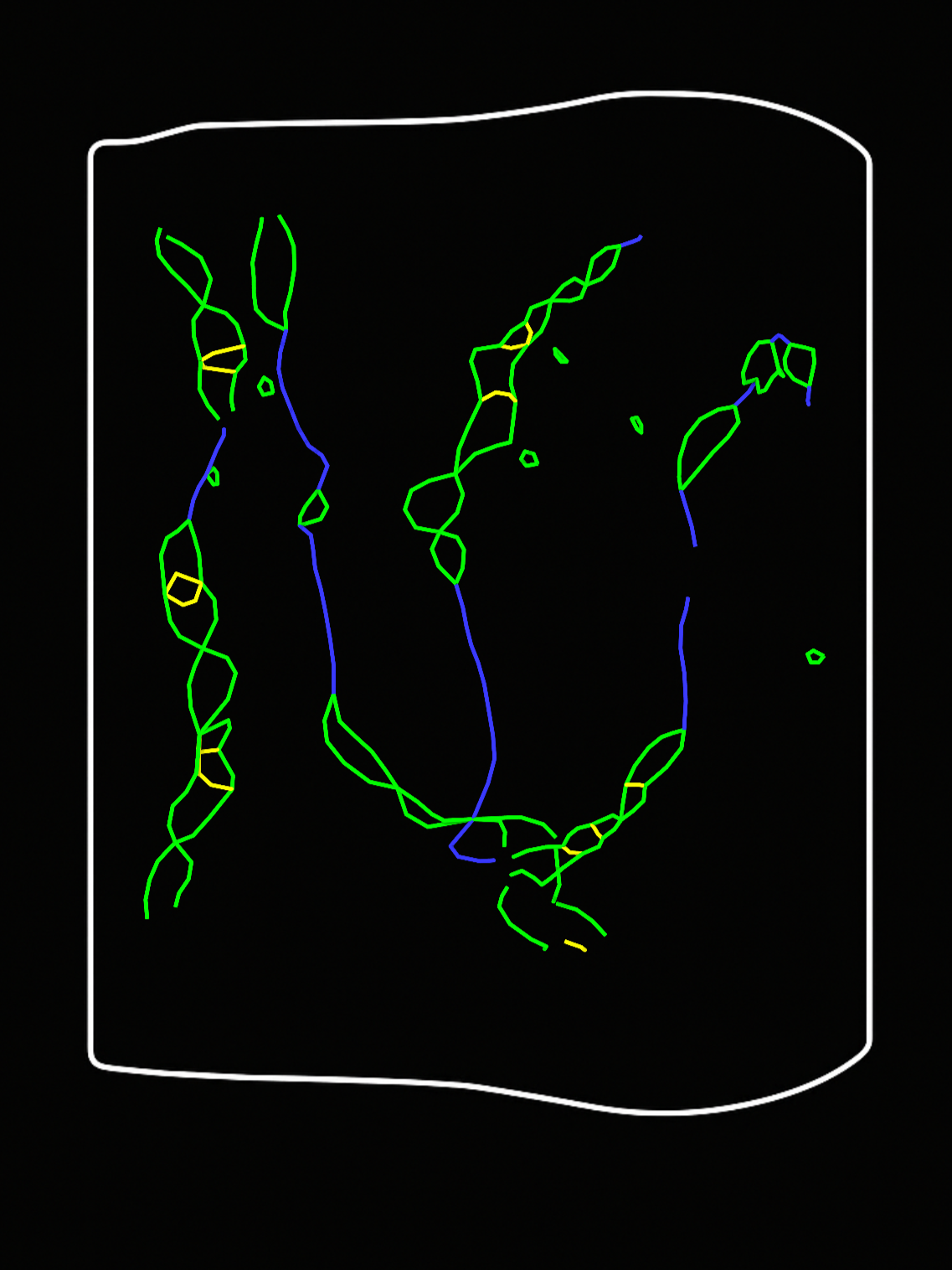}}
        \caption{$9 \, \%$ strain.}
    \end{subfigure}
    \caption{Evolution of dislocation microstructure in slow time, at different strains for $20 \, nm$ sample under uniaxial compression. Different types of dislocations are indicated using the colors shown in the legend.}
    \label{fig:mean_atoms_comp}
\label{fig:snap_dxa_compression}    
\end{figure}

\begin{figure}[h!]
    \centering
    \begin{subfigure}[b]{0.65\linewidth}
        \centering
        \includegraphics[width=\linewidth]{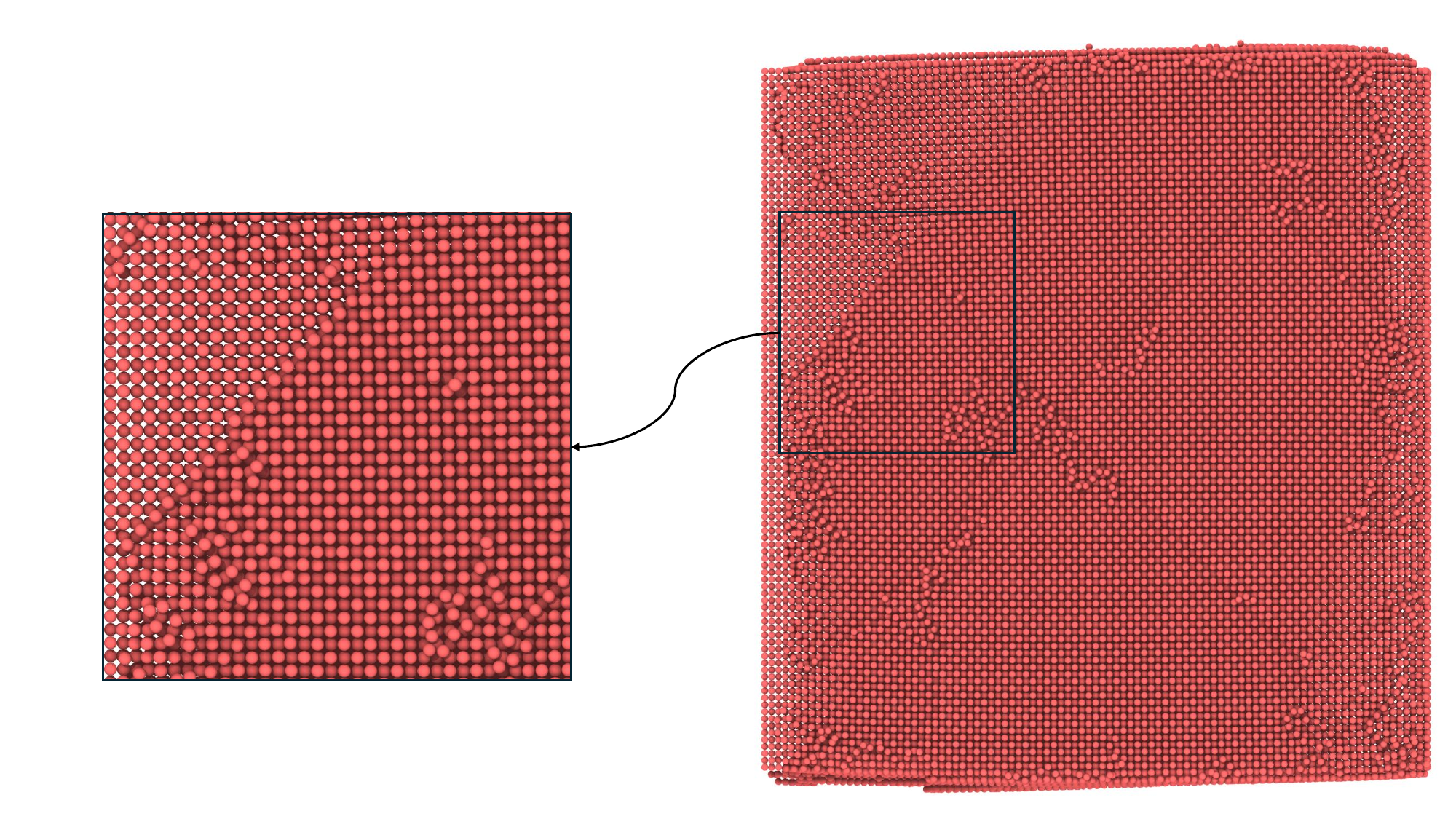}
        \caption{Mean atomic positions at $5 \, \%$ strain for $20 \, nm$ sample under uniaxial compression. A defective zone of the sample is zoomed in to show slip and plastic deformation.}
    \end{subfigure}
    \hfill
    \begin{subfigure}[b]{0.3\linewidth}
        \centering
        \includegraphics[width=\linewidth]{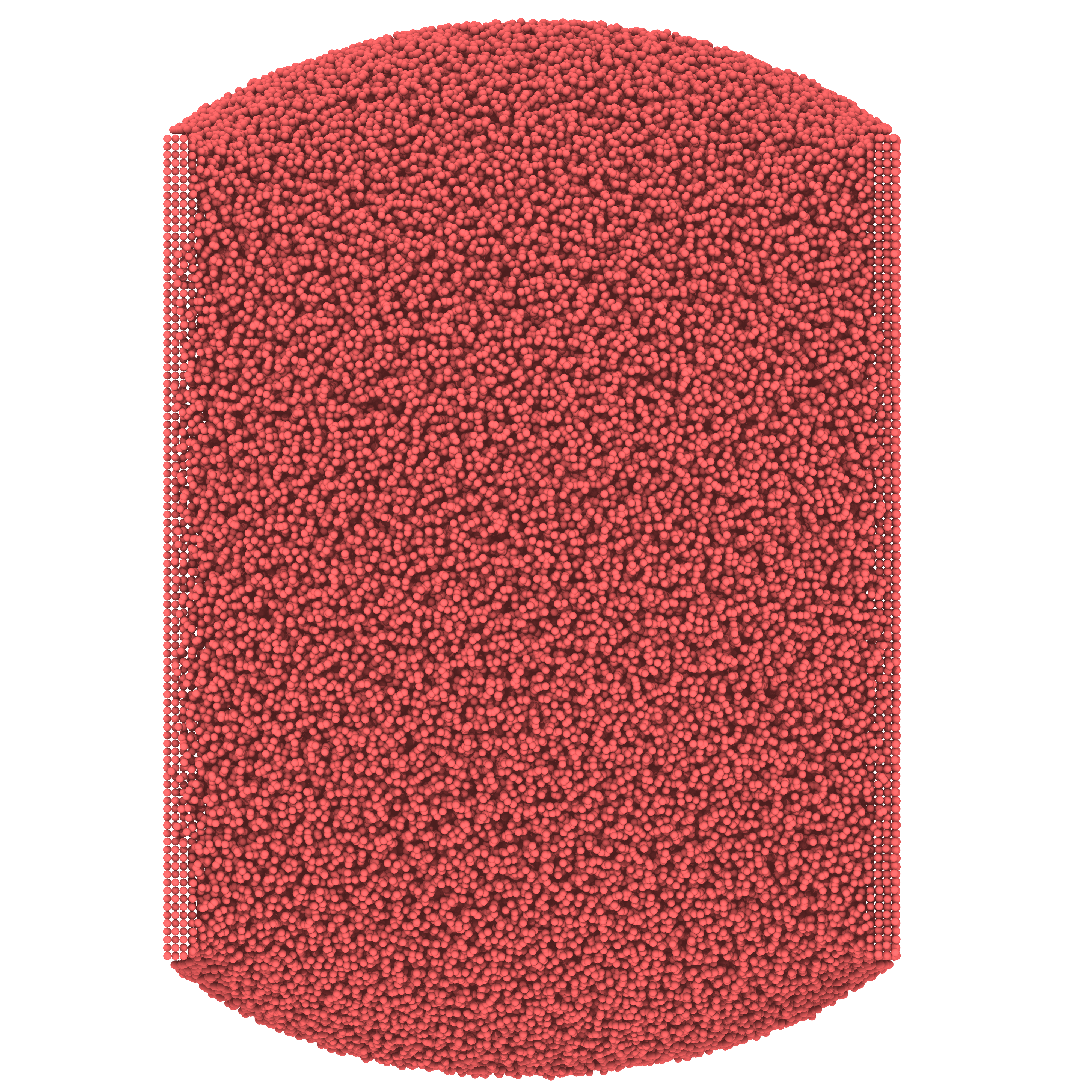}
        \caption{Mean atomic positions at $11 \, \%$ strain for $20 \, nm$ sample under uniaxial compression. We can observe that the sample has liquefied at this strain.}
        \label{fig:comp_liquefy}
    \end{subfigure}
    \caption{Mean atomic positions at $5 \, \%$ and $11 \, \%$ strain for $20 \, nm$ sample under uniaxial compression.}
    \label{fig:atoms_comp}
\end{figure}

\begin{figure}[h!]
    \centering
    \includegraphics[width=0.9\linewidth]{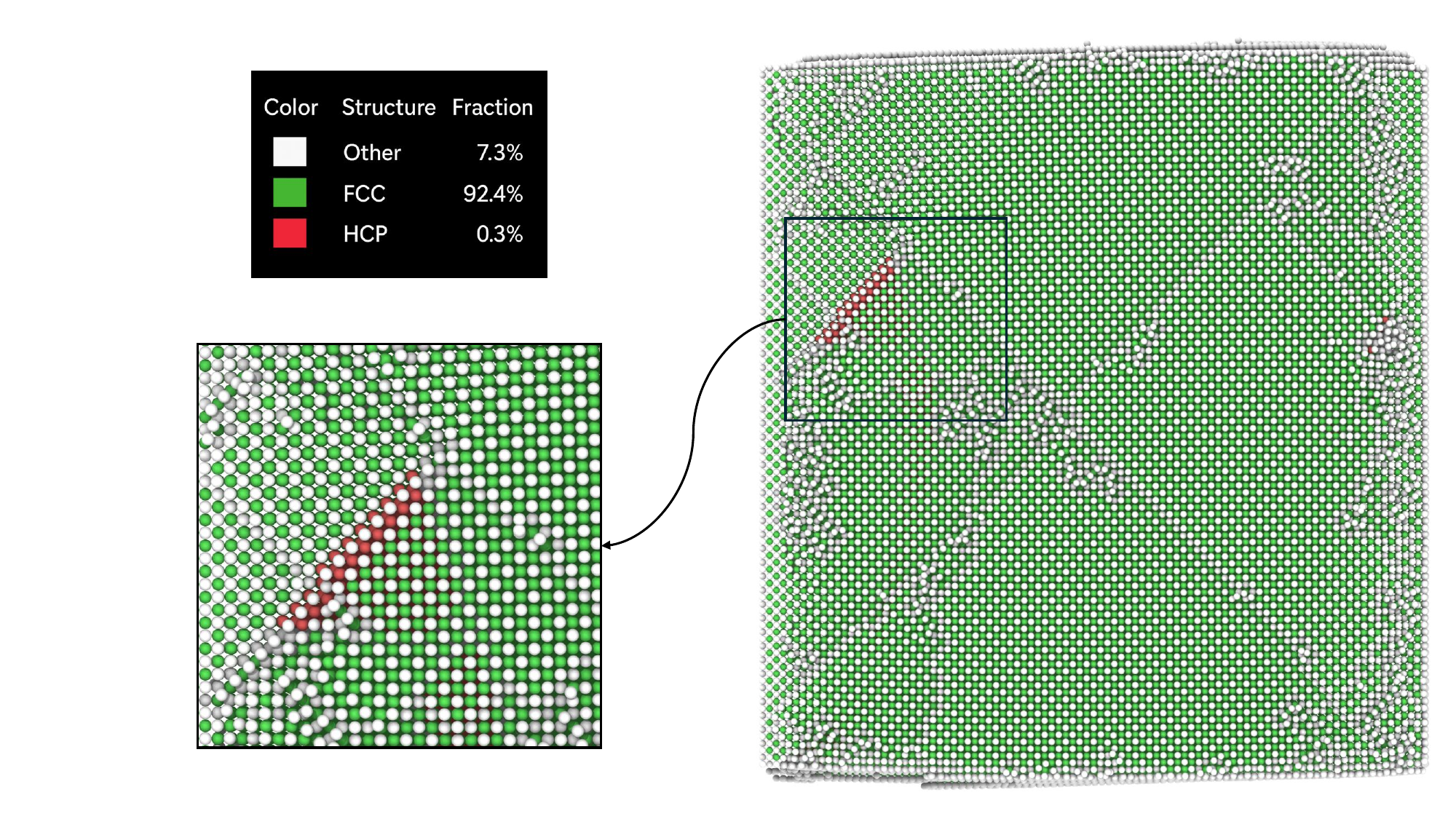}
    \caption{Mean atomic position at $5 \, \%$ strain for $20 \, nm$ sample under uniaxial compression. Atoms are colored as FCC, HCP or others. Only FCC atoms in green are at their regular lattice positions. Atoms which are not in their lattice positions such as the HCP atoms in red and other atoms in white result from slip caused by dislocation motion. A defective zone of the crystal is zoomed in to show the atoms which are out of their lattice positions.}
    \label{fig:atoms_comp_colored}
\end{figure}

\subsection{Speedup in compute time}

The speedup $S$ in computer time between the pure MD and PTA calculations is calculated as follows. Let $T^{CPU}_f$ and $T^{CPU}_{PTA}$ be the compute time to reach strain $\epsilon_f$ and $\epsilon_{PTA}$ using pure MD and PTA, respectively. The speedup $S$ is defined as:
\begin{equation}
\label{eq:speedup}
    S = \frac{\left(\frac{T^{CPU}_f}{\epsilon_f}\right)} {\left(\frac{T^{CPU}_{PTA}}{\epsilon_{PTA}}\right)}. \\
\end{equation}

We have run MD using PTA and pure MD on an $8 \, nm$ sample. The compute time for pure MD ($T^{CPU}_f$) to reach a strain of $6 \times 10^{-11}\%$ strain at an applied strain rate of $10^{-3}\, s^{-1}$ is 43290 seconds. In comparison, it took a compute time ($T^{CPU}_{PTA}$) of 3268 seconds for PTA calculations to reach  slow time $0.5\%$ strain at the same applied strain rate. We used 8 MPI processes and 1 GPU node (the hardware specifications have been provided earlier in Section \ref{sec:MD}) for both the cases. The speed up $S$ in compute time between the the two methods, as defined in Eq.~\eqref{eq:speedup} comes out to be:
\begin{align*}
    S = \frac{\left(\frac{46290}{6 \times 10^{-11}}\right)} {\left(\frac{3268}{0.5}\right)}
      = 1.195 \times 10^{9}.
\end{align*}
This shows that PTA is significantly faster than pure MD for the problem. Note that this is a very conservative estimate of the speedup since MD is run only till very small strain, while starting from a defect-free state. In this regime, the lattice structure is close to that of a perfect crystal and only mild lattice stretching is observed. At higher strains, the dynamics is much more computationally expensive due to generation and evolution of defects. This means the compute time for pure MD is expected to increase even more compared to PTA at higher strains. This makes reaching large strains using pure MD alone practically impossible. However, the increase in compute time for PTA will not be  significant since it involves only running short bursts of MD. 

\subsection{Verification with fine dynamics}
In this section, we demonstrate a verification of the evolution of the slow variable $\textnormal{v}_m$ with PTA using Eq. \eqref{eq:pta_pred} (which we denote as $\textnormal{v}^{PTA}_m$ in this section) by comparing it to the value of the slow variable $\textnormal{v}^{m,fine}_d$ using Eq. \eqref{eq:measure} obtained by running the fine dynamics for the entire slow time interval of the simulation (which we denote as $\textnormal{v}^{m,fine}_d$). Note that $\textnormal{v}^{m,fine}_d$ is different from $\textnormal{v}^m_d$ used in PTA to accept the measure in Step 3 of Section \ref{sec:pta_algo}. The calculation of $\textnormal{v}^{m,fine}_d$ does not involve approximations such as the usage of Simpson's rule in Eq. \eqref{eq:simp} and, more importantly, the fine initial guesses as described in Step 4 of Section \ref{sec:pta_algo}. Hence, the evolution of $\textnormal{v}^{m,fine}_d$ is the exact evolution of the slow variable in the slow time interval that is considered for this verification, corresponding to an MD trajectory. Since the applied strain rate is in the quasi-static regime, we can only reach a very small strain range using the fine dynamics. It is due to the difficulty associated with running MD over large strains that a direct verification with fine dynamics over large strains with PTA results is currently not possible. We consider an $8~nm$ sample and apply uniaxial tensile loading for this verification. We divide the discussion into two parts. In Section~\ref{sec:defect_free}, we load the sample from a defect-free initial state, while in Section~\ref{sec:defective}, we load the sample from a defective initial state. The defect-free initial state corresponds to the pre-yield regime, whereas the defective state corresponds to the post-yield regime. 
We then calculate the relative error between the two using the following:
\begin{equation}
\label{eq:rel_err}
    Error,\textnormal{v}(t) = \left| \frac{\textnormal{v}_m^{PTA}(t) - \textnormal{v}^{m,fine}_d(t)}{\textnormal{v}^{m,fine}_d(t)} \right| .
\end{equation}

The error as defined above includes the component of the error arising due to the fine initial conditions used to calculate $\textnormal{v}^{PTA}_m$ using PTA (as discussed in Step 4 of Section \ref{sec:pta_algo}), since the fine initial condition used in the calculation of $\textnormal{v}^{m,fine}_d$ is exact as it is obtained by running the fine dynamics in the entire slow time interval.

\subsubsection{Defect-free initial-state}
\label{sec:defect_free}
The evolution of the relative errors of the slow variables—namely, the averaged potential energy $\overline{U}^{PTA}$, the averaged kinetic energy $\overline{K}^{PTA}$, and the averaged normal stress $\overline{T}^{PTA}_{x}$—with respect to their corresponding fine-scale values $\overline{U}^{fine}$, $\overline{K}^{fine}$, and $\overline{T}^{fine}_{x}$, calculated using Eq.~\eqref{eq:rel_err}, for the $8~\mathrm{nm}$ sample are summarized in Table~\ref{tab:defect_free}.
For the stress plot shown in Figure~\ref{fig:stress_tension}, we start from a state at $3.1 \, \%$ strain, in the pre-yield regime, and load it under uniaxial tension. We run the fine, molecular dynamics for $2.5 \times 10^7$ steps, in which $5$ jumps in the slow time-scale (denoted as $h$ in Section \ref{sec:pta_algo}) are involved. The number of MD steps and the relative errors between slow variables obtained using $PTA$ and $fine$  dynamics corresponding to each slow jump are reported in Table \ref{tab:defect_free}. We observed that the relative error for the slow variables, averaged potential energy, averaged kinetic energy, and averaged normal stress, are within  $5.87 \times 10^{-3} \, \%$,  $0.78 \, \%$, and $0.27 \, \%$, respectively, at all points. This indicates that the slow evolution with PTA using the guess for the fine initial conditions is in close agreement with that obtained by running the fine dynamics alone, for the strain range that we considered.

\begin{table}[h!]
\centering
\begin{tabular}{|c|c|c|c|}
\hline
$MD \, steps \, (\times 10^7$) & $Error, \overline{U} \, (\times 10^{-3}) \, (\%)$ & $Error, \overline{K} \, (\%)$ & $Error, \overline{T}_x \, (\%)$  \\
\hline
  0.5 & $5.74$ &  $0.77$ & $0.14$ \\
  \hline
  1.0 & $3.35$ &  $0.44$ &  $0.27$ \\
  \hline
  1.5 & $2.94$ &  $0.39$ &  $0.04$ \\
  \hline
  2.0 & $5.87$ &  $0.78$ &  $0.21$ \\
  \hline
  2.5 & $3.29$ &  $0.43$ &  $0.24$ \\
\hline
\end{tabular}
\caption{Evolution of relative errors: $Error, \overline{U}$, $Error, \overline{K}$ and $Error,{\overline{T}_{x}}$ (in \%) with MD steps for $8 \, nm$ sample in uniaxial tension, starting with a defect-free initial-state at $3.1 \, \%$ strain.}
\label{tab:defect_free}
\end{table}

\subsubsection{Defective initial-state}
\label{sec:defective}
In this section, we evolve the same slow variables as discussed in Section~\ref{sec:defect_free}, but we start the simulations with a defective state of the sample. For the stress plot shown in Figure~\ref{fig:stress_tension}, we start from a state at $8.1 \, \%$ strain, in the post-yield regime, and load it under uniaxial tension. Similar to the defect-free case, as discussed in the previous section, there are $4$ jumps in the slow time-scale involved in the $2.5 \times 10^7$ MD steps. The relative errors between slow variables obtained using $PTA$ and $fine$  dynamics at each slow jump are reported in Table \ref{tab:defective}. We observe that the relative error for the slow variables, averaged potential energy, averaged kinetic energy, and averaged normal stress, are within  $20.8 \times 10^{-3} \, \%$,  $0.32 \, \%$ and $0.74\, \%$, respectively, at all points, which indicates that the slow evolution with PTA using the guess for the fine initial conditions is in close agreement with that obtained by running the fine dynamics alone, for the strain range that we considered. 

\begin{table}[h!]
\centering
\begin{tabular}{|c|c|c|c|}
\hline
$MD \, steps \, (\times 10^7$) & $Error, \overline{U} (\times 10^{-3}) \, (\%)$ & $Error, \overline{K} \, (\%)$ & $Error, \overline{T}_x \, (\%)$  \\
\hline
  0.5 & $2.5$ &  $0.19$ & $0.74$ \\
  \hline
  1.0 & $2.8$ &  $0.21$ &  $0.30$ \\
  \hline
  1.5 & $12.5$ &  $0.32$ &  $0.19$ \\
  \hline
  2.0 & $13.6$ &  $0.23$ &  $0.46$ \\
  \hline
  2.5 & $20.8$ &  $0.31$ &  $0.31$ \\
  
\hline
\end{tabular}
\caption{Evolution of relative errors: $Error, \overline{U}$, $Error, \overline{K}$ and $Error,{\overline{T}_{x}}$ (in \%) with MD steps for $8 \, nm$ sample in uniaxial tension, starting with a defective initial-state at $8.1 \, \%$ strain.}
\label{tab:defective}
\end{table}

\subsection{Comparison with experimental results}
\label{sec:exp_uchic}
In this section, we discuss the experimental results of \cite{uchic2004sample} for comparison with our model predictions. They performed uniaxial compression tests on $Ni_{3}Al\text{--}Ta$ micropillars, which were fabricated directly on a bulk crystal using focused-ion beam (FIB) techniques. The samples were prepared in the size range of 0.5 to 20 $\mu$m in diameter and tested using a conventional nanoindentation device equipped with a flat-punch indentation tip. They observed that the stress-strain curve become harder as the sample size decreases. 
We observe similar \emph{smaller is harder} size effect in our simulation results, for both tension (in Fig. \ref{fig:tension_size_effect}) and compression (in Fig. \ref{fig:comp_size_effect}) cases. 
They also observed that increasing the diameter of the micropillars led to a decrease in flow stress. We observe similar inverse-relation between yield strength and sample size under uniaxial tension in Fig. \ref{fig:tension_yield_strength}. The initial elastic slope of the stress–strain curve is not dependent on the sample size, which is consistent with our results for the uniaxial compression case in Fig. \ref{fig:comp_size_effect}. One point to note is that the sample sizes considered in the experiments and in our simulations are around 2-3 orders of magnitude different. In the experimental work of \cite{uchic2004sample}, micron and submicron sized samples were tested while the samples used in our simulations are in the order of a few nanometers upto $20 \, nm$. Thus, the dislocation nucleation mechanisms and the plastic deformation that follows in the two cases are different, which leads to much higher yield stress and harder response in our model predictions. Inspite of these differences, we still observed similar trends related to size effect in our results. 

In order to compare the predictions with experiments conducted on larger samples which are relevant for engineering purposes, one possible approach is to fit the stress-strain data of samples (of size 0.1 micron or above, so that it can capture the operating dislocation mechanisms at those length scales) using our approach to a surrogate material model. This model can then serve as the constitutive model for upscaled Finite Element simulations, with sample sizes (possibly) orders of magnitude larger than that used in MD. However, there are challenges in coupling time-averaged data from microscopic models to macroscale plasticity models, owing to the vast separation in time-scales that govern the microscale and macroscale models. One such approach using time-averaged Discrete Dislocation Dynamics response serving as inputs to a mesoscale plasticity model (and the challenges therewith) is provided in our previous work \cite{chatterjee2020plasticity};  of relevance here is also the recent work \cite{mielke2025deriving}. Although challenging, such coupled multiscale models have the potential of being truly predictive and microstructure sensitive.


\section{Conclusion}

We demonstrate the first application of Practical Time Averaging (PTA) framework to study the macroscopic response of three-dimensional Aluminum samples of sizes upto 20 nanometers, under quasi-static loading rates in uniaxial tension and compression, upto appreciable values of strain. The combination of such small loading rates and large three-dimensional sample sizes is beyond the practical reach of current conventional MD, and remains challenging even for existing accelerated-dynamics and PES-exploration-based approaches, particularly under strain-controlled loading. Our study is not limited to the macroscopic stress-strain response but also to understand the manifestation of the underlying dislocation-mediated plastic deformation on the macroscopic response, such as the effect  of sample size, applied strain rate and initial temperature. We also show comparison of finite temperature response with that of molecular statics. 
Although our results cannot be directly validated with experiments due to the difficulties associated with conducting experiments at such small length scale, the predictions are in qualitative agreement. A summary of the predictions are listed below: 

\begin{enumerate}
    \item Serrations are observed in the stress-strain plots of the $8 \, nm$ block under both uniaxial tension and compression. This behavior is due to the low surface-to-volume ratio of the $8 \, nm$ block. In addition, an increase in internal potential energy is observed with increase in applied strain. This is because work is done on the sample during deformation. During instances where a drop in stress occurs, a sharp increase in kinetic energy is observed. This happens because dislocation nucleation and motion creates lattice waves. Moreover, dislocation nucleation correspond to very fast events which result in jumps even in an averaged quatity like the support of an invariant measures, which consequently result in jumps in the slow variables.

    \item A significant size effect is observed in the tensile properties of the material. Both the yield strength and elastic modulus decrease as the sample size increases from $4 \, nm$ and $30 \, nm$. In this range, the yield strength dropped from $7.05 \, GPa$ for $4 \,nm$ sample to $2.62 \, GPa$ for $30 \, nm$ sample, while the elastic modulus decreased from $101.31 \, GPa$ for $4 \, nm$ sample to $68.39 \, GPa$ for $30 \, nm$ sample. As the sample size increases, the material behavior transitions toward that of bulk aluminum, and the elastic modulus approaches values predicted by continuum models. The observed ``smaller is stronger'' behavior aligns with the experimental and modeling results in the literature for submicron sized samples. 

    \item The overall stress response beyond yielding is lower for the larger $(20 \, nm)$ sample compared to the smaller $(8 \, nm)$ sample. This occurs because, in smaller samples with a higher surface-to-volume ratio, dislocations can escape to the surface more easily, resulting in a higher stress response Moreover, the dislocation sources in smaller samples are limited and they activate at higher stresses, leading to harder response compared to larger samples. Prediction of mechanical behaviour and size effects of nanometer-sized samples at small strain-rates (at the scales employed in this work) is extremely challenging using existent approaches. Our method provides a viable alternative for conducting such studies.
    
    Our approach is also a starting point for understanding behaviour of materials at a larger scale, where the mechanisms we observed may no longer be relevant. But our work in this paper is a test of methodology for the (smaller) scales that we have been able to explore. The algorithm used in our approach will remain unchanged even for larger samples. However, since the number of atoms will be much higher for such samples compared to the sizes we have considered, it will take significantly more computational resources (higher number of cores, efficient hardware architechture, increased wall-clock time) and optimized LAMMPS settings in order to do so.

    \item The statistical variation in stress is smaller for the $20 \, nm$ block compared to the $8 \, nm$ block. The standard deviation of the stress–strain curve was $0.3549 \, GPa$ and $0.2188 \, GPa$ for the $8 \, nm$ and  $20 \, nm$ samples under tension, and $0.4285 \, GPa$ and $0.1087 \, GPa$ under compression, respectively. This indicates that larger samples exhibit a smoother stress response compared to smaller samples.

    \item Plastic deformation due to dislocation nucleation and evolution depends on the applied strain rate as expected. The stress–strain curve corresponding to the higher strain rate remained consistently above that of the lower strain rate along the stress axis. This is because at higher strain rates, to reach the same strain the material has less time for dislocations  to move, thereby leading to higher flow stress. 

    \item The simulations reveal the sensitivity of the system to its initial state of the system. Although the initial temperature of the sample is kept same, changing the initial atomic velocity distribution leads to variations in the stress–strain response, demonstrating the stochasticity in the response. 

    \item Our approach is able to predict the evolution of the microstructure of the sample in slow time by tracking the averaged position of every atom in the  assembly. This allows us to identify the defective regions in the crystal with more plastic deformation. 

    \item We also observe the evolution of the dislocation microstructure at different strains in slow time and are also able to categorize them into different types, like Shockley partials, Stair-rod and full dislocations. 
\end{enumerate}

In order to validate the results with experiments, a possible approach is to fit the stress-strain data from micron-sized samples using our approach into a surrogate material model which will act as constitutive equation for upscaled FEM simulations on larger samples, as mentioned previously. However, such coupled, multiscale approaches are challenging due to the significant separation in time-scales of the fast, microscopic dynamics and that of the mesoscale model under quasi-static strain rates. This will be a topic for future work. 

Another future application of our method is in advanced structural materials like High Entropy Alloys, BCC refractory metal alloys and Mg Alloys, among others, for which traditional Crystal Plasticity Finite Element Method (CPFEM) has resulted in very limited success. The mechanics of dislocation and slip for such materials is complicated, which makes formulation of constitutive flow rules at slow strain rates very challenging. The mechanical response predicted using our approach does not involve phenomenological assumptions and is a direct outcome of the underlying microstructural evolution from time-averaged MD. Hence, it does not suffer from such limitations and may potentially prove to be useful for studying the plastic deformation of such materials. It may be used to predict response of new materials as a function of temperature, orientation and strain-rate from \emph{first principles} and also for understanding complicated dislocation interactions in such materials at slow time scales. However, it is essential that the statistical/representative volume element (RVE) size used in MD to generate the surrogate material model must be of sufficient size (approximately $100nm$ or larger) in order to capture the dislocation microstructure evolution that occurs in the bulk of the material.

\section*{CRediT authorship contribution statement}

\textbf{S.Baruah:} Formal analysis, Investigation, Methodology, Validation, Visualization, Writing – original draft, Writing – review and editing; \textbf{S.Chatterjee:} Conceptualization, Investigation, Methodology, Project administration, Resources, Supervision, Validation, Visualization, Writing – review and editing; \textbf{A.Acharya:} Conceptualization, Investigation, Methodology, Validation, Writing – review and editing; \textbf{G.Wang:} Investigation, Validation, Visualization, Software, Writing – review and editing. 

\section*{Declaration of Competing Interest}
The authors declare that they have no known competing financial interests or personal relationships that could have appeared to
influence the work reported in this paper.

\section*{Data Availability}
Data will be made available on request. 

\section*{Acknowledgements} 

\textbf{S.Baruah} acknowledges the Institute Fellowship received from Indian Institute of Technology Delhi and the High Performance Computing facility at Indian Institute of Technology Delhi. \textbf{S. Chatterjee} acknowledges the New Faculty Seed Grant and Equipment Matching Grant received from Indian Institute of Technology Delhi, the financial support received from the Anusandhan National Research Foundation (ANRF, erstwhile SERB) via grant no. SRG/2022/001328 and the financial support received from the Department of Applied Mechanics at Indian Institute of Technology Delhi. 


\appendix

\section{Convergence criteria for the running time average of state function}\label{app:convergence}
We define $\hat{R}^m_s$ as the running time average of state function $m$ at slow time $s$, given as:
\[
\hat{R}^m_s (i) = \frac{1}{i} \sum_{i=1}^{i} m_i.
\]
The successive values of $m_i$ are obtained by running the fast dynamics with the initial conditions obtained from step 4. 
We denote $R^m_s$ as the converged value of $\hat{R}^m_s$, defined as:
\begin{equation}
R^m_s = \frac{1}{N_s} \sum_{i=1}^{N_s} m_i,
\label{R_m}
\end{equation}
where $N_s$ denotes the number of fine steps required for $\hat{R^m_s}$ to converge upto a specified value of tolerance. The maximum value of $N_s$ is $N_{max}$ ($N_s < N_{max}$). We define the tolerance as
\begin{equation}
\label{eq:R_m_conv}
 e^m (i) = \left| \frac{\hat{R^m_s}(i) - \hat{R^m_s}(i-k)}{\hat{R^m_s}(i)} \right|. 
\end{equation}
Here, $(i)$ and $(i-k)$ are step numbers in fast time. If $e^m(i) < tol_m$ we declare $R^m_s=\hat{R^m_s}(i)$ and $\hat{R^m_s}$ has converged at $N_s=i$. The values of $tol_m$ used for our simulation purposes are defined in Table.~\ref{table_1}. In our problem, we have the state functions as $U(\sigma_i)$ and $R_x(\sigma_i)$. In the following figures, we have shown the plots of the state functions and their running time averages. 
\begin{figure}[H]
    \centering
    \begin{subfigure}{0.49\linewidth}
        \centering
        \includegraphics[width=\linewidth]{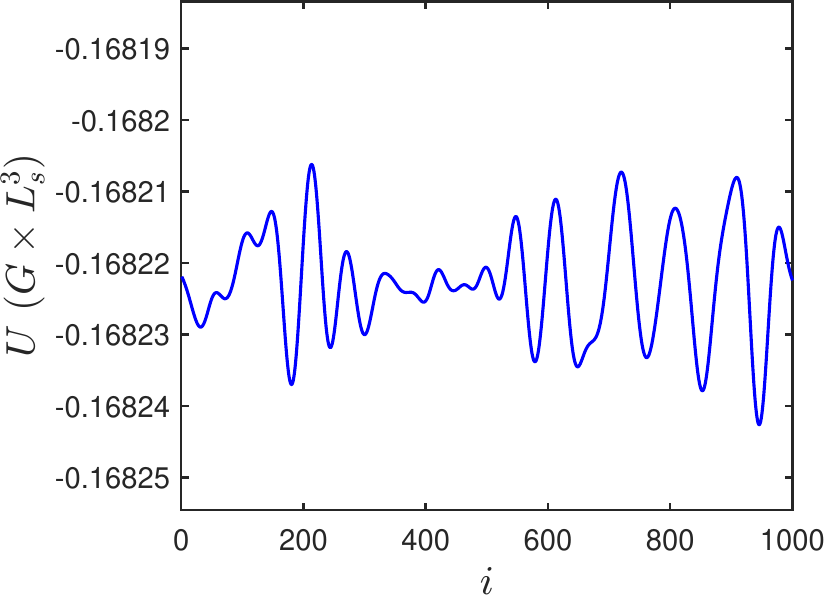}
        \caption{}
    \end{subfigure}
    \hfill
    \begin{subfigure}{0.49\linewidth}
        \centering
        \includegraphics[width=\linewidth]{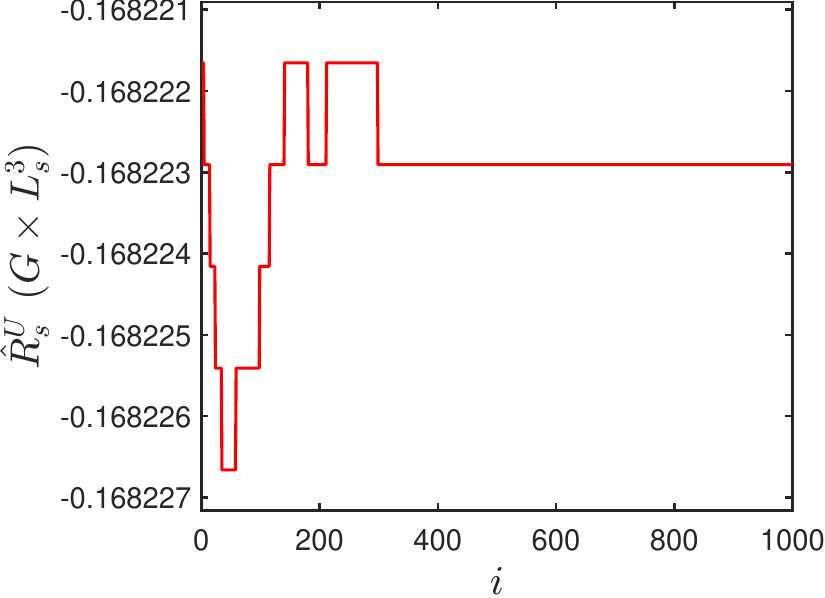}
        \caption{}
    \end{subfigure}
    \caption{Potential energy $(U)$ and its running time average $(\hat{R}^U_s)$.}
\end{figure} 

\begin{figure}[H]
    \centering
    \begin{subfigure}{0.49\linewidth}
        \centering
        \includegraphics[width=\linewidth]{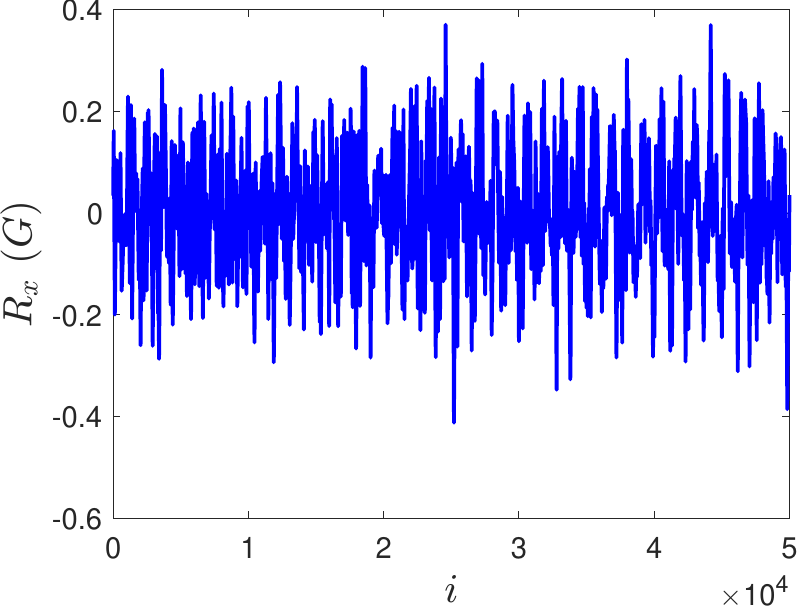}
        \caption{}
    \end{subfigure}
    \hfill
    \begin{subfigure}{0.48\linewidth}
        \centering
        \includegraphics[width=\linewidth]{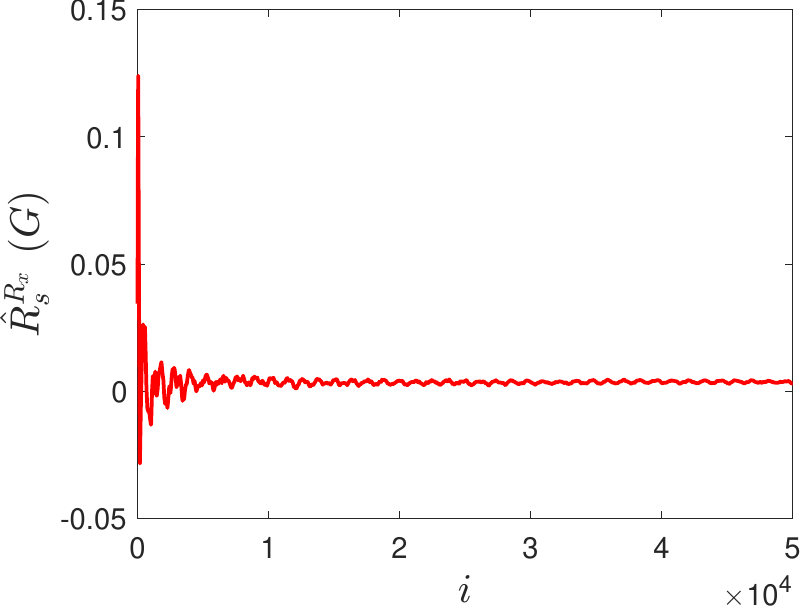}
        \caption{}
    \end{subfigure}
    \caption{Reaction force $(R_x)$ and its running time average $(\hat{R}^{R_x}_s)$.}
\end{figure} 

\bibliography{reference}

@article{Voter1997,
  title = {Hyperdynamics: Accelerated Molecular Dynamics of Infrequent Events},
  volume = {78},
  ISSN = {1079-7114},
  url = {http://dx.doi.org/10.1103/PhysRevLett.78.3908},
  DOI = {10.1103/physrevlett.78.3908},
  number = {20},
  journal = {Physical Review Letters},
  publisher = {American Physical Society (APS)},
  author = {Voter,  Arthur F.},
  year = {1997},
  month = may,
  pages = {3908–3911}
}

@article{Sugita1999,
  title = {Replica-exchange molecular dynamics method for protein folding},
  volume = {314},
  ISSN = {0009-2614},
  url = {http://dx.doi.org/10.1016/S0009-2614(99)01123-9},
  DOI = {10.1016/s0009-2614(99)01123-9},
  number = {1–2},
  journal = {Chemical Physics Letters},
  publisher = {Elsevier BV},
  author = {Sugita,  Yuji and Okamoto,  Yuko},
  year = {1999},
  month = nov,
  pages = {141–151}
}

@article{Sorensen2000,
  title = {Temperature-accelerated dynamics for simulation of infrequent events},
  volume = {112},
  ISSN = {1089-7690},
  url = {http://dx.doi.org/10.1063/1.481576},
  DOI = {10.1063/1.481576},
  number = {21},
  journal = {The Journal of Chemical Physics},
  publisher = {AIP Publishing},
  author = {Sorensen,  Mads R. and Voter,  Arthur F.},
  year = {2000},
  month = jun,
  pages = {9599–9606}
}

@article{AS06,
  title={On a computational approach for the approximate dynamics of averaged variables in nonlinear ODE systems: toward the derivation of constitutive laws of the rate type},
  author={Acharya, Amit and Sawant, Aarti},
  journal={Journal of the Mechanics and Physics of Solids},
  volume={54},
  number={10},
  pages={2183--2213},
  year={2006},
  publisher={Elsevier}
}

@book{bornemann1998homogenization,
  title={Homogenization in time of singularly perturbed mechanical systems},
  author={Bornemann, Folkmar},
  number= {1687},
  volume={Lecture Notes in Mathematics},
  year={1998},
  publisher={Springer-Verlag Berlin Heidelberg}
}

@article{neishtadt2019mechanisms,
  title={On mechanisms of destruction of adiabatic invariance in slow--fast Hamiltonian systems},
  author={Neishtadt, Anatoly},
  journal={Nonlinearity},
  volume={32},
  number={11},
  pages={R53--R76},
  year={2019},
  publisher={IOP Publishing}
}

@article{bornemann1997homogenization,
  title={Homogenization of {H}amiltonian systems with a strong constraining potential},
  author={Bornemann, Folkmar A and Sch{\"u}tte, Christof},
  journal={Physica D: Nonlinear Phenomena},
  volume={102},
  number={1-2},
  pages={57--77},
  year={1997},
  publisher={Elsevier}
}

@article{klar2021second,
  title={Second-order fast--slow dynamics of non-ergodic Hamiltonian systems: Thermodynamic interpretation and simulation},
  author={Klar, Matthias and Matthies, Karsten and Reina, Celia and Zimmer, Johannes},
  journal={Physica D: Nonlinear Phenomena},
  volume={428},
  pages={133036},
  year={2021},
  publisher={Elsevier}
}

@article{mielke2025deriving,
  title={Deriving a GENERIC system from a Hamiltonian system: Mielke, Peletier and Zimmer},
  author={Mielke, Alexander and Peletier, Mark A and Zimmer, Johannes},
  journal={Archive for Rational Mechanics and Analysis},
  volume={249},
  number={5},
  pages={62},
  year={2025},
  publisher={Springer}
}

@article{artstein2002singularly,
  title={On singularly perturbed ordinary differential equations with measure-valued limits},
  author={Artstein, Zvi},
  journal={Proceedings of Equadiff 10},
  pages={15--26},
  year={2002},
  publisher={Masaryk University}
}

@article{Laio2002,
  title = {Escaping free-energy minima},
  volume = {99},
  ISSN = {1091-6490},
  url = {http://dx.doi.org/10.1073/pnas.202427399},
  DOI = {10.1073/pnas.202427399},
  number = {20},
  journal = {Proceedings of the National Academy of Sciences},
  publisher = {Proceedings of the National Academy of Sciences},
  author = {Laio,  Alessandro and Parrinello,  Michele},
  year = {2002},
  month = sep,
  pages = {12562–12566}
}

@article{Izvekov2006,
  title = {Modeling real dynamics in the coarse-grained representation of condensed phase systems},
  volume = {125},
  ISSN = {1089-7690},
  url = {http://dx.doi.org/10.1063/1.2360580},
  DOI = {10.1063/1.2360580},
  number = {15},
  journal = {The Journal of Chemical Physics},
  publisher = {AIP Publishing},
  author = {Izvekov,  Sergei and Voth,  Gregory A.},
  year = {2006},
  month = oct 
}

@article{Hijn2010,
  title = {Mori–Zwanzig formalism as a practical computational tool},
  volume = {144},
  ISSN = {1364-5498},
  url = {http://dx.doi.org/10.1039/B902479B},
  DOI = {10.1039/b902479b},
  journal = {Faraday Discuss.},
  publisher = {Royal Society of Chemistry (RSC)},
  author = {Hijón,  Carmen and Español,  Pep and Vanden-Eijnden,  Eric and Delgado-Buscalioni,  Rafael},
  year = {2010},
  pages = {301–322}
}

@article{Miao2015,
  title = {Gaussian Accelerated Molecular Dynamics: Unconstrained Enhanced Sampling and Free Energy Calculation},
  volume = {11},
  ISSN = {1549-9626},
  url = {http://dx.doi.org/10.1021/acs.jctc.5b00436},
  DOI = {10.1021/acs.jctc.5b00436},
  number = {8},
  journal = {Journal of Chemical Theory and Computation},
  publisher = {American Chemical Society (ACS)},
  author = {Miao,  Yinglong and Feher,  Victoria A. and McCammon,  J. Andrew},
  year = {2015},
  month = jul,
  pages = {3584–3595}
}

@article{Zamora2016,
  title = {The Modern Temperature-Accelerated Dynamics Approach},
  volume = {7},
  ISSN = {1947-5446},
  url = {http://dx.doi.org/10.1146/annurev-chembioeng-080615-033608},
  DOI = {10.1146/annurev-chembioeng-080615-033608},
  number = {1},
  journal = {Annual Review of Chemical and Biomolecular Engineering},
  publisher = {Annual Reviews},
  author = {Zamora,  Richard J. and Uberuaga,  Blas P. and Perez,  Danny and Voter,  Arthur F.},
  year = {2016},
  month = jun,
  pages = {87–110}
}

@article{Bolhuis2021,
  title = {Transition Path Sampling as {M}arkov {C}hain {M}onte {C}arlo of Trajectories: Recent Algorithms,  Software,  Applications,  and Future Outlook},
  volume = {4},
  ISSN = {2513-0390},
  url = {http://dx.doi.org/10.1002/adts.202000237},
  DOI = {10.1002/adts.202000237},
  number = {4},
  journal = {Advanced Theory and Simulations},
  publisher = {Wiley},
  author = {Bolhuis,  Peter G. and Swenson,  David W. H.},
  year = {2021},
  month = mar 
}

@article{Shi2023,
  title = {Coarse‐grained molecular dynamics simulation of polymers: Structures and dynamics},
  volume = {13},
  ISSN = {1759-0884},
  url = {http://dx.doi.org/10.1002/wcms.1683},
  DOI = {10.1002/wcms.1683},
  number = {6},
  journal = {WIREs Computational Molecular Science},
  publisher = {Wiley},
  author = {Shi,  Rui and Qian,  Hu‐Jun and Lu,  Zhong‐Yuan},
  year = {2023},
  month = aug 
}

@article{Mohr2024,
  title = {Enhanced sampling strategies for molecular simulation of <scp>DNA</scp>},
  volume = {14},
  ISSN = {1759-0884},
  url = {http://dx.doi.org/10.1002/wcms.1712},
  DOI = {10.1002/wcms.1712},
  number = {2},
  journal = {WIREs Computational Molecular Science},
  publisher = {Wiley},
  author = {Mohr,  Bernadette and van Heesch,  Thor and Pérez de Alba Ortíz,  Alberto and Vreede,  Jocelyne},
  year = {2024},
  month = mar 
}

@article{Noid2024,
  title = {Rigorous Progress in Coarse-Graining},
  volume = {75},
  ISSN = {1545-1593},
  url = {http://dx.doi.org/10.1146/annurev-physchem-062123-010821},
  DOI = {10.1146/annurev-physchem-062123-010821},
  number = {1},
  journal = {Annual Review of Physical Chemistry},
  publisher = {Annual Reviews},
  author = {Noid,  W.G. and Szukalo,  Ryan J. and Kidder,  Katherine M. and Lesniewski,  Maria C.},
  year = {2024},
  month = jun,
  pages = {21–45}
}

@article{Berezhkovskii2011,
  title = {Time scale separation leads to position-dependent diffusion along a slow coordinate},
  volume = {135},
  ISSN = {1089-7690},
  url = {http://dx.doi.org/10.1063/1.3626215},
  DOI = {10.1063/1.3626215},
  number = {7},
  journal = {The Journal of Chemical Physics},
  publisher = {AIP Publishing},
  author = {Berezhkovskii,  Alexander and Szabo,  Attila},
  year = {2011},
  month = aug 
}

@article{kabir2024size,
  author    = {Kabir, Hadi and Aghdam, Mohammad Mohammadi and Samandari, Saeed Saber and Moeini, Mohsen},
  title     = {A molecular dynamics study on the size effects of {Fe\textsubscript{3}O\textsubscript{4}} nanoparticles on the mechanical characteristics of polypyrrole/{Fe\textsubscript{3}O\textsubscript{4}} nanocomposite},
  journal   = {Molecular Simulation},
  volume    = {50},
  number    = {7-9},
  pages     = {493--505},
  year      = {2024}
}

@article{vogl2021effect,
  title={Effect of size and shape on the elastic modulus of metal nanowires},
  author={Vogl, Lilian Maria and Schweizer, Peter and Richter, Gunther and Spiecker, Erdmann},
  journal={MRS Advances},
  volume={6},
  number={27},
  pages={665--673},
  year={2021},
  publisher={Springer}
}

@article{yu2013study,
  title={Study on Size-Dependent Young’s Modulus of a Silicon Nanobeam by Molecular Dynamics Simulation},
  author={Yu, H and Sun, C and Zhang, WW and Lei, SY and Huang, QA},
  journal={Journal of Nanomaterials},
  volume={2013},
  number={1},
  pages={319302},
  year={2013},
  publisher={Wiley Online Library}
}

@article{wan2021size,
  title={The size effects of point defect on the mechanical properties of monocrystalline silicon: A molecular dynamics study},
  author={Wan, Wei and Tang, Changxin and Qiu, An and Xiang, Yongkang},
  journal={Materials},
  volume={14},
  number={11},
  pages={3011},
  year={2021},
  publisher={MDPI}
}

@article{van2003quantifying,
  title={Quantifying the early stages of plasticity through nanoscale experiments and simulations},
  author={Van Vliet, Krystyn J and Li, Ju and Zhu, Ting and Yip, Sidney and Suresh, Subra},
  journal={Physical Review B},
  volume={67},
  number={10},
  pages={104105},
  year={2003},
  publisher={APS}
}

@article{uchic2004sample,
  title={Sample dimensions influence strength and crystal plasticity},
  author={Uchic, Michael D and Dimiduk, Dennis M and Florando, Jeffrey N and Nix, William D},
  journal={Science},
  volume={305},
  number={5686},
  pages={986--989},
  year={2004},
  publisher={American Association for the Advancement of Science}
}

@article{CHANG2017348,
title = {Molecular dynamics study of strain rate effects on tensile behavior of single crystal titanium nanowire},
journal = {Computational Materials Science},
volume = {128},
pages = {348-358},
year = {2017},
issn = {0927-0256},
author = {Le Chang and Chang-Yu Zhou and Lei-Lei Wen and Jian Li and Xiao-Hua He}
}

@article{PhysRevB.59.3393,
  title = {Interatomic potentials for monoatomic metals from experimental data and ab initio calculations},
  author = {Mishin, Y. and Farkas, D. and Mehl, M. J. and Papaconstantopoulos, D. A.},
  journal = {Phys. Rev. B},
  volume = {59},
  issue = {5},
  pages = {3393--3407},
  numpages = {0},
  year = {1999},
  month = {Feb},
  publisher = {American Physical Society},
}

@article{chatterjee2018computing,
  title={Computing singularly perturbed differential equations},
  author={Chatterjee, Sabyasachi and Acharya, Amit and Artstein, Zvi},
  journal={Journal of Computational Physics},
  volume={354},
  pages={417--446},
  year={2018},
  publisher={Elsevier}
}

@article{chatterjee2020plasticity,
  title={Plasticity without phenomenology: a first step},
  author={Chatterjee, Sabyasachi and Po, Giacomo and Zhang, Xiaohan and Acharya, Amit and Ghoniem, Nasr},
  journal={Journal of the Mechanics and Physics of Solids},
  volume={143},
  pages={104059},
  year={2020},
  publisher={Elsevier}
}

@article{tan2014modeling,
  title={Modeling of slow time-scale behavior of fast molecular dynamic systems},
  author={Tan, Likun and Acharya, Amit and Dayal, Kaushik},
  journal={Journal of the Mechanics and Physics of Solids},
  volume={64},
  pages={24--43},
  year={2014},
  publisher={Elsevier}
}

@article{komanduri2001molecular,
  title={Molecular dynamics (MD) simulation of uniaxial tension of some single-crystal cubic metals at nanolevel},
  author={Komanduri, R and Chandrasekaran, N and Raff, LM},
  journal={International Journal of Mechanical Sciences},
  volume={43},
  number={10},
  pages={2237--2260},
  year={2001},
  publisher={Elsevier}
}

@article{slemrod2012time,
  title={Time-averaged coarse variables for multi-scale dynamics},
  author={Slemrod, Marshall and Acharya, Amit},
  journal={Quarterly of applied mathematics},
  volume={70},
  number={4},
  pages={793--803},
  year={2012}
}

@article{ovito,
Author = {Stukowski, Alexander},
Title = {{Visualization and analysis of atomistic simulation data with OVITO-the
   Open Visualization Tool}},
Journal = {{Modelling and Simulation in Materials Science and Engineering}},
Year = {{2010}},
Volume = {{18}},
Number = {{1}},
pages={{015012}},
DOI = {{10.1088/0965-0393/18/1/015012}},
Article-Number = {{015012}},
ISSN = {{0965-0393}},
EISSN = {{1361-651X}},
ResearcherID-Numbers = {{Stukowski, Alexander/G-9695-2017}},
ORCID-Numbers = {{Stukowski, Alexander/0000-0001-6750-3401}},
Unique-ID = {{ISI:000272791800012}},
}

@article{roy_acharya_2005,
title = {Finite element approximation of field dislocation mechanics},
journal = {Journal of the Mechanics and Physics of Solids},
volume = {53},
number = {1},
pages = {143-170},
year = {2005},
issn = {0022-5096},
doi = {https://doi.org/10.1016/j.jmps.2004.05.007},
url = {https://www.sciencedirect.com/science/article/pii/S0022509604001097},
author = {Anish Roy and Amit Acharya}
}

@article{acharya_roy_2006_1,
title = {Size effects and idealized dislocation microstructure at small scales: Predictions of a Phenomenological model of Mesoscopic Field Dislocation Mechanics: Part I},
journal = {Journal of the Mechanics and Physics of Solids},
volume = {54},
number = {8},
pages = {1687-1710},
year = {2006},
issn = {0022-5096},
doi = {https://doi.org/10.1016/j.jmps.2006.01.009},
url = {https://www.sciencedirect.com/science/article/pii/S0022509606000238},
author = {Amit Acharya and Anish Roy}}

@article{roy_acharya_2006_2,
title = {Size effects and idealized dislocation microstructure at small scales: Predictions of a Phenomenological model of Mesoscopic Field Dislocation Mechanics: Part II},
journal = {Journal of the Mechanics and Physics of Solids},
volume = {54},
number = {8},
pages = {1711-1743},
year = {2006},
issn = {0022-5096},
doi = {https://doi.org/10.1016/j.jmps.2006.01.012},
url = {https://www.sciencedirect.com/science/article/pii/S0022509606000226},
author = {Anish Roy and Amit Acharya}}

@article{rarora_2020_ijss,
title = {Dislocation pattern formation in finite deformation crystal plasticity},
journal = {International Journal of Solids and Structures},
volume = {184},
pages = {114-135},
year = {2020},
issn = {0020-7683},
doi = {https://doi.org/10.1016/j.ijsolstr.2019.02.013},
url = {https://www.sciencedirect.com/science/article/pii/S0020768319300927},
author = {Rajat Arora and Amit Acharya}}

@article{rarora_2020_jmps,
title = {A unification of finite deformation J2 Von-Mises plasticity and quantitative dislocation mechanics},
journal = {Journal of the Mechanics and Physics of Solids},
volume = {143},
pages = {104050},
year = {2020},
issn = {0022-5096},
doi = {https://doi.org/10.1016/j.jmps.2020.104050},
url = {https://www.sciencedirect.com/science/article/pii/S0022509620302830},
author = {Rajat Arora and Amit Acharya}}

@article{frenkel1926theorie,
  title={Zur theorie der elastizit{\"a}tsgrenze und der festigkeit kristallinischer k{\"o}rper},
  author={Frenkel, JA},
  journal={Zeitschrift f{\"u}r Physik},
  volume={37},
  number={7},
  pages={572--609},
  year={1926},
  publisher={Springer}
}

@manual{lammps_nve,
  title        = {LAMMPS Documentation: fix nve},
  author       = {{LAMMPS Developers}},
  organization = {Sandia National Laboratories},
  year         = {2024},
  note         = {Velocity-Verlet time integration},
  url          = {https://docs.lammps.org/fix_nve.html},
  urldate      = {2026-01-09}
}

@manual{lammps_verlet,
  title        = {LAMMPS Documentation: run\_style verlet},
  author       = {{LAMMPS Developers}},
  organization = {Sandia National Laboratories},
  year         = {2024},
  note         = {Velocity form of the St{\"o}rmer--Verlet integrator},
  url          = {https://docs.lammps.org/run_style.html},
  urldate      = {2026-01-09}
}

@article{Kushima2009,
  author  = {Kushima, Akihiro and Lin, Xi and Li, Ju and Eapen, Jacob and Mauro, John C. and Qian, Xiaofeng and Diep, Phong and Yip, Sidney},
  title   = {Computing the viscosity of supercooled liquids},
  journal = {The Journal of Chemical Physics},
  volume  = {130},
  number  = {22},
  pages   = {224504},
  year    = {2009},
  doi     = {10.1063/1.3139006}
}

@article{Yan2016Review,
  author  = {Yan, Xin and Cao, Penghui and Tao, Weiwei and Sharma, Pradeep and Park, Harold S.},
  title   = {Atomistic modeling at experimental strain rates and timescales},
  journal = {Journal of Physics D: Applied Physics},
  volume  = {49},
  number  = {49},
  pages   = {493002},
  year    = {2016},
  doi     = {10.1088/0022-3727/49/49/493002}
}

@article{Fan2013PNAS,
  author  = {Fan, Yue and Osetskiy, Yuri N. and Yip, Sidney and Yildiz, Bilge},
  title   = {Mapping strain rate dependence of dislocation-defect interactions by atomistic simulations},
  journal = {Proceedings of the National Academy of Sciences},
  volume  = {110},
  number  = {44},
  pages   = {17756--17761},
  year    = {2013},
  doi     = {10.1073/pnas.1310036110}
}

@article{YanSharma2016,
  author  = {Yan, Xin and Sharma, Pradeep},
  title   = {Time-Scaling in Atomistics and the Rate-Dependent Mechanical Behavior of Nanostructures},
  journal = {Nano Letters},
  volume  = {16},
  number  = {6},
  pages   = {3487--3492},
  year    = {2016},
  doi     = {10.1021/acs.nanolett.6b00117}
}

@article{Barkema1996,
  author  = {Barkema, G. T. and Mousseau, Normand},
  title   = {Event-Based Relaxation of Continuous Disordered Systems},
  journal = {Physical Review Letters},
  volume  = {77},
  number  = {21},
  pages   = {4358--4361},
  year    = {1996},
  doi     = {10.1103/PhysRevLett.77.4358}
}

@article{Henkelman1999,
  author  = {Henkelman, Graeme and J{\'o}nsson, Hannes},
  title   = {A dimer method for finding saddle points on high dimensional potential surfaces using only first derivatives},
  journal = {The Journal of Chemical Physics},
  volume  = {111},
  number  = {15},
  pages   = {7010--7022},
  year    = {1999},
  doi     = {10.1063/1.480097}
}
\bibliographystyle{apalike}

\end{document}